\documentclass[nofootinbib,prd,aps,showpacs,superscriptaddress]{revtex4-2}
\usepackage[utf8]{inputenc}
\usepackage[T1]{fontenc}
\usepackage{lmodern}
\usepackage{graphicx}
\usepackage{enumerate}
\usepackage{titlesec, blindtext, color}
\usepackage{hyperref}
\usepackage{amsmath}
\usepackage{amssymb}
\usepackage{amsthm}
\usepackage{ulem}
\usepackage{bm}
\usepackage {tensor}
\usepackage{multirow}
\usepackage{etoolbox}
\usepackage{enumitem}
\usepackage[english]{babel}
\usepackage{dsfont}
\usepackage{float}
\usepackage{slashed,braket}
\usepackage{stackengine}
\usepackage{mathrsfs}
\usepackage{verbatim}
\usepackage{array}
\usepackage{bbold}
\usepackage{upgreek}

\newcommand{\beq}{\begin{equation}}
\newcommand{\eeq}{\end{equation}}
\newcommand{\beqa}{\begin{eqnarray}}
\newcommand{\eeqa}{\end{eqnarray}}

\newcommand{\Tr}{\operatorname{Tr}}

\newcommand{\mz}[1]{{#1}}

\newcommand{\mzc}[1]{{#1}}

\begin{document}

\title{Effective Lagrangian for the macroscopic motion of Weyl fermions in $^3$He-A}

\author{M. Selch}
\email{maik.selch@t-online.de}
\affiliation{Physics Department, Ariel University, Ariel 40700, Israel}

%\author{Ruslan A.~Abramchuk}
%\email{abramchuk@phystech.edu}
%\affiliation{Physics Department, Ariel University, Ariel 40700, Israel}
%\affiliation{On leave of absence from Kurchatov Complex for Theoretical and Experimental Physics, B. Cheremushkinskaya 25, Moscow, 117259, Russia}

\author{M. A. Zubkov}
\email{mikhailzu@ariel.ac.il}
\affiliation{Physics Department, Ariel University, Ariel 40700, Israel}

\date{\today}

\begin{abstract}
We consider macroscopic motion of the normal component of superfluid $^3$He - A in global thermodynamic equilibrium within the context of the Zubarev statistical operator method. We formulate the corresponding effective theory in the language of the functional integral. The effective Lagrangian comprising macroscopic motion of fermionic excitations is calculated explicitly for the emergent relativistic fermions of the superfluid $^3$He - A phase immersed in a non-trivial bosonic background due to a space and time dependent matrix-valued vierbein featuring nonzero torsion as well as the Nieh-Yan anomaly. \mz{We \mzc{do not consider} the dynamics of the superfluid component itself and thereby its backreaction effects due to normal component macroscopic flow. It is being treated as an external background within which the emergent relativistic fermions of the normal component move. The matrix-valued vierbein formulation comprises an additional two dimensional internal spin space for the two axially charged Weyl fermions living at the Fermi points which may be replaced by one featuring a Dirac fermion doublet with a real valued vierbein, an axial Abelian gauge field and a spin connection gauge field mixing the Dirac and internal spin spaces. We carry out this change of description in detail and determine the constraints on the superfluid background as well as the the normal component motion as determined from the Zubarev statistical operator formalism in global thermodynamic equilibrium.} As an application of the developed theory we consider macroscopic rotation around the axis of pure integer mass vortices. The corresponding thermodynamic quantities of the normal component are analyzed. \mz{Our formulation incorporates both superfluid background flow and macroscopic motion flow of the normal component and thereby enables an analysis of their interrelation.}\newline\newline

\mz{\large{Keywords}: \normalsize{Superfluidity in $^3$He, Emergent relativistic invariance, Zubarev statistical operator}}
\end{abstract}


\maketitle

\tableofcontents

\section{Introduction}

{At low temperatures and appropriate external pressure the 3D Fermi liquid $^3$He undergoes a phase transition to the superfluid phase. The superfluid region is bipartite featuring the so - called $^3$He - A and $^3$He - B phases (see \cite{Volovik2003,VollhardtWolfle1990} and references therein). \mz{The superfluid phases represent a flow of two coupled components, the so-called superfluid and normal components. While the $^3$He - A phase is gapless, the $^3$He - B phase exhibits a finite energy gap in the normal component. The superfluid component hosts a number of Nambu - Goldstone bosons originating from spontaneously broken symmetries in the superfluid phase.}
One of the most salient properties of the $^3$He - A superfluid is that it allows for the simulation of phenomena associated with high energy physics in the laboratory. Namely, emergent relativistic Weyl fermions \cite{VZ2014} appear locally around the two Fermi points of the (time reversal breaking) $^3$He - A phase representing the normal component. They behave similar to elementary particles. \mz{The superfluid component is due to the order parameter dynamics. It features an emergent matrix-valued vierbein and a chiral gauge field which are minimally coupled to the Weyl fermions. \cite{Volovik2003,VolovikVachaspati1996,VolovikKhazan1982,Volovik1990}. The physics of the $A$-phase is reminiscent of standard model physics above the electroweak scale.}
Collective modes in the $^3$He - B superfluid are similar to the Higgs modes of particle physics \cite{Nambu1985} \mz{which appear below the eletroweak scale.} 
%The same refers also to such modes in $^3$He - A, and even to the 2D phases of $^3$He. 
A Nambu sum rule exists that relates the energy gaps of bosonic modes to the fermion "gap" \cite{VZ2015,VolovikZubkov2014,VolovikZubkovHiggs}. 
%The Nambu sum rule in a similar form also exists in relativistic Nambu - Jona - Lasinio models \cite{VolovikZubkovHiggs}.
}

{The behavior of superfluid helium in the presence of macroscopic motion is of great interest both in condensed matter and high energy physics. Superfluid $^3$He - A simulates, to a certain extent, the quark - gluon plasma in the presence of macroscopic motion, which is relevant for the physics of heavy ion collisions, and for certain astrophysical applications (say, if we are speaking of the description of matter inside neutron stars). Both $^3$He - A and quantum chromodynamics feature Weyl or Dirac fermions coupled to (emergent) (non-) Abelian gauge and/or vielbein fields subject to a host of anomalous transport phenomena. \mzc{It is well known that responses to external fields may be related to quantum field theory anomalies \cite{Z2016_1,ZZ2019_0,ZZ2021,ZW2019,SZZ2020}.} Of recent interest within $^3$He - A is the so - called Nieh Yan anomaly \cite{NiehYan1982} which appears as a result of torsion of the emergent vielbein field
\cite{HuangHanStone2020,KhaidukovZubkov2018,NissinenVolovik2019,NissinenVolovik2020,Nissinen2020,LaurilaNissinen2020}.}	

{In the present paper we focus on the $^3$He - A superfluid phase and analyse the dynamics of the emergent Weyl fermions in the presence of macroscopic motion based on the Zubarev statistical operator method \cite{Zubarev1979} in the regime of emergent relativistic invariance. We allow for a nontrivial flow of the superfluid component as well, but restrict attention to the case without superfluid dynamics. In this approximation the superfluid component is treated as an inert background above which the normal component moves. We appeal to the path integral formulation. \mz{Earlier we applied the machinery used to the analysis of the quark - gluon plasma which is a non-confined but still strongly interacting phase within the theory of quantum chromodynamics \cite{AZZ2023}. The Zubarev statistical operator method has recently found extensive use in the description of the physics of the quark - gluon plasma produced during heavy ion collisions. We aim to extend the scope of application of this method to $^3$He - A and therefore discuss the benefits of it in view of its successful application within the context of the quark - gluon plasma.}}

{The functional integral representation of the BCS theory of $^3$He is widely used in condensed matter physics in order to describe superfluid phases (see, e. g., \cite{He3,He3gauss,He3B,BrusovPopov1980}). This approach has been summarized in \cite{Brusovs}.}

{In Zubarev's approach the macroscopic motion and varying temperature are encoded in the so-called frigidity vector field, and in chemical potentials varying in space and time. The Zubarev operator allows to deal with non-equilibrium systems as well. However, in the present paper we are interested in the description of $^3$He - A in global thermodynamic equilibrium \cite{Zubarev1979,Buzzegoli2017cqy,Becattini2020qol,Buzzegoli2020ycf}, or for the consideration of the same system in the hydrodynamic approximation, where it remains in quasi - equilibrium locally. Thermodynamic equilibrium imposes restrictive conditions on the type of macroscopic motion permitted. These types include motion with constant uniform velocity, rotation, a certain type of uniform accelerated motion, and combinations of these. The corresponding frigidity vector field of macroscopic motion is parametrized by a constant vector field and a constant anti-symmetric tensor field of thermal vorticity. The axial part of the vorticity tensor is proportional to the angular velocity, while the polar part is proportional to linear uniform acceleration.}

{The frigidity four vector field has been extracted from a simulation of the quark - gluon plasma that appears during heavy ion collisions (see, e.g., \cite{Bravina2021arj}). The same might be done also for the $^3$He - A superfluid. In each grain of the substance the motion can be considered as being in quasi - equilibrium with macroscopic motion that consists of straight uniform motion with constant velocity, rotation, and uniformly accelerated motion. Especially interesting is the description of quasiparticle dynamics in the presence of the macroscopic motion associated with the superfluid velocity existing in the presence of various vortices.} 

{\mz{Linear uniform motion has proven to be of use in the discussion of physics of quantum Hall fluids \cite{selch2025nonrenormalizationfractionalquantumhall}. The so-called Hall conductivity may be found to be topological under several circumstances. The quantum Hall effect features a current orthogonal to an external magnetic field and an external electric field in the laboratory frame. In the boosted Hall fluid frame comoving with the current the external electric field vanishes and the analysis of the topological response is simplified.}}

%{It is instructive to compare the effective action obtained using our approach and the effective action for the rotating system that was derived based on the transition to the rotating reference frame. The latter action has been considered by several authors, and was used for lattice Monte-Carlo simulations of QCD  \cite{Yamamoto2013zwa,Braguta2020biu,Braguta2021jgn,Chernodub2022veq,Braguta2023kwl,Braguta2023yjn,Landsteiner2020xuv}. In lattice simulations the imaginary angular velocity was used, and analytical continuation to real values was performed. Then the vacuum of the system is rotated. A non-trivial metric was also introduced for various types of macroscopic motion \cite{H1,H2,Chernodub2021nff}. Our approach to rotation is to define the Zubarev statistical operator of a rotating system \cite{Zubarev1979}. Notice that in \cite{VilenkinCVE} rigid rotation has been considered within a similar approach, see also  \cite{Landsteiner2018era,Landsteiner2020xuv}. In this approach the vacuum remains at rest while the excitations above the vacuum are rotated. For the duality between these two descriptions of rotation see also \cite{Stone1999gi}. It is known that the rotating vacuum is identical to the static vacuum for free Dirac fermions \cite{Abramchuk2018jhd}. In \cite{AZZ2023} it was demonstrated that the effective action for the gauge field interacting with fermions is the same as when it is obtained using the two different approaches for the description of rotation mentioned above (see also \cite{Yamamoto2013zwa}).}

{\mz{A rotating fireball is produced during non-central heavy ion collisions. Quantum chromodynamics under rotation in these conditions has been considered by several authors, and was used for lattice Monte-Carlo simulations of QCD  \cite{Yamamoto2013zwa,Braguta2020biu,Braguta2021jgn,Chernodub2022veq,Braguta2023kwl,Braguta2023yjn,Landsteiner2020xuv}.}}

{The third type of equilibrium macroscopic motion is linear uniform acceleration at the initial moment. At later stages of motion the acceleration is not kept constant, but depends on time in a certain way. This kind of motion has been investigated analytically (see, for example, \cite{Prokhorov2023dfg,Khakimov2023emy}). Equilibrium quantum chromodynamics with such kind of macroscopic motion has been investigated recently using lattice simulations \cite{Chernodubconf}.}

{\mz{Our work is organized as follows. We begin with a review of the basics of the phase transition of the Fermi liquid $^3$He to superfluid $^3$He - A in section \ref{reviewon3hea}. We proceed with the standard formalism to parametrize the emergent relativistic fermionic action in section \ref{emergentphysics} followed by a reparametrization. We consider this reparametrization useful as it is both not commonly employed and it features a universal vierbein coupled to the emergent relativistic Weyl fermions as well as combines the entire spin dynamics into an emergent spin connection gauge field. The Zubarev statistical operator method allows for a convenient inclusion of macroscopic motion in general into a Lagrangian formulation. We first review this method in a simple context. For superfluid $^3$He - A we first need to identify possible currents which may enter the Zubarev statistical operator. We derive conditions to be fulfilled by macroscopic motion in $^3$He - A in order for the substance to be in global thermodynamic equilibrium subsequently. This program is outlined in section \ref{zubarevformalism} and whose details depend on the actual symmetries present. The relativistic representation of the statistical operator allows us to identify a Hamiltonian density comprising macroscopic motion. In section \ref{pathintegralformulation} we convert the Hamiltonian density (which is being integrated over a spacelike hypersurface) into a Lagrangian via the introduction of an ''emergent'' time direction along which the fields, but not the macroscopic motion variables, are being evolved. We finally provide a concise parametrization of the macroscopic motion variables within the Lagrangian. In particular, we find that macroscopic rotation is admitted in the presence of vortices. Here we follow closely the procedure proposed earlier in \cite{AZZ2023} within the path integral formulation of quantum field theory. A similar approach has been developed in \cite{H1,H2} for truely relativistic systems. We then analyse the thermodynamical quantities of the normal component of the superfluid that rotates around the axis of the pure mass vortices in section \ref{thermoequisolutions}. We subsequently collect our main findings in the discussion section \ref{SectDisc} and conclude our work with an outline of future research directions in section \ref{conclusions}.}}

\section{The emergence and characteristics of the $^3$He - A superfluid}
\label{reviewon3hea}

\subsection{The action of $^3$He with omission of spin - orbit interaction}
\label{SectIIA}
We set $\hbar =1$ throughout our calculations. The symmmetry group of liquid $^3$He is given by 
\begin{align}
H=U(1)\times SO_3.
\end{align}
Since the spin orbit coupling in liquid $^3$He (the dipole-dipole interaction) is relatively small, its omission yields a reasonable approximation of the description of $^3$He. According to \cite{alonsopopov} $^3$He without spin-orbit term may be described by the effective theory with action
\begin{align}
S=\sum_{p,s}\overline{a}_s(p)\epsilon (p)a_s(p)-\frac{g}{\beta V}\sum_{p;i,\alpha =1,2,3}\overline{J}_{i\alpha}(p)J_{i\alpha}(p)
\label{3heaction}
\end{align}
whereby
\begin{align}
&p=(\omega ,k),\,\,\,\, \hat{k}=\frac{k}{|k|}, ,\,\,\,\,\epsilon (p)=i\omega -(\frac{k^2}{2M_3}-\mu )\approx i\omega -v_F(|k|-k_F),\\
&J_{i\alpha}(p)=\frac{1}{2}\sum_{p_1+p_2}(\hat{k}_1^i-\hat{k}_2^i)a_A(p_2)[\sigma_\alpha ]^C_Ba_C(p_1)\epsilon^{AB},\,\,\,\, \epsilon^{-+}=-\epsilon^{+-}=1.
\end{align}
Here $V$ is the 3D volume, while $\beta=\frac{1}{T}$ with temperature $T$. Both $V$ and $\beta$ should be set to infinity at the end of the calculations. $a_{\pm}(p)$ is the fermion variable in momentum space with hermitian conjugate $\overline{a}_{\pm}(p)$. $M_3$ is the mass of $^3$He atoms, $\mu$ is their chemical potential. The energy density function $\epsilon$ is expanded around the Fermi surface. The parameters $k_F$ and $v_F$ are Fermi momentum and velocity, respectively, while $g$ is the coupling constant. The Pauli matrices are denoted by $\sigma_{\alpha}$. Notice that we will throughout this work denote hermitian conjugation with an overline instead of a superscript dagger.\par
The neglect of the spin-orbit term enhances the symmetry group to 
\begin{align}
G=U(1)\times SO_3^L\times SO_3^S
\end{align}
such that spin and orbital rotation groups, $SO_3^S$ and $SO_3^L$, may be considered independently. Eq. (\ref{3heaction}) is invariant under the action of the group $G$. The relevant symmetry group $G$ of physical laws, which is spontaneously broken in superfluid phases of $^3$He, contains the subgroup $U(1)$ which is responsible for the conservation of particle number as well as the group of rotations $SO_3^J$. The order parameter - the high-energy Higgs field - belongs to the representation $S=1$ and $L=1$ of the $SO_3^S$ and $SO_3^L$ groups and is represented by a $3\times 3$-complex matrix $A$ with components $A_{i\alpha}$. This matrix therefore comprises $18$ real components. We proceed with bosonization. We make use of the Hubbard-Stratonovich formula in the form
\begin{align}
e^{\frac{g}{\beta V}\sum_{p,i,\alpha}\overline{J}_{i\alpha}(p)J_{i\alpha}(p)}=N\Pi_{p,i,\alpha}\int D\overline{A}_{i\alpha}(p)DA_{i\alpha}(p)e^{-\frac{1}{g}\overline{A}_{i\alpha}A_{i\alpha}(p)-\frac{1}{\sqrt{\beta V}}\overline{A}_{i\alpha}(p)J_{i\alpha}(p)-\frac{1}{\sqrt{\beta V}}\overline{J}_{i\alpha}(p)A_{i\alpha}(p)}.
\end{align}
The parameter $N$ is a normalization constant which we leave undetermined. The $A_{i\alpha}(p)$ are bosonic variables. These variables may be considered as the field of Cooper pairs, which serves as the analog of the Higgs field in relativistic theories. The resulting action is quadratic in fermionic fields. Gaussian integration over fermionic degrees of freedom produces the effective bosonic action
\begin{align}
S_{eff}=\frac{1}{g}\sum_{p,i,\alpha}\overline{A}_{i\alpha}(p)A_{i\alpha}(p)+\frac{1}{2}log\Big(Det(G^{-1}[\overline{A},A])\Big)\label{Szero}
\end{align}
for the bosonic fields $A$ and $\overline{A}$ where
\begin{align}
G^{-1}[\overline{A},A]=\begin{pmatrix}
(i\omega -v_F(|k|-k_F))\delta_{p_1p_2} & \frac{1}{2\sqrt{\beta V}}[(\hat{k}_1^i+\hat{k}_2^i)A_{i\alpha}(p_1-p_2)]\sigma_{\alpha}\\
\frac{1}{2\sqrt{\beta V}}[(\hat{k}_1^i+\hat{k}_2^i)\overline{A}_{i\alpha}(-p_1+p_2)]\sigma_{\alpha} & (i\omega +v_F(|k|-k_F))\delta_{p_1p_2}
\end{pmatrix}
\end{align}
is the inverse Fermion Green function in the basis of Nambu - Gorkov spinors
\begin{align}
\Psi (p)=\begin{pmatrix}
\chi^A(p)\\
\epsilon^{BA}\overline{\chi}^B(-p)
\end{pmatrix}
=\begin{pmatrix}
a_+(p) \\
a_-(p) \\
\overline{a}_-(-p) \\
-\overline{a}_+(-p) 
\end{pmatrix}.
\label{fundamental1}
\end{align}

\subsection{Vacuum of $^3$He - A in the London limit for inhomogeneous fields}

\mz{The bosonic fields $A_{i\alpha}$ are a priori arbitrary dynamical degrees of freedom. They parametrize the superfluid component within the superfluid phases of $^3$He. The fermionic content may then be viewed has a second fluid ''normal'' to the superfluid background. From a thermodynamic point of view the bosonic fields may approximately freeze to give rise to a superfluid condensate whose characteristics are determined by the minimization of the corresponding thermodynamic potential. The latter thereby selects the superfluid order parameter form for temperature $T$, pressure $p$ possibly in the presence of other external fields (like, e. g., a magnetic field). In the absence of external fields the $^3$He - A phase only emerges at finite external pressures as the Fermi liquid $^3$He is being cooled down. In the process the Fermi surface destabilizes in the presence of fermionic fluctuations.}\par 
\mz{We proceed to describe the form of the superfluid order parameter for the $^3$He - A superfluid. The residual dynamics of the bosonic fields $A_{i\alpha}$ may be organized in a derivative expansion \cite{VollhardtWolfle1990}. We note again that we will consider the superfluid component approximately as an external field which couples to the fermionic normal component without consideration of the backreaction of the fermions on the superfluid background.}

The values of $A_{i\alpha}$ in the London limit have the form
\begin{align}
A_{i\alpha}=\sqrt{\beta V}\Delta_0(\bold{m}_i-i\bold{n}_i)d_{\alpha}=\sqrt{\beta V}k_Fv_{\perp}(\bold{m}_i-i\bold{n}_i)d_{\alpha}, \,\,\,\, i,\alpha =1,2,3.
\end{align}
Here $\bold{d}$ is a unit vector as are $\bold{m}$ and $\bold{n}$ which satisfy $\bold{m}\cdot \bold{n}=0$. We define $\bold{l}=\bold{m}\times\bold{n}$ and we have $v_{\perp}=\frac{\Delta_0}{k_F}$. The Landau order parameter for condensation into superfluid $^3$He-A breaks parity and time reversal symmetry. It belongs to the representation $S=1$ and $L=1$ of the $SO_3^S$ and $SO_3^L$ groups which implies the representation by a $3\times 3$-complex matrix $A$ with components $A_{i\alpha}$. It also transforms with charge two under the $U(1)$ group. The implied symmetry breaking scheme is given by
\begin{align}
G\to H_A=U(1)^{L_{\bold{l}}-\frac{N}{2}}\times U(1)^{S_{\bold{d}}}\times\mathbb{Z}_2.
\end{align}
The first $U(1)$ implies the invariance of the order parameter under simultaneous rotations in space around the $\bold{l}$-axis by elements of $SO_3^L$ and rotation under the particle number $U(1)$ group with the indicated proportions. The second $U(1)$ implies invariance under rotations around the spin $\bold{d}$-axis. We denote the discrete symmetry by $P=U(1)^S_{(\pi ,\bold{d}_{\perp})}\cdot U(1)^L_{(\pi ,\bold{l})}$ which implies Landau order parameter invariance under simultaneous rotations around a spin axis perpendicular to $\bold{d}$ by an angle $\pi$ and an orbital rotation around the frame vector field $\bold{l}$ by an angle $\pi$.\par 
We represent the fermionic effective action in \mzc{real time} in position space as follows
\begin{align}
S_{eff}=\frac{1}{2}\int d^4x\overline{\Psi}[i\partial_t+(\frac{\Delta_{\psi}}{2M_3}+\mu)\tau^3+\frac{i}{2}v_{\perp}(\bold{d}\boldsymbol{\sigma})(\bold{m}\overset{\leftrightarrow}{\bold{\nabla}}_{\psi})\tau^1+\frac{i}{2}v_{\perp}(\bold{d}\boldsymbol{\sigma})(\bold{n}\overset{\leftrightarrow}{\bold{\nabla}}_{\psi})\tau^2]\Psi
\end{align}
Here $\tau^a$ are the Pauli matrices corresponding to Bogolyubov spin and $\Delta$ is the 3D Laplace operator. The derivatives are meant to act only on the fermion fields as indicated by the subscript. We use the symbol $\nabla$ to denote an ordinary derivative. The linearization of the energy density function around $(0,0,k_F)$ with $k_F>0$, that means in the proximity of the two Fermi points, gives rise to the consistency condition
\begin{align}
\frac{(|k|-k_F)^2}{2M_3}\ll v_{\perp}(|k|-k_F)
\end{align}
for a Taylor expansion to be valid. This implies for typical length scales $a$ and time scales $\tau$ the following conditions
\begin{align}
a\sim (|k|-k_F)^{-1}\gg \frac{v_F}{v_{\perp}k_F},\,\,\,\, \tau \gg \frac{1}{v_{\perp}k_F}.
\label{timespaceconstraints}
\end{align}
The three vector fields $\bold{m}$, $\bold{n}$, $\bold{l}$ form an orthonormal triad which depends on position space and time
\begin{align}
\bold{m}(\bold{r},t)=\Omega (\bold{r},t)\bold{m}(0),\,\,\,\, \bold{n}(\bold{r},t)=\Omega (\bold{r},t)\bold{n}(0),\,\,\,\, \bold{l}=\Omega (\bold{r},t)\bold{l}(0),\,\,\,\, \Omega (\bold{r},t)=e^{\phi^a(\bold{r},t)\hat{\Sigma}_a}
\end{align}
where $\Omega (\bold{r},t)\in O_3\, \forall\, \bold{r},t$ and $\hat{\Sigma}^a\in o_3$. Similarly the vector field $\bold{d}$ depends on the position in space and time
\begin{align}
\bold{d}(\bold{r},t)=\Lambda (\bold{r},t)\bold{d}(0),\,\,\,\, \Lambda (\bold{r},t)=e^{\lambda^a(\bold{r},t)\hat{\Sigma}_a}.
\end{align}
For derivatives of basis vectors we introduce the notation
\begin{align}
\nabla_{\mu}\bold{m}=\bold{B}_{\mu}\times\bold{m},\,\,\,\, \nabla_{\mu}\bold{n}=\bold{B}_{\mu}\times\bold{n},\,\,\,\, \nabla_{\mu}\bold{l}=\bold{B}_{\mu}\times\bold{l},\,\,\,\, B_{\mu}^a\hat{\Sigma}_a\equiv [\nabla_{\mu}\Omega (\bold{r},t)]\Omega^{-1}(\bold{r},t).
\end{align}
A superfluid ''chemical potential'' $\mu_s$ and as well as a superfluid "velocity" $v_s$ induced by varying $\bold{\phi}$ may be defined as 
\begin{align}
2\mu_s=\bold{l}\bold{B}_t,\,\,\,\, 2M_3\bold{v}^k_s=\bold{l}\bold{B}_k.
\label{chempotvel}
\end{align}
These two quantities may be expressed directly through $\bold{m}$ and $\bold{n}$ 
\begin{align}
\frac{1}{2}\Big(\bold{n}\nabla_{\mu}\bold{m}-\bold{m}\nabla_{\mu}\bold{n}\Big)=\frac{1}{2}\Big((\bold{n}[\bold{B}_{\mu}\times \bold{m}])-(\bold{m}[\bold{B}_{\mu}\times \bold{n}])\Big)=\bold{l}\bold{B}_{\mu}.
\end{align}
Therefore we have
\begin{align}
2\mu_s=\frac{1}{2}\Big(\bold{n}\nabla_t\bold{m}-\bold{m}\nabla_t\bold{n}\Big),\,\,\,\, 2M_3\bold{v}_s^k=\frac{1}{2}\Big(\bold{n}\nabla_k\bold{m}-\bold{m}\nabla_k\bold{n}\Big).\label{musvs}
\end{align}
We furthermore assume the set of inequalities
\begin{align}
v_s<v_c\ll v_{\perp},v_{\parallel}
\label{velocityinequalities}
\end{align}
to hold where we used $v_s=|\bold{v}_s|$ and $v_{\parallel}=v_F$. This corresponds to the case of slowly varying $\Omega$. Throughout the upcoming sections we assume that the reference frame under consideration is related to the superfluid container, and the normal velocity $v_n$ is zero (due to $v_s<v_c$, where $v_c$ is the critical velocity of the Landau instability condition). These assumptions allow us to avoid additional complications which would otherwise arise in the actual superfluid $^3$He - A phase. 

\section{Low energy effective theory with emergent relativistic invariance}
\label{emergentphysics}

\mz{In this section we first present the standard formulation and explain the characteristics of the fermionic fields within superfluid $^3$He - A. Afterwards we perform a reparametrization of the theory. The advantages of this reformulation is that it features a universal vierbein coupled to the emergent relativistic Weyl fermions and combines the entire spin dynamics into an emergent spin connection gauge field.}

\subsection{Fermionic action near the Fermi points}

Near the Fermi points $K^i_{R,L}=K^i_{\pm}=\pm k_Fl^i$ we define  
\begin{align}
\psi_R(p) \equiv \psi_R(\delta p+{\cal A})=\Psi (K_++\delta p)=\begin{pmatrix}
\chi (K_++ \delta p) \\
-\chi^C(K_--\delta p)
\end{pmatrix},\,\,\,\, \psi_L( p ) \equiv \psi_L(\delta p-{\cal A})=\tau^3\Psi (K_-+\delta p)=\begin{pmatrix}
\chi (K_-+ \delta p) \\
\chi^C(K_+- \delta p)
\end{pmatrix}
\label{fundamental2}
\end{align}
with $\chi^C=-i\sigma^2\chi^{\ast}$. Here $\boldsymbol{{\cal A}} =  \bold{K}_+ =  k_F \bold{l}$ with $\mathcal{A}=(\mathcal{A}_0,\boldsymbol{\mathcal{A}})$ and $\mathcal{A}_0=0$ is the emergent axial gauge field originating from the Fermi points that may change their position in space and time. This defines momentum $\delta p = p \mp {\cal A}$ relative to the Fermi point position. (The upper sign is chosen for the right - handed Fermi point $K_+$ while the lower sign is chosen for $K_-$.)  The effective electric and magnetic fields within superfluid $^3$He - A are defined by
\begin{align}
\bold{E}=-\nabla_t\boldsymbol{\mathcal{A}}=-k_F\nabla_t\bold{l},\,\,\,\, \bold{H}=\mzc{\boldsymbol{\nabla}\times\boldsymbol{\mathcal{A}}=k_F\boldsymbol{\nabla}\times \bold{l}}.
\end{align}
{At this point we define the fields $\psi_{R}$, $\psi_L$ in coordinate space by Fourier transformation with respect to $p$:
$$
\psi_{R,L}(x) = \int \frac{d^4 p}{ (2 \pi)^4}e^{i p x} \psi_{R,L}(p)
$$}
Consequently the effective fermion field action for $^3$He - A with relativistic invariance reads
\begin{align}
\nonumber S_{eff}=&\frac{1}{4}\int d^4x\bold{e}[\overline{\psi}_Li\bold{e}_b^{\mu}(x)\overline{\tau}^b\nabla_{\mu}\psi_L-[\nabla_{\mu}\overline{\psi}_L]i\bold{e}_b^{\mu}(x)\overline{\tau}^b\psi_L+\overline{\psi}_Ri\bold{e}_b^{\mu}(x)\tau^b\nabla_{\mu}\psi_R-[\nabla_{\mu}\overline{\psi}_R]i\bold{e}_b^{\mu}(x)\tau^b\psi_R]\\
\equiv &\int d^4x\bold{e}\mathcal{L}=\int d^4x\bold{e}(\mathcal{L}_L+\mathcal{L}_R)
\label{weylaction}
\end{align}
with covariant derivative $\nabla_{\mu} = \partial_{\mu} - i {\cal A}_{\mu}\gamma^5$ (where $\gamma^5 \psi_{R/L} = \pm \psi_{R/L}$ is the chirality matrix). From now on $\nabla_{\mu}$  is meant to be a covariant derivative when acting on fermion fields. The Grassmann variables $\psi_{R,L}$ obey
\begin{align}
\psi_R(\delta p+{\cal A})=i\tau^1\sigma^2\psi^{\ast}_L(-\delta p+{\cal A}),\,\,\,\,\psi_L(\delta p-{\cal A})=-i\tau^1\sigma^2\psi^{\ast}_R(-\delta p-{\cal A})
\nonumber
\end{align}
that is
\begin{align}
	\psi_R(p)=i\tau^1\sigma^2\psi^{\ast}_L(-p),\,\,\,\,\psi_L(p)=-i\tau^1\sigma^2\psi^{\ast}_R(-p)
	\label{majoranaconstraint}
\end{align}

and the generalized vierbein $\bold{e}_a^{\mu}$ that belongs to the Lie algebra $u(2)$ and has components
\begin{align}
&1=\bold{e}\bold{e}_0^0,\,\,\,\, 0=\bold{e}\bold{e}_0^i=\bold{e}\bold{e}_i^0,\\
&v_{\perp}(\bold{m}^i-i\bold{n}^i)(\bold{d}\boldsymbol{\sigma})=\bold{e}(\bold{e}_1^i-i\bold{e}_2^i),\,\,\,\, v_{\parallel}l^i= \bold{e}\bold{e}_3^i,\\
&\bold{e}=(v_{\parallel}v_{\perp}^2)^{\frac{1}{3}}
\end{align}
and we denote $v_{\parallel}=v_F$ and $\bold{e}=det(\bold{e}_{\mu}^a)$. Moreover $\tau =(1,\boldsymbol{\sigma})$, $\overline{\tau}=(1,-\boldsymbol{\sigma})$. In matrix notation the vierbein may be written in the form
\begin{align}
\bold{e}_a^{\mu}=\bold{e}^{-1}\begin{pmatrix}
1 & 0 \\
0 & v_{\perp}\bold{m}(\bold{d}\boldsymbol{\sigma}) \\
0 & v_{\perp}\bold{n}(\bold{d}\boldsymbol{\sigma}) \\
0 & v_{\parallel}\bold{l}
\end{pmatrix},\,\,\,\, a,\mu =0,1,2,3
\end{align}
with inverse vierbein
\begin{align}
\bold{e}_{\mu}^a= \bold{e}\begin{pmatrix}
1 & 0 & 0 & 0 \\
0 & \frac{1}{v_{\perp}}\bold{m}(\bold{d}\boldsymbol{\sigma}) & \frac{1}{v_{\perp}}\bold{n}(\bold{d}\boldsymbol{\sigma}) & \frac{1}{v_{\parallel}}\bold{l}
\end{pmatrix},\,\,\,\, a,\mu =0,1,2,3.
\end{align}
\mzc{The physical meaning of the vierbein field in the language of the conventional theory of superfluid $^3$He can be read off from Eq. (\ref{musvs}), where expressions for the superfluid velocity and superfluid chemical potential are given through vectors $\bf m$ and $\bf n$.}
{Notice that the axial gauge field $\mathcal{A}$ and the vierbein $\bold{e}_a^{\mu}$ are not independent but related by
\begin{align}
\boldsymbol{\mathcal{A}}=\mzc{k_F\bold{l}=k_F\Big(\frac{v_{\perp}}{v_{\parallel}}\Big)^{\frac{2}{3}}\bold{e}_3}.
\end{align}}
The constraint in Eq. (\ref{majoranaconstraint}) implies (after transposition and using the Grassmann-valuedness of the spinors) the momentum space identity
\begin{align}
\overline{\psi}_L(p)\bold{e}_b^{\mu}\overline{\tau}^bp_{\mu}\psi_L(p)=\overline{\psi}_R(-p)\bold{e}_b^{\mu}\tau^b(-p_{\mu})\psi_R(-p).
\end{align}
As a consequence, left- and right-handed spinors are not independent, if both positive and negative momenta are being summed over. {We will throughout the following stick to the convention of treating left- and right-handed spinors as independent for all momenta. We may ultimately just enforce the constraint of Eq. (\ref{majoranaconstraint}) in an explicit calculation. This simple rule will turn out to work both classically and quantum mechanically, as we will elaborate further below. We consider the consequences of imposing the constraint a priori in Appendix \ref{app1} (see also Eqs. (\ref{majoranaconstraint2}) and (\ref{constraintchiralcomponents}) below)}. 
%This ultimately implies to take into account a factor of $\frac{1}{2}$ mutliplying any quantity proportional to the number of degrees of freedom. Alternatively, the normalization of states by the number density charge $j^0_V$ (to be introduced below)) may be chosen to incorporate the reduction of degrees of freedom by half. \par

%\mzo{Please adopt everywhere further the covariant derivatives with respect to the axial gauge field $\cal A$. This is what we missed in our previous consideration.}

We may as well write the effective action in Eq. (\ref{weylaction}) in a more compact way by introducing Dirac spinors. The Weyl representation with notation
\begin{align}
&\gamma^a=\begin{pmatrix} 0 & \tau^a \\ \overline{\tau}^a & 0 \end{pmatrix},\,\,\,\, \gamma^5=i\gamma^0\gamma^1\gamma^2\gamma^3=\begin{pmatrix} -\mathbb{1} & 0 \\  0 & \mathbb{1} \end{pmatrix},\,\,\,\, \{\gamma^a,\gamma^b\}=2\eta^{ab},\,\,\,\, a=0,1,2,3,\\
&\psi_L=\frac{1}{2}(1-\gamma^5)\psi ,\,\,\,\, \psi_R=\frac{1}{2}(1+\gamma^5)\psi \,\,\Rightarrow \psi=\begin{pmatrix} \psi_L \\ \psi_R \end{pmatrix}
\end{align}
is employed. The effective action in Dirac spinor notation takes the form
\begin{align}
S_{eff}=&\frac{1}{4}\int d^4x\bold{e}[\overline{\psi}i\bold{e}_b^{\mu}(x)\gamma^0\gamma^b\nabla_{\mu}\psi -[\nabla_{\mu}\overline{\psi}]i\bold{e}_b^{\mu}(x)\gamma^0\gamma^b\psi ]\equiv \int d^4x\bold{e}\mathcal{L}.
\label{diracaction}
\end{align}

Several differences between the action in Eq.(\ref{weylaction}) (Eq. (\ref{diracaction})) and that of relativistic Weyl (Dirac) fermions exist:

\begin{itemize}[label=(\arabic*)]
\item[1)] The vierbein $\bold{e}_a^{\mu}$ and its inverse are matrix valued due to the term $\bold{d}\boldsymbol{\sigma}$. We define scalar valued vierbeins by
\begin{align}
(e^{\pm})_a^{\mu}=e^{-1}\begin{pmatrix}
1 & 0 \\
0 & \pm v_{\perp}\bold{m} \\
0 & \pm v_{\perp}\bold{n} \\
0 & v_{\parallel}\bold{l}
\end{pmatrix},\,\,\,\, a,\mu =0,1,2,3
\end{align}
with inverse scalar valued vierbein
\begin{align}
(e^{\pm})_{\mu}^a= e\begin{pmatrix}
1 & 0 & 0 & 0 \\
0 & \pm\frac{1}{v_{\perp}}\bold{m} & \pm\frac{1}{v_{\perp}}\bold{n} & \frac{1}{v_{\parallel}}\bold{l}
\end{pmatrix},\,\,\,\, a,\mu =0,1,2,3.
\end{align}
We use bold letters in the case of matrix valued vierbein comprising the term $\bold{d}\boldsymbol{\sigma}$ and the unbolded notation, if it is absent. Moreover, the scalar vierbein determinants coincide with that of the matrix valued vierbein $e\equiv e^{\pm}=\bold{e}$. The motivation to consider these scalar vierbeins 
arises due to consideration of the eigenspaces of the matrix valued operator $(\boldsymbol{\sigma}\bold{d})$. We will subsequently suppress the superscript $s=\pm$ on the scalar valued vierbein and consider it only implicitly. Further below be will consider the choice $s=+$.
We define the projection operators
\begin{align}
\mathbb{P}^+=\frac{1+(\bold{d}\boldsymbol{\sigma})}{2},\,\,\,\,\mathbb{P}^-=\frac{1-(\bold{d}\boldsymbol{\sigma})}{2}
\end{align}
We have the common notation $\mathbb{P}^s = \frac{1+s ({\bf d} \boldsymbol{\sigma})}{2}$. The normalized eigenspinors of this operator for $s=\pm$ are
	$$ 
	\eta^{\pm} =\frac{1}{\sqrt{2(1 \mp d_3)}} \left(\begin{array}{c}  \mp (d_1 - i d_2)\\ \pm d_3 -1 \end{array} \right)
	$$
We introduce the two - component spinors $\Psi_{L/R}^\pm$ as follows
$$
\psi_{L/R} = \sum_{s=\pm}\psi_{L/R}^s,\,\,\,\,\psi^{\pm}_{L/R}=\Psi^\pm_{L/R}\otimes \eta^\pm
$$ 
(In the following we will omit the symbol $\otimes$ of the tensor product for brevity.)
Then
\begin{align}
\Psi_{L/R}^+\eta^+=\mathbb{P}^+\psi_{L/R}=\psi_{L/R}^+,\,\,\,\, \Psi_{L/R}^-\eta^-=\mathbb{P}^-\psi_{L/R}=\psi^-_{L/R}.
\end{align}
We may further employ the projection operators $\mathbb{P}^s$ ($s=\pm$) in order to represent the 8 - component Dirac spinors in terms of the 4 - component spinors
\begin{align}
\Psi^+\eta^+=\mathbb{P}^+\psi =\psi^+,\,\,\,\, \Psi^-\eta^-=\mathbb{P}^-\psi =\psi^- ,\,\, \Rightarrow\, \psi =\psi^++\psi^-=\Psi^+\eta^++\Psi^-\eta^-.
\end{align}

\item[2)] The vierbein $e_a^{\mu}$ (as well as $\bold{e}_a^{\mu}$) is not orthonormal with respect to the Minkowski metric but instead fulfills
\begin{align}
e_a^{\mu}e_b^{\nu}g_{\mu\nu}=diag(1,-1,-1,-1)\equiv\eta_{ab},\,\,\,\, \bold{e}_a^{\mu}\bold{e}_b^{\nu}g_{\mu\nu}=diag(1,-1,-1,-1)\equiv\eta_{ab}\mathbb{1}
\end{align}
with metric (for a diagonalizing coordinate frame respecting the inherent anisotropy)
\begin{align}
g_{\mu\nu}=e^2\cdot diag\Big(1,-\frac{1}{v_{\perp}^2},-\frac{1}{v_{\perp}^2},-\frac{1}{v_{\parallel}^2}\Big)=(v_{\parallel}v_{\perp}^2)^{\frac{2}{3}}\cdot diag\Big(1,-\frac{1}{v_{\perp}^2},-\frac{1}{v_{\perp}^2},-\frac{1}{v_{\parallel}^2}\Big).
\end{align}
This metric is a natural measure of distance within superfluid $^3$He-A. We use $g_{\mu\nu}$ and its inverse to raise or lower spacetime indices (Greek letters) and $\eta_{ab}$ and its inverse to raise or lower Lorentz indices (Latin letters a,b,c...). Spatial spacetime indices are labeled by Latin letters i,j,k.... The spacetime we are working on is flat as a consequence of the constancy of $g_{\mu\nu}$. We will furthermore work with the definition $\epsilon_{0123}=1$ for the $\epsilon$-symbol where the indices refer to the local Lorentz frame.

\item[3)] The action has vanishing spin connection gauge field $\omega^{\mu}_{ab}$. Its presence is required in the standard relativistic theory in order to ensure local Lorentz invariance. Instead we are only given the gauge field $e_a^{\mu}$ of translations with its curvature 
\begin{align}
T^{\mu}_{ab}=-(e_a^{\nu}\nabla_{\nu}e_b^{\mu}-e_b^{\nu}\nabla_{\nu}e_a^{\mu}).
\end{align}
The minus sign is in line with the standard definition 
\begin{align}
T^{a}_{\mu\nu}=\nabla_{\mu}e^a_{\nu}-\nabla_{\nu}e^a_{\mu}.
\end{align}
Using $\nabla_\nu e^a_\mu e^\mu_b = 0$ we obtain $e^\mu_b \nabla_\nu e^a_\mu  = - e^a_\mu\nabla_\nu  e^\mu_b $ and $\nabla_\mu e^a_\nu  = -e^b_\nu e^a_\rho\nabla_\mu  e^\rho_b$ . This leads to
\begin{align}
T^{a}_{\mu\nu}=-e^b_\nu e^a_\rho\nabla_\mu  e^\rho_b + e^b_\mu e^a_\rho\nabla_\nu  e^\rho_b \nonumber
\end{align}
and finally
\begin{align}
e_c^{\mu}e_a^{\rho}e_b^{\sigma}T^{c}_{\rho\sigma}=-e_c^{\mu}e_a^{\rho}e_b^{\sigma}e^d_{\sigma}e^c_{\nu}\nabla_{\mu} e^{\nu}_d + e_c^{\mu}e_a^{\rho} e_b^{\sigma}e^d_{\rho}e^c_{\nu}\nabla_{\sigma}e^{\nu}_d = -e_a^{\nu}\nabla_{\nu}e^{\mu}_b+e_b^{\nu}\nabla_{\nu}e^{\mu}_a.\nonumber
\end{align}
The tensor field $T^{\mu}_{ab}$ is also known as the torsion tensor field. In absence of the spin connection, we will not require that the vierbein is covariantly constant in general. In this context covariant constancy of the vierbein is equivalent to a constant vierbein. The action then features global translation and Lorentz invariance. We will nevertheless introduce the spin connection in order to derive the spin tensor below. In the relativistic theory of Dirac fermions the spin connection enters via the covariant derivative (which is diagonal in the internal spin space due to $(\boldsymbol{\sigma}\bold{d})$)
\begin{align}
D_{\mu}\psi^{\pm} =(\nabla_{\mu}+\frac{1}{8}[\gamma^a,\gamma^b]\omega_{\mu}^{ab})\psi^{\pm} ,\,\,\,\, \overline{\psi}^{\pm}\gamma^0\overset{\leftarrow}{D}_{\mu}=\overline{\psi}^{\pm}\gamma^0(\overset{\leftarrow}{\nabla}_{\mu}-\frac{1}{8}[\gamma^a,\gamma^b]\omega_{\mu}^{ab})
\end{align}
(note that our overline does not comprise $\gamma^0$). In the relativistic theory of Weyl fermions the spin connection enters via the covariant derivative
\begin{align}
D_{\mu}\psi^{\pm}_{L/R} =(\nabla_{\mu}+\frac{1}{8}\tau^{ab}\omega_{\mu}^{ab})\psi^{\pm}_{L/R} ,\,\,\,\, \overline{\psi}^{\pm}_{L/R}\overset{\leftarrow}{D}_{\mu}=\overline{\psi}^{\pm}_{L/R}(\overset{\leftarrow}{\nabla}_{\mu}-\frac{1}{8}\bar{\tau}^{ab}\omega_{\mu}^{ab})
\end{align}
where the generators of the internal Lorentz group act on the left/right - handed spinors as follows
$$
\tau^{ab}\psi^{\pm}_L = (\tau^a \bar{\tau}^b-\tau^b\bar{\tau}^a) \psi^{\pm}_L, \quad \tau^{ab}\psi^{\pm}_R = (\bar{\tau}^a {\tau}^b-\bar{\tau}^b\tau^a) \psi^{\pm}_R,\quad \bar{\psi}^{\pm}_R\bar{\tau}^{ab} = \bar{\psi}^{\pm}_R(\tau^a \bar{\tau}^b-\tau^b\bar{\tau}^a) , \quad \bar{\psi}^{\pm}_L\bar{\tau}^{ab} = \bar{\psi}^{\pm}_L(\bar{\tau}^a {\tau}^b-\bar{\tau}^b\tau^a).
$$
%Correspondingly, the Lorentz scalars are $\bar{\psi}^s_L\psi^t_R + \bar{\psi}^s_L\psi^t_R$ while the pseudo - scalars are $\bar{\psi}^s_L\psi^t_R - \bar{\psi}^s_L\psi^t_R$, where $s,t=\pm$ denote spin indices. \par
The above expressions hold in the same way when written in terms of $\Psi^{\pm}$ and $\overline{\Psi}^{\pm}$. We will modify the definition in the latter case, though by replacing $\Psi^-$ simultaneously by $e^{\frac{\pi}{4}[\gamma^1,\gamma^2]}\Psi^-$. This definiton will be motivated shortly and gives rise to a different spin tensor. This will be remarked again further below.
\end{itemize}
The effective action of $^3$He-A may now be written in the Weyl form
\begin{align}
\nonumber S_{eff}=&\frac{1}{4}\sum_{r,s=\pm}\int d^4xe[\overline{\Psi}^r_Li\bar{\eta}^r\bold{e}_b^{\mu}(x)\overline{\tau}^b\nabla_{\mu}\eta^s\Psi^s_L-[\nabla_{\mu}\bar{\eta}^s\overline{\Psi}^s_L]i\bold{e}_b^{\mu}(x)\overline{\tau}^b\eta^r\Psi^r_L\\
&+\overline{\Psi}^r_Ri\bar{\eta}^r\bold{e}_b^{\mu}(x)\tau^b\nabla_{\mu}\eta^s\Psi^s_R-[\nabla_{\mu}\overline{\eta}^s\overline{\Psi}^s_R]i\bold{e}_b^{\mu}(x)\tau^b\eta^r\Psi^r_R]\nonumber\\
\overset{\nabla_{\mu}\bold{d}\equiv 0}{=}&\frac{1}{4}\sum_{s=\pm}\int d^4xe[\overline{\Psi}^s_Li(e^s)_b^{\mu}(x)\overline{\tau}^b\nabla_{\mu}\Psi^s_L-[\nabla_{\mu}\overline{\Psi}^s_L]i(e^s)_b^{\mu}(x)\overline{\tau}^b\Psi^s_L+\overline{\Psi}^s_Ri(e^s)_b^{\mu}(x)\tau^b\nabla_{\mu}\Psi^s_R-[\nabla_{\mu}\overline{\Psi}^s_R]i(e^s)_b^{\mu}(x)\tau^b\Psi^s_R]\nonumber\\
\equiv &\int d^4xe(\mathcal{L}^+_L+\mathcal{L}^-_L+\mathcal{L}^+_R+\mathcal{L}^-_R).
\label{weylaction2}
\end{align}
Notice that the two eigenstates of $(\bold{d}\boldsymbol{\sigma})$ couple to each other if and only if $\nabla_{\mu}\bold{d}\neq 0$. We will therefore often consider the situation of homogeneous and inhomogeneous $\bold{d}$ separately.\par
The effective action in Dirac spinor notation is given by
\begin{align}
\nonumber S_{eff}=&\frac{1}{4}\int d^4xe\sum_{r,s=\pm}[\overline{\Psi}^ri\bar{\eta}^r\bold{e}_b^{\mu}(x)\gamma^0\gamma^b\nabla_{\mu}\eta^s\Psi^s -[\nabla_{\mu}\bar{\eta}^s\overline{\Psi}^s]i\bold{e}_b^{\mu}(x)\gamma^0\gamma^b\eta^r\Psi^r]\nonumber\\
\overset{\nabla_{\mu}\bold{d}\equiv 0}{=}&\frac{1}{4}\int d^4xe\sum_{s=\pm}[\overline{\Psi}^si(e^s)_b^{\mu}(x)\gamma^0\gamma^b\nabla_{\mu}\Psi^s -[\nabla_{\mu}\overline{\Psi}^s]i(e^s)_b^{\mu}(x)\gamma^0\gamma^b\Psi^s]\nonumber\\
=&\frac{1}{4}\int d^4xe\sum_{s=\pm}[\overline{\Psi}^si(e^+)_b^{\mu}(x)\gamma^0\Gamma^{s,b}\nabla_{\mu}\Psi^s -[\nabla_{\mu}\overline{\Psi}^s]i(e^+)_b^{\mu}(x)\gamma^0\Gamma^{s,b}\Psi^s]\nonumber\\
\equiv &\int d^4xe(\mathcal{L}^++\mathcal{L}^-)
\label{diracaction2}
\end{align}
with
\begin{align}
\Gamma^{s,b}=\gamma^b \text{\,\, for\,\,}s=+,\,b=0,1,2,3,\,\,s=-,\,b=0,3,\,\,\,\, \Gamma^{s,b}=-\gamma^b \text{\,\,for\,\,} s=-,\,b=1,2.
\end{align}
We will now introduce a notation tailored towards the geometry implied by the additional spin space. 

\subsection{A reparametrization - universal vierbein field and spin connection gauge field}

The internal spin space gives rise to two choices for a scalar valued vierbein according to
\begin{align}
\bold{e}_a^{\mu}\eta^s=(e^s)_a^{\mu}\eta^s,\,s=\pm .
\end{align}
No preference between either of these two exists. Let us introduce the $8$ - component spinor
\begin{align}
\Psi=\begin{pmatrix} \Psi^+, & e^{\frac{\pi}{4}[\gamma^1,\gamma^2]}\Psi^-
\end{pmatrix}^T.
\label{phaserotation}
\end{align}
%This implies that we perform the transformation $\Psi^-\to e^{\frac{\pi}{4}[\gamma^1,\gamma^2]}\Psi^-$ in the previous expressions comprising $\Psi^-$.
The phase factor in Dirac space manifests a preference of scalar valued vierbein, namely we will write $e_a^{\mu}=(e^+)_a^{\mu}$ (and identically for the inverse). The additional phase factor may be moved to the other component with simultaneous change of choice for the scalar valued vierbein. We may then rewrite the effective action for $^3$He-A in the relativistic regime as
\begin{align}
S_{eff}=\frac{1}{4}\int d^4xe[\overline{\Psi}i\gamma^0\gamma^be_b^{\mu}D_{\mu}\Psi -[\overline{\Psi}\gamma^0\overset{\leftarrow}{D}_{\mu}]i\gamma^be_b^{\mu}\Psi ]
\label{relaction}
\end{align}
with covariant derivative
\begin{align}
D_{\mu}=\nabla_{\mu}-i\mathcal{B}_{\mu} = \partial_\mu - i {\cal A}_\mu \gamma^5 -i\mathcal{B}_{\mu} ,\,\,\,\, \mathcal{B}_{\mu}^{rs}=i(\overline{\eta}^r\nabla_{\mu}\eta^s)(\delta^{rs}\mathbb{1}+i\epsilon^{rs}\frac{\pi}{8}[\gamma^1,\gamma^2])
\end{align}
and $\epsilon^{-+}=-\epsilon^{+-}=1$. The gauge field $\mathcal{B}_{\mu}$ may be written in matrix form as
\begin{align}
\mathcal{B}_{\mu}=\begin{pmatrix}
b^+_{\mu} & \frac{1}{8}\omega_{\mu 12}[\gamma^1,\gamma^2] \\
\frac{1}{8}\omega^{\ast}_{\mu 12}[\gamma^1,\gamma^2] & b^-_{\mu}
\end{pmatrix}
\end{align}
with Abelian Berry connections 
\begin{align}
b^s_{\mu}=i\overline{\eta}^s\nabla_{\mu}\eta^s
\end{align}
and spin connection
\begin{align}
\omega_{\mu 12}=2\pi i\overline{\eta}^+\nabla_{\mu}\eta^-.
\end{align}
The non-Abelian gauge field $\mathcal{B}_{\mu}$ implies a mixing of Dirac and internal spin spaces and is nonzero if and only if $\nabla_{\mu}\bold{d}\neq 0$. It comprises two Abelian Berry connections which refer to the respective eigenfunctions of $(\boldsymbol{\sigma}\bold{d})$ on the internal spin space as well as a spin connection on the combined Dirac spinor and internal spin space. It fulfills the relations
\begin{align}
[\mathcal{B}_{\mu},\gamma^a]=0,\,a=0,3,5,\,\,\,\,[\mathcal{B}_{\mu}\rvert_{\omega_{\mu 12}=0},\gamma^a]=0,\,\,\,\, \{\mathcal{B}_{\mu}\rvert_{b^+=b^-},\gamma^a\}=0, \,a=1,2.
\end{align}
The gauge field $\mathcal{B}_{\mu}$ may as well be decomposed as
\begin{align}
\mathcal{B}_{\mu}=\frac{b^+_{\mu}+b^-_{\mu}}{2}\mathbb{1}_D\mathbb{1}+\frac{1}{8}Re(\omega_{\mu 12})[\gamma^1,\gamma^2]\sigma^1-\frac{1}{8}Im(\omega_{\mu 12})[\gamma^1,\gamma^2]\sigma^2+\frac{b^+_{\mu}-b^-_{\mu}}{2}\mathbb{1}_D\sigma^3.
\end{align}
The Berry connections as well as the spin connection comprising the overall gauge field may be expressed in terms of the components of the spin vector $\bold{d}$ and its first derivatives as follows
\begin{align}
&b_{\mu}^+=\frac{1}{2(1-d_3)}[d_1\nabla_{\mu}d_2-d_2\nabla_{\mu}d_1],\label{bplus}\\
&b_{\mu}^-=\frac{1}{2(1+d_3)}[d_1\nabla_{\mu}d_2-d_2\nabla_{\mu}d_1],\label{bminus}\\
&\omega_{\mu 12}=\frac{\pi}{\sqrt{d_1^2+d_2^2}}[d_2\nabla_{\mu}d_1-d_1\nabla_{\mu}d_2+i\nabla_{\mu}d_3].
\end{align}
We will make a final refinement by performing a field redefinition within the $(s=-)$-component of $\Psi$ such that $\Psi =(\Psi^+,\Psi^-)$. In terms of the spinor $\Psi$ and its projections under $\mathbb{P}^{s}$ ($s=\pm$) and $\mathbb{P}_{C}$ ($C=L/R$) the constraint of Eq. (\ref{majoranaconstraint}) may be brought, employing Eq. (\ref{constraintchiralcomponents}) in Appendix \ref{app1}, into the form
\begin{align}
(\Psi^+)^{\ast}=\hat{d}^{\ast}\gamma^2\Psi^-,\,\,\,\, (\Psi^-)^{\ast}=\hat{d}^{\ast}\gamma^2\Psi^+\,\,\Leftrightarrow \,\, (\Psi^+_{L/R})^{\ast}=\hat{d}^{\ast}(\tau^2/\overline{\tau}^2)\Psi^-_{R/L},\,\,\,\, (\Psi^-_{L/R})^{\ast}=\hat{d}^{\ast}(\tau^2/\overline{\tau}^2)\Psi^+_{R/L}.\label{MajoranaCond}
\end{align}
In position space all spinors are functions of the spacetime coordinate $x$, while in momentum space one spinor is evaluated at four momentum $p$ with the other one evaluated at $-p$. This constraint is to be imposed when evaluating, e. g., correlation functions in the quantum theory, unless one keeps track of necessary corrections, especially taking proper account of the number of degrees of freedom. \par
The redefinition of the $\Psi^-$ component corresponds to a rotation by an angle $\pi$ around the orbital $\bold{l}$-direction which is precisely the second component of the discrete symmetry $P=U(1)^S_{(\pi ,\bold{d}_{\perp})}\cdot U(1)^L_{(\pi ,\bold{l})}$. We may undertake another rotation by an angle $\pi$  around an axis orthogonal to the spin vector $\bold{d}$ in spin space which rotates $\Psi^-$ into $\tilde{\Psi}^+$ according to
\begin{align}
\begin{pmatrix}
0 \\
\Psi^-
\end{pmatrix}=
e^{-i\frac{\pi}{2}(\bold{a}\boldsymbol{\sigma})}\begin{pmatrix}
\tilde{\Psi}^+ \\
0
\end{pmatrix}=
-i(\bold{a}\boldsymbol{\sigma})\begin{pmatrix}
\tilde{\Psi}^+ \\
0
\end{pmatrix},\,\,\,\,
\bold{a}=\begin{pmatrix}
cos(\phi ) \\
sin(\phi ) \\
0
\end{pmatrix}.
\end{align}
We parametrize the freedom of choice of the axis of rotation by a space and time dependent angle $\phi\in [0,2\pi ]$. This transformation implies a shift in the spinor description from the doublet of $(\bold{d}\boldsymbol{\sigma})=\pm 1$ projected spinor components to the doublet with $(\bold{d}\boldsymbol{\sigma})=+1$ spinor components. This rotation implies a transformation not only of the spinors but also of $\mathcal{B}_{\mu}$ to $\tilde{\mathcal{B}}_{\mu}$ for which we find
\begin{align}
\tilde{\mathcal{B}}_{\mu}=\begin{pmatrix}
b_{\mu}^+\mathbb{1}_D & \frac{1}{8}Re(\omega_{\mu 12})(-ie^{i\phi})[\gamma^1,\gamma^2] \\
-\frac{1}{8}Im(\omega_{\mu 12})(ie^{-i\phi})[\gamma^1,\gamma^2] & (b_{\mu}^--\nabla_{\mu}\phi )\mathbb{1}_D
\end{pmatrix},\,\,\,\,\Psi =\begin{pmatrix}
\Psi^+ \\
\tilde{\Psi}^+
\end{pmatrix}.
\end{align}
The angle $\phi$ is pure gauge and may be absorbed by a local phase rotation of the $\tilde{\Psi}^+$ component. Alternatively we may choose the gauge $\phi =\frac{\pi}{2}$. This makes $\mathcal{B}_{\mu}$ and $\tilde{\mathcal{B}}_{\mu}$ identical in form, though they act on formally different vector spaces. Due to this isomorphism we will stick to $\mathcal{B}_{\mu}$ and $\Psi =(\Psi^+,\Psi^-)^T$ from now on. \mz{We will consider the field $\mathcal{B}_{\mu}$ a spin connection gauge field, as it comprises spin related effects both of the internal spin space as well as the Dirac spin space}.\par
In the following we will make use of the equivalent forms of the actions defined above. We will further work with units such that $e=\bold{e}=(v_{\parallel}v_{\perp}^2)^{\frac{1}{3}}=1$ and refer to the action in Eq. (\ref{relaction}) as the geometric formulation. Our choice of units are compared to the usual natural units in Appendix \ref{app3}.

\section{The Zubarev statistical operator formalism}
\label{zubarevformalism}

\mz{We intend to describe the physics of the superfluid $^3$He - A phase simultaneously in the presence of a non-trivial superfluid component flow as well as macroscopic motion of the fermionic normal component. This may be conveniently achieved within the Zubarev statistical operator method which we outline subsequently. The most general context where it may applied in is local thermodynamic equilibrium together within a hydrodynamic approximation of the substance under consideration. To begin with we review the essential ingredients of this method. We then proceed to apply it to $^3$He - A in the case of global thermodynamic equilibrium. This requires us to first identify the currents which appear in the superfluid phase followed by an analysis to identify which of these currents combine into macroscopically conserved currents under the additional assumption of global thermodynamic equilibrium.}

\subsection{Essentials of the Zubarev statistical operator method}

\mz{Following \cite{Zubarev1979} we present the relativistically covariant form of the statistical operator which provides a candidate for a proper description of macroscopic motion of a substance, for which a continuous medium or hydrodynamic approximation is valid, in (global) thermodynamic equilibrium. A pedagogical treatment of the Zubarev statistical operator formalism may be found in \cite{Buzzegoli2020ycf}. Note that we assume here full Poincar\'e symmetry of a theory with several conserved global currents in flat Minkowski spacetime. The logarithm of the statistical operator $\hat{\rho}$ may be expressed as
\begin{align}
\log\hat \rho =-\log Z -\int d\Sigma \beta n_{\nu}\Big(\hat{T}^{\nu\rho}u_{\rho}-\sum_i\mu_i\hat{j}_i^{\nu}\Big).
\label{statisticalZub}
\end{align}}
\mz{The constant $Z$ ensures normalization of the statistical operator $Tr(\hat{\rho})=1$. Here integration is over a $3$-dimensional spacelike hypersurface $\Sigma$. By $d \Sigma$ we denote the hypersurface element of integration. The four vector field $n_{\nu}(x)$ is orthogonal to the surface $\Sigma$, while $u_{\rho}$ may be interpreted as the macroscopic four velocity of a substance. These vectors obey the normalization conditions
$n^{\mu}(x)n_{\mu}(x)=u^{\mu}(x)u_{\mu}(x)=1$. The function $\beta(x)$ may be interpreted as inverse temperature depending on coordinates. The combination \(\beta_\mu(x)=\beta(x)u_\mu(x)\) is termed frigidity vector field. By $\hat{T}^{\nu\rho}$ we denote the gravitational (or, Belinfante - Rosenfeld) energy momentum tensor operator, while $\hat{j}_i^{\nu}$ represent conserved current operators with associated chemical potentials $\mu_i$. The spacetime considered here is flat Minkowski spacetime, which admits a foliation into $3$-dimensional spacelike hypersurfaces $\Sigma (\sigma)$ depending on the parameter $\sigma$. We consider the evolution of the system in the parameter $\sigma$ with its initial value $\sigma_i$ and final value $\sigma_f$.}

\mz{The boundary conditions for the evolution are the total translational, angular and boost momentum and charges (as expectation values of their associated operators, but the same relations apply for the operators themselves)
\begin{align}
P^\mu_{i,f} = \Big\langle\int d\Sigma_\nu T^{\mu\nu}|_{\sigma_{i,f}}\Big\rangle,\,\,\,\, M^{\mu\nu}_{i,f} = \Big\langle\int d\Sigma_\rho (x^{\mu} T^{\rho\nu}-x^{\nu} T^{\rho\mu})|_{\sigma_{i,f}}\Big\rangle,\,\,\,\,Q_{i,f}^k = \Big\langle\int d\Sigma_\nu j^{\nu}_k|_{\sigma_{i,f}}\Big\rangle .
\label{conservedcharges}
\end{align}}
\mz{The gravitational (or, Belinfante - Rosenfeld) energy momentum tensor operator in the Zubarev statistical operator comprises all ten Poincar\'e charges, the four translational charges with canonical energy momentum tensor operator $\hat T^{\mu\nu}_{can}$ as Noether current and Lorentz transformation charges with canonical Lorentz transformation tensor operator 
\begin{align}
\hat M^{\mu\nu\lambda}_{can}=(x^{\nu} \hat T^{\mu\lambda}_{can\,}-x^{\lambda}\hat T^{\mu \nu}_{can\,})+\hat S^{\mu \nu \lambda}_{\,\,\,\,}.
\end{align}}
\mz{The former term comprises angular and boost momentum contributions, while the latter term is the spin current operator. The spin current is related to the antisymmetric part of the canonical energy momentum tensor operator by
\begin{align}
D_{\mu}\hat S^{\mu\nu\lambda}=\hat T_{can}^{\lambda\nu}-\hat T_{can}^{\nu\lambda}
\end{align}}
\mz{with covariant derivative $D_{\mu}$. The (symmetric) gravitational (or, Belinfante - Rosenfeld) energy momentum tensor operator may then be expressed in terms of the canonical energy momentum tensor and the spin current by
\begin{align}
\hat T^{\mu\nu}=\hat T_{can}^{\mu\nu}+\frac{1}{2}D_{\lambda}(\hat S^{\mu\nu\lambda}+\hat S^{\nu\mu\lambda}-\hat S^{\lambda\nu\mu}).
\end{align}}
\mz{For the Zubarev statistical operator to properly describe a macroscopically moving medium we assume that a continuous medium, hydrodynamic description applies to the physical system under consideration. Due to the formulation in terms of thermodynamic quantities it is assumed that the physical system is at least in local thermodynamic equilibrium. In global thermodynamic equilibrium the Poincar\'e and current charges of Eq. (\ref{conservedcharges}) are conserved and thereby their initial and final values are equal.}\par
\mz{The statistical operator may be derived from the maximum entropy principle with constraints (here with just one vector current)
\begin{align}
n_{\mu}(x)\Tr(\hat{\rho}\hat{T}^{\mu\nu}(x))=n_{\mu}(x)T^{\mu\nu}_{cm}(x),\,\,\,\, n_{\mu}(x)\Tr(\hat{\rho}\hat{j}^{\mu}(x))=n_{\mu}(x)j^{\mu}_{cm}(x)
\end{align}
where $cm$ is short for continuous medium. Either $\hat{T}^{\mu\nu}\equiv \hat{T}^{\mu\nu}_{BR}$ or $\hat{T}^{\mu\nu}\equiv \hat{T}^{\mu\nu}_{can}$ with inclusion of the Lorentz transformation tensor operator $\hat{M}^{\mu\nu\rho}_{can}$. The resulting local thermodynamic equilibrium (LTE) statistical operator is precisely the Zubarev statistical operator 
\begin{align}
\hat{\rho}_{LTE}=\frac{1}{Z_{LTE}}\exp\Big[-\int_{\Sigma (\sigma)}d\Sigma (\sigma ) n_{\mu}(\hat{T}^{\mu\nu}(x)\beta_{\nu}(x)-\hat{j}^{\mu}(x)\zeta (x))\Big],\,\,\,\,Tr(\hat{\rho}_{LTE})=1
\end{align}
which fulfills 
\begin{align}
n_{\mu}(x)T^{\mu\nu}_{LTE}[\beta^{\rho} ,\zeta ,u](x)=n_{\mu}(x)T^{\mu\nu}_{cm}(x),\, n_{\mu}(x)j^{\mu}_{LTE}[\beta^{\rho},\zeta ,u](x)=n_{\mu}(x)j^{\mu}_{cm}(x)
\end{align}
from which $\beta^{\rho}$ and $\zeta$ may be determined with hydrodynamic local equilibrium energy momentum tensor and current operator
\begin{align}
T^{\mu\nu}_{LTE}[\beta^{\rho},\zeta ,u](x)=\Tr(\hat{\rho}_{LTE}\hat{T}^{\mu\nu}(x)),\,\,\,\, j^{\mu}_{LTE}[\beta^{\rho},\zeta ,u](x)=\Tr(\hat{\rho}_{LTE}\hat{j}^{\mu}(x))
\end{align}}
\mz{A preliminary Zubarev statistical operator constructed out of a set of general currents is projected by the global thermodynamic equilibrium condition onto the subspace of conserved currents (by constraining the current coefficients). The stationarity condition $\frac{d\hat\rho}{d\sigma} =0$ implies global thermodynamic equilibrium and requires the integrand to be divergence free \cite{Zubarev1979}. In order to have equilibrium we should require that the expression of Eq. (\ref{statisticalZub}) does not depend on $\sigma$. For this to be valid it is sufficient for the right-hand side of Eq. (\ref{statisticalZub}) not to depend on the form of $\Sigma$ at all. The requirement of global thermodynamic equilibrium is equivalent to 
\begin{align}
0=\partial_{\nu}\beta\Big(\hat T^{\nu\rho}u_{\rho}-\sum_i\mu_i\hat j_i^{\nu}\Big)=\hat T^{\nu\rho}\partial_{\nu}\beta u_{\rho}-\sum_i\hat j_i^{\nu}\partial_\nu\beta\mu_i.\label{Zub2}
\end{align}
We assumed conserved energy momentum and current operators that vanish at spacelike infinity and applied the Stokes theorem. This equation is satisfied by 
\begin{align}
\beta\mu_i=\zeta_i={\rm const},\,\,\,\,  \beta_\rho = \beta u_{\rho} &= b_\rho + \bar\omega_{\rho\sigma} x^\sigma ,\,\,\,\,\bar\omega_{\rho\sigma}=-\bar\omega_{\sigma\rho}
\label{Zub3}
\end{align}
with constant antisymmetric tensor $\bar\omega_{\rho\sigma}$, the thermal vorticity.}
\mz{Then
\begin{equation}
\beta(x) = \sqrt{b^2 + g^{\mu\rho}\bar\omega_{\mu\nu}\bar\omega_{\rho\sigma} x^\nu x^\sigma + 2 g^{\mu\rho}b_\mu  \bar\omega_{\rho\sigma} x^\sigma},\,\,\,\, u_\alpha (x) = \frac{b_\alpha + \bar\omega_{\alpha\lambda} x^\lambda}{\sqrt{b^2 + g^{\mu\rho}\bar\omega_{\mu\nu}\bar\omega_{\rho\sigma} x^\nu x^\sigma + 2 g^{\mu\rho}b_\mu  \bar\omega_{\rho\sigma} x^\sigma}}
\end{equation}
One can define the uniform four velocity 
\begin{align}
v_{\mu}=\frac{1}{\beta (x)}b_{\mu},
\end{align}
the four acceleration
\begin{equation}
a_\alpha =\frac{1}{\beta (x)}\bar\omega_{\alpha\lambda}u^\lambda = \frac{\bar\omega_{\alpha\lambda}(b^\lambda + \bar\omega^\lambda_{\,\,\,\,\gamma} x^\gamma)}{b^2 + g^{\mu\rho}\bar\omega_{\mu\nu}\bar\omega_{\rho\sigma} x^\nu x^\sigma + 2 g^{\mu\rho}b_\mu  \bar\omega_{\rho\sigma} x^\sigma} \label{Zub3}
\end{equation}
and the angular vorticity
\begin{align}
\omega_\alpha = -\frac{1}{2\beta (x)}\epsilon_{\alpha\beta\gamma\delta}u^\beta \bar\omega^{\gamma\delta}=-\frac{\epsilon_{\alpha\beta\gamma\delta}(b^\beta + \bar\omega^\beta_{\,\,\,\,\tau} x^\tau) \bar\omega^{\gamma\delta}}{2({b^2 + g^{\mu\rho}\bar\omega_{\mu\nu}\bar\omega_{\rho\sigma} x^\nu x^\sigma + 2 g^{\mu\rho}b_\mu  \bar\omega_{\rho\sigma} x^\sigma})}
\end{align}
implying
\begin{align}
\bar\omega_{\mu\nu}=\beta (\epsilon_{\mu\nu\rho\sigma}\omega^{\rho}u^{\sigma}+a_{\mu}u_{\nu}-a_{\nu}u_{\mu}).
\end{align}
The chemical potential receives the form
\begin{equation}
\mu_i(x)=\frac{\zeta_i}{\beta(x)} = \frac{\zeta_i}{\sqrt{b^2 + g^{\mu\rho}\bar\omega_{\mu\nu}\bar\omega_{\rho\sigma} x^\nu x^\sigma + 2 g^{\mu\rho}b_\mu  \bar\omega_{\rho\sigma} x^\sigma}}.
\end{equation}}

\begin{figure}
\begin{center}
\includegraphics[scale=0.64]{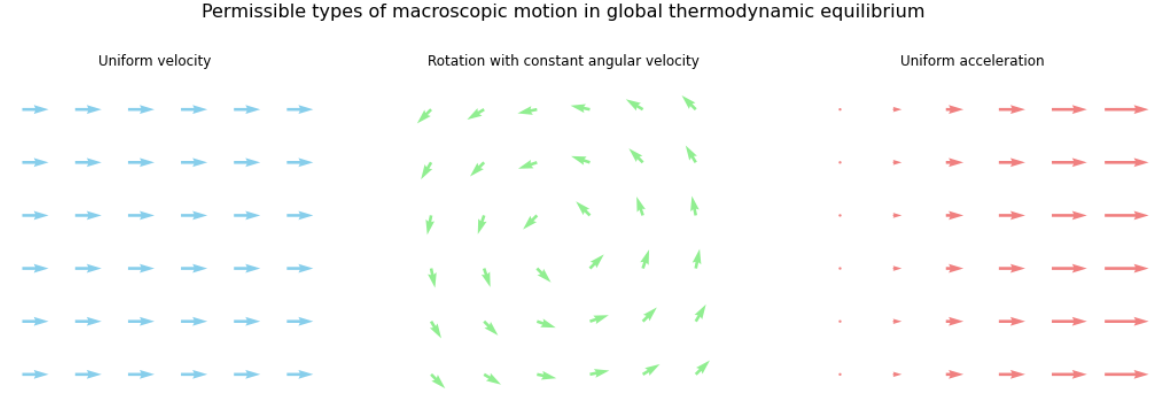}
\end{center}
\caption{The allowed types of macroscopic motion of a substance in global thermodynamic equilibrium at constant inverse temperature $\beta =\beta_0$ for the case of full Poincar\'e symmetry. On the left-hand side we illustrate the case of uniform velocity as parametrized by $v^{\mu}$. The central part of the figure depicts the case of rotation as parametrized by $\omega^{\mu}$. On the right-hand side we illustrate the case of uniform acceleration as parametrized by $a^{\mu}$. The most general type of allowed macroscopic motion in global thermodynamic equilibrium is a superposition of these three types of motion.}
\label{allowedtypesmacromotion}
\end{figure}

\mz{We illustrate the allowed types of macroscopic motion in the presence of full Poincar\'e symmetry in Fig (\ref{allowedtypesmacromotion}). These comprise uniform motion (as parametrized by $v^{\mu}$ at constant $\beta =\beta_0$), rotation (as parametrized by $\omega^{\mu}$ at constant $\beta =\beta_0$) and uniform acceleration (as parametrized by $a^{\mu}$ at constant $\beta =\beta_0$). In general these types of motion may be present simultaneously.}\par
\mz{We may prove the equivalence (up to a timelike boundary term at spacelike infinity where we assume the current operators to vanish) of using either the canonical Poincar\'e currents or the Belinfante - Rosenfeld energy momentum tensor under the assumptions of current conservation and global thermodynamic equilibrium as follows. Assume first that $D_{\mu}T^{\mu\nu}_{can}=D_{\mu}M^{\mu\nu\rho}_{can}=0$. Then $D_{\mu}T^{\mu\nu}=0$ and
\begin{align}
\nonumber &\int d\Sigma n_{\mu}[T^{\mu\nu}_{can}b_{\nu}+\frac{1}{2}M^{\mu\nu\rho}_{can}\overline{\omega}_{\nu\rho}]\\
\nonumber &=\int d\Sigma n_{\mu}[T^{\mu\nu}_{can}(b_{\nu}+\overline{\omega}_{\rho\nu}x^{\rho})+\frac{1}{2}S^{\mu\nu\rho}\overline{\omega}_{\nu\rho}]\\
\nonumber &=\int d\Sigma n_{\mu}[T^{\mu\nu}\beta_{\nu}-\frac{1}{2}D_{\rho}(S^{\mu\nu\rho}+S^{\nu\mu\rho}-S^{\rho\nu\mu})\beta_{\nu}+\frac{1}{2}S^{\mu\nu\rho}\overline{\omega}_{\nu\rho}]\\
\nonumber &=\int d\Sigma n_{\mu}[T^{\mu\nu}\beta_{\nu}+\frac{1}{2}(S^{\mu\nu\rho}+S^{\nu\mu\rho}-S^{\rho\nu\mu})\partial_{\rho}\beta_{\nu}+\frac{1}{2}S^{\mu\nu\rho}\overline{\omega}_{\nu\rho}]\\
\nonumber &=\int d\Sigma n_{\mu}[T^{\mu\nu}\beta_{\nu}+\frac{1}{2}S^{\mu\nu\rho}\partial_{\rho}\beta_{\nu}+\frac{1}{2}S^{\mu\nu\rho}\overline{\omega}_{\nu\rho}]\\
&=\int d\Sigma n_{\mu}T^{\mu\nu}\beta_{\nu}
\end{align}}
\mz{with $\beta_{\nu}=b_{\nu}+\overline{\omega}_{\rho\nu}x^{\rho}$ and $b_{\nu},\overline{\omega}_{\rho\nu}=const.$ (the latter due to canonical current conservations). Reversely from $D_{\mu}T^{\mu\nu}=0$ and
\begin{align}
0=T^{[\mu\nu ]}=T^{[\mu\nu ]}_{can}-\frac{1}{2}D_{\rho}S^{\rho\nu\mu},\,\,\,\,T^{[\mu\nu ]}_{(can)}=\frac{1}{2}(T^{\mu\nu}_{(can)}-T^{\nu\mu}_{(can)})
\end{align}}
\mz{we obtain $D_{\mu}T^{\mu\nu}_{can}=D_{\mu}M^{\mu\nu\rho}_{can}=0$ and $\beta_{\nu}=b_{\nu}+\overline{\omega}_{\rho\nu}x^{\rho}$ with $b_{\nu},\overline{\omega}_{\rho\nu}=const.$ from the Killing equation for the frigiditiy vector field $\beta_{\nu}$ in global thermodynamic equilibrium from the Belinfante - Rosenfeld energy momentum tensor conservation. We may then simply go through the above manipulations in reverse.}\par 
\mz{The following integrals of motion enter the expression for the statistical operator
\begin{align}
\hat P^\mu =\int d\Sigma n_\nu \hat T^{\nu \mu},\,\,\,\,\hat J^{\mu\nu}=\int d\Sigma n_\rho (\hat x^{\mu}T^{\rho \nu}-x^{\nu}\hat T^{\rho\mu}),\,\,\,\,\hat Q_i=\int d\Sigma n_\nu \hat j_i^\nu .
\end{align}
We obtain
\begin{equation}
\rho = \frac{1}{Z}e^{-b_\mu \hat P^\mu + \frac{1}{2}\bar{\omega}_{\mu\nu} \hat J^{\mu\nu} + \sum_i \zeta_i \hat Q_i}\label{rho00}.
\end{equation}
The tensor operator $\hat{J}^{\mu\nu}$ may be decomposed as
\begin{equation}
	\hat J^{\mu\nu} = \epsilon_{\mu\nu\alpha\beta}\hat J^\alpha u^\beta - \hat K_\mu u_\nu + \hat K_\nu u_\mu .
\end{equation}
Here $K_\mu$ is the generator of boosts, while $J_\nu$ is generator of rotation (both are taken in the comoving reference frame). In terms of these generators we obtain the following expression for the statistical operator
\begin{equation}
\hat	\rho = \frac{1}{Z}e^{-\beta (v_\mu \hat P^\mu + a_\mu \hat K^\mu  - {\omega}_{\mu} \hat J^{\mu} - \sum_i \mu_i \hat Q_i)}.\label{rho0}
\end{equation}
In the macroscopic motion rest frame $v^{\mu}=(1,0,0,0)$. In this frame and in absence of macroscopic rotation and acceleration ($a_{\mu}=\omega_{\mu}=0$) we recover the form familiar from statistical physics (with Hamilton operator $\hat{H}=\hat{P}^0$)
\begin{equation}
\hat	\rho = \frac{1}{Z}e^{-\beta (\hat{H}-\sum_i \mu_i \hat Q_i)}.
\end{equation}}

\subsection{The Zubarev statistical operator within superfluid $^3$He - A}

We devote this part of the section to the calculation of several quantities within the theory of massless (chiral) fermions in $^3$He - A in order to motivate the form of the global equilibrium Zubarev statistical operator which we introduce in the next section. \par
We will make use of the short-hand notations $T_{C=L/R}$ and $T^{s=\pm}$ in order to indicate that we only consider the $C=R/L$- or $s=\pm$-part of a tensor field $T$.

\subsubsection{Equations of motion}
The Euler-Lagrange equations of motion, derived from the variation of the action $\delta S/\delta \psi (x) = \delta S/\delta \bar{\psi}(x) = 0$, take the form
\begin{align}
0=&\frac{\partial\mathcal{L}}{\partial \psi }-\nabla_{\mu}\frac{\partial\mathcal{L}}{\partial (\nabla_{\mu}\psi )}=-2(\overline{\psi}\gamma^0\overset{\leftarrow}{\nabla}_{\mu})i\bold{e}_a^{\mu}\gamma^a-\overline{\psi}\gamma^0i(\nabla_{\mu}\bold{e}_a^{\mu})\gamma^a\\
0=&\frac{\partial\mathcal{L}_{L}}{\partial \psi_L}-\nabla_{\mu}\frac{\partial\mathcal{L}_L}{\partial (\nabla_{\mu}\psi_L)}=-2(\overline{\psi}_L\overset{\leftarrow}{\nabla}_{\mu})i\bold{e}_{a}^{\mu}\overline{\tau}^a-\overline{\psi}_Li\nabla_{\mu}\bold{e}_a^{\mu}\overline{\tau}^a\\
0=&\frac{\partial\mathcal{L}}{\partial \overline{\psi}}-\nabla_{\mu}\frac{\partial\mathcal{L}}{\partial (\nabla_{\mu}\overline{\psi})}=2i\gamma^0\bold{e}_a^{\mu}\gamma^a\nabla_{\mu}\psi +i\gamma^0(\nabla_{\mu}\bold{e}_a^{\mu})\gamma^a\psi \\
0=&\frac{\partial\mathcal{L}_{L}}{\partial \overline{\psi}_L}-\nabla_{\mu}\frac{\partial\mathcal{L}_L}{\partial (\nabla_{\mu}\overline{\psi}_L)}=2i\bold{e}_{a}^{\mu}\overline{\tau}^a\nabla_{\mu}\psi_L+i(\nabla_{\mu}\bold{e}_a^{\mu})\overline{\tau}^a\psi_L
\end{align}
and equivalently for the right-handed case with simultaneous exchange $\tau\leftrightarrow\overline{\tau}$. The equations of motion feature an extra term as compared to the case of massless relativistic Weyl or Dirac fermions due to the divergence of the vierbein. In terms of the geometric formulation the equations of motion read
\begin{align}
0=\frac{\partial\mathcal{L}}{\partial\Psi}-D_{\mu}\frac{\partial\mathcal{L}}{\partial (D_{\mu}\Psi)}=&-\overline{\Psi}\gamma^0\overset{\leftarrow}{D}_{\mu}i\gamma^be_b^{\mu}-\overline{\Psi}\gamma^0i\gamma^be_b^{\mu}\overset{\leftarrow}{D}_{\mu}\nonumber\\
=&-2\overline{\Psi}\gamma^0\overset{\leftarrow}{\nabla}_{\mu}i\gamma^be_b^{\mu}+\overline{\Psi}\gamma^0\gamma^be_b^{\mu}\mathcal{B}_{\mu}+\overline{\Psi}\gamma^0\mathcal{B}_{\mu}\gamma^be_b^{\mu}-\overline{\Psi}\gamma^0i\gamma^b(\nabla_{\mu}e_b^{\mu}) \label{eomwithoutconstraint2}\\
0=\frac{\partial\mathcal{L}}{\partial\overline{\Psi}}-D_{\mu}\frac{\partial\mathcal{L}}{\partial (D_{\mu}\overline{\Psi})}=&i\gamma^0\gamma^be_b^{\mu}D_{\mu}\Psi+\gamma^0D_{\mu}i\gamma^be_b^{\mu}\Psi\nonumber\\
=&2i\gamma^0\gamma^be_b^{\mu}\nabla_{\mu}\Psi+\gamma^0\gamma^be_b^{\mu}\mathcal{B}_{\mu}\Psi +\gamma^0\mathcal{B}_{\mu}\gamma^be_b^{\mu}\Psi +i\gamma^0\gamma^b(\nabla_{\mu}e_b^{\mu})\Psi .
\label{eomwithoutconstraint}
\end{align}
%\mz{${\rm det}\, e$ depends on coordinates, at least in the core of the topological defects. Therefore, its might be important as well. It may be useful to use representation of Eq. (2.33) in https://arxiv.org/pdf/hep-th/0103093.pdf in order to obtain Dirac equation for $\psi$ immediately via variation with respect to $\bar{\psi}$.}
{Enforcement of the constraint $\overline{\Psi}=\Psi^T\hat{U}\hat{d}^{\ast}$ implies equivalence of the two equations of motion for $\Psi$ and $\overline{\Psi}$ with respect to each other (see also Eq. (\ref{eomwithconstraint}) in Appendix \ref{app1} following Eq. (\ref{dderivative}).
The equations of motion imply the relations
\begin{align}
\mathcal{L}\rvert_{EOM}=\mathcal{L}^s\rvert_{EOM}=\mathcal{L}_C\rvert_{EOM}=\mathcal{L}^s_C\rvert_{EOM}=0,\,\,\,\,C=L/R,\,\,s=\pm
\end{align} 
where we indicated that the respective Lagrangians are to be evaluated on solutions of the equations of motion. 

\subsubsection{Conserved and non-conserved currents}
The effective action represented in Eqs. (\ref{weylaction}), (\ref{diracaction}) and (\ref{relaction}) exhibits both spacetime and internal symmetries. Some of the former may be explicitly broken due to inhomogeneous vierbein. After making the transition from a matrix valued vierbein to a scalar valued vierbein accompanied by a non-Abelian gauge field, the inhomogeneity may arise either from the scalar valued vierbein or the non-Abelian gauge field or both. Let us consider the vector current
\begin{align}
j^{\mu}_V=\frac{1}{2}\overline{\Psi}\gamma^0e_a^{\mu}\gamma^a\Psi =\frac{1}{2}\overline{\psi}\gamma^0\bold{e}_a^{\mu}\gamma^a\psi
\label{vectorcurrent}
\end{align}
as well as the axial current
\begin{align}
j^{\mu}_A=\frac{1}{2}\overline{\Psi}\gamma^0e_a^{\mu}\gamma^5\gamma^a\Psi =\frac{1}{2}\overline{\psi}\gamma^0\bold{e}_a^{\mu}\gamma^5\gamma^a\psi .
\label{axialcurrent}
\end{align}
Both are conserved as a consequence of the equations of motion. 
%As Noether currents they emerge from the symmetry operations
%\begin{align}
%\psi_L\to e^{i\alpha}\psi_L,\,\,\,\, \overline{\psi}_L\to \overline{\psi}_Le^{-i\alpha},\,\,\,\, \psi_R\to e^{i\alpha}\psi_R,\,\,\,\, \overline{\psi}\to\overline{\psi}e^{-i\alpha}
%\end{align}
%for the vector current and 
%\begin{align}
%\psi_L\to e^{i\alpha}\psi_L,\,\,\,\, \overline{\psi}_L\to \overline{\psi}_Le^{-i\alpha},\,\,\,\, \psi_R\to e^{-i\alpha}\psi_R,\,\,\,\, \overline{\psi}\to\overline{\psi}e^{i\alpha}
%\end{align}
%for the axial current. 
At the quantum level, only the vector current remains conserved, while the axial current is anomalous. We may instead consider a modification of the axial current and arrive at the quantum conserved total current $j^{\mu}_{CS}$ which takes the form
\begin{align}
j^{\mu}_{CS}=j^{\mu}_A+K^{\mu} + K_{\cal A}^\mu
\end{align}
with axial Chern - Simons vector field $K_{\cal A}^{\mu}$ (as well as axial field strength tensor $\mathcal{F}_{\mu\nu}$) 
\begin{align}
K^{\mu}_{\mathcal{A}}=\frac{1}{16\pi^2}\epsilon^{\mu\nu\rho\sigma}\mathcal{A}_{\nu}\mathcal{F}_{\rho\sigma},\,\,\,\, \mathcal{F}_{\mu\nu}=\partial_{\mu}\mathcal{A}_{\nu}-\partial_{\nu}\mathcal{A}_{\mu}
\label{axialchernsimonscurrent}
\end{align}
and torsional Chern - Simons vector field
\begin{align}
K^{\mu}=-\frac{\Lambda^2}{4\pi^2}\frac{1}{4}Tr(\epsilon^{\mu\nu\rho\sigma}\bold{e}^a_{\nu}\bold{T}^b_{\rho\sigma}\bold{e}^m_a\bold{e}_{bm})=-\frac{\Lambda^2}{4\pi^2}\frac{1}{2}\epsilon^{\mu\nu\rho\sigma}e_{a\nu}T^a_{\rho\sigma}.
\label{chernsimonscurrent}
\end{align}
%\mzo{Please take into account axial Chern - Simons field, which will lead to the ordinary term in chiral anomaly proportional to $\epsilon^{\mu\nu\rho\sigma} \partial_{\mu}{\cal A}_\nu \partial_{\rho}{\cal A}_\sigma$. }
The parameter $\Lambda$ represents a UV cutoff scale for the anomaly at zero temperature. Further discussion and corrections of the anomaly in the context of $^3$He-A may be found in \cite{NissinenVolovik2019,NissinenVolovik2020,Nissinen2020,LaurilaNissinen2020}. The quantity
\begin{align}
\bold{T}^a_{\mu\nu}=\nabla_{\mu}\bold{e}^a_{\nu}-\nabla_{\nu}\bold{e}^a_{\mu}
\end{align}
is the matrix-valued torsion tensor field. The current $j_{CS}^{\mu}$ is conserved, but not gauge invariant. Only its global charge is both conserved and gauge invariant. The divergence of the vector field $K^{\mu}$, $\nabla_{\mu}K^{\mu}$, is known as the Nieh-Yan form that, together with the corresponding current divergence $\nabla_{\mu}K_{\mathcal{A}}^{\mu}$, obstructs the conservation of the classically conserved axial current in the quantum theory.
The axial current in Eq. (\ref{axialcurrent}) with matrix-valued vierbein $\bold{e}_a^{\mu}$ can be represented equivalently in terms of a scalar-valued vierbein $e_a^{\mu}$ comprising two left- and right-handed fermion species, the eigenstates of $\bold{d}\boldsymbol{\sigma}$ for either chirality. Both species yield the same contribution to the anomaly. 
%The states $(L,s=\pm )$ enter with the same sign into the anomaly which is opposite to that of the states $(R,s=\pm )$. 
It may be checked explicitly that $(\bold{d}\boldsymbol{\sigma})$ drops out of Eq. (\ref{chernsimonscurrent}), as the final equality indicates. This means that the emergent gauge field $\mathcal{B}_{\mu}$, which has been fully considered by $K^{\mu}$, is not relevant for the chiral anomaly.

Let us define
\begin{align}
a=-\frac{1}{8\pi^2}\epsilon^{\mu\nu\rho\sigma}\nabla_{\mu}T_{\nu\rho\sigma}=-\frac{1}{16\pi^2}\epsilon^{\mu\nu\rho\sigma}T^a_{\mu\nu}T^b_{\rho\sigma}\eta_{ab},\,\,\,\, T_{\nu\rho\sigma}=e_{a\nu}T^a_{\rho\sigma}
\end{align}
such that
\begin{align}
\nabla_{\mu}K^{\mu}=\Lambda^2a.
\end{align}
The form $a$ is called the Nieh - Yan form, while its integral over space - time is known as the Nieh - Yan topological invariant. Its value is not changed if the vierbein is modified smoothly. 
%\mzo{There is another form of $a$, which is proportional to $\epsilon^{\mu\nu\rho\sigma}T^a_{\mu\nu}T^b_{\rho\sigma}\eta_{ab}$. Please present this form as well.}
Two additional conserved currents appear due to the eigenspinor separation with respect to the spin operator $(\bold{d}\boldsymbol{\sigma})$ in the case of vanishing Abelian Berry curvatures $b^+_{\mu}=b^-_{\mu}=0$. These have the form
\begin{align}
j^{\mu}_{sA}=\frac{1}{2}\overline{\Psi}\gamma^0e_a^{\mu}\gamma^a\sigma^3\Psi
\end{align}
as well as
\begin{align}
j^{\mu}_{M}=\frac{1}{2}\overline{\Psi}\gamma^0e_a^{\mu}\gamma^5\gamma^a\sigma^3\Psi .
\end{align}
We use the subscript $sA$ as a short-hand notation for spin-axial and the subscript $M$ for mixed. The currents with subscripts $V,A,sA,M$ may be obtained by variation of the action with respect to corresponding fictitious gauge fields $A^i_{\mu}$ and charges $q^i$ living in chiral and spin subspaces with indices $i\in\{(L,+),(L,-),(R,+),R,-)\}$ and fictitious covariant derivatives $D^i_{\mu}=\nabla_{\mu}-iq^iA^i_{\mu}$, respectively. We may address the individual currents by the choices of charges $q^{L/R,\pm}=1$ for $j_V$, $q^{L,\pm}=1,q^{R,\pm}=-1$ for $j_A$, $q^{L,+},q^{R,+}=1,q^{L,-}=q^{R,-}=-1$ for $j_{sA}$ and $q^{L,+}=q^{R,-}=1,q^{L,-}=q^{R,+}=-1$ for $j_M$, respectively. The currents may then be obtained from the action by variation according to
\begin{align}
j_i^{\mu}(x)=\frac{1}{e}\frac{\delta S}{\delta (q^iA^i_{\mu}(x))}
\end{align}
where $S$ is expressed in terms of the covariant derivative $D^i_{\mu}$. Four more currents which are non-diagonal in the internal spin space may be constructed which are conserved if $b^+=b^-$ and either $Re(\omega_{\mu 12})=0$ or $Im(\omega_{\mu 12})=0$. These may again be derived from the action by introducing appropriate fictitious gauge fields. Collectively, the currents may be denoted as
\begin{align}
j^{a\mu}_V=&\frac{1}{2}\overline{\Psi}\gamma^0e_b^{\mu}\gamma^b\sigma^a\Psi \label{generalizedvectorcurrents}\\
j^{a\mu}_A=&\frac{1}{2}\overline{\Psi}\gamma^0e_b^{\mu}\gamma^5\gamma^b\sigma^a\Psi .
\label{vectorandaxialcurrents}
\end{align}
The divergences of these currents are given by 
\begin{align}
\nabla_{\mu}j^{a\mu}_V=&-\frac{1}{4}\overline{\Psi}\gamma^0ie_b^{\mu}[\sigma^a,\{\mathcal{B}_{\mu},\gamma^b\}]\Psi =-\frac{1}{4}\overline{\Psi}\gamma^0ie_b^{\mu}\{[\sigma^a,\mathcal{B}_{\mu}],\gamma^b\}\Psi \\
\nabla_{\mu}j^{a\mu}_A=&-\frac{1}{4}\overline{\Psi}\gamma^0i\gamma^5e_b^{\mu}[\sigma^a,\{\mathcal{B}_{\mu},\gamma^b\}]\Psi =-\frac{1}{4}\overline{\Psi}\gamma^0i\gamma^5e_b^{\mu}\{[\sigma^a,\mathcal{B}_{\mu}],\gamma^b\}\Psi .
\end{align}
The conservation condition reduces thereby to the algebraic condition
\begin{align}
\{[\sigma^a,\mathcal{B}_{\mu}],\gamma^b\}=0
\label{algebraiccondition}
\end{align}
in both vector and axial currents. This is trivially fulfilled in the case $\nabla_{\mu}\bold{d}=0$ and more specifically for the individual currents under the conditions mentioned for each case. As has been discussed in the previous section, the relativistic effective action of $^3$He - A is distinct from the ordinary relativistic action for massless Weyl fermions by the lack or vanishing of the Dirac space spin connection which is the gauge field arising from local Lorentz invariance. Its absence leaves us with a not necessarily symmetric energy momentum tensor arising from the gauge field of translation, namely the vierbein,
\begin{align}
T_a^{\mu}(x)=-\frac{1}{e}\frac{\delta S}{\delta e^a_{\mu}(x)}=&\frac{1}{4}\overline{\Psi}\gamma^0ie_b^{\mu}\gamma^be_a^{\nu}D_{\nu}\Psi -\frac{1}{4}\overline{\Psi}\gamma^0\overset{\leftarrow}{D}_{\nu}e_a^{\nu}ie_b^{\mu}\gamma^b\Psi =\frac{1}{4}[\overline{\psi}\gamma^0i\bold{e}_b^{\mu}\gamma^b e_a^{\nu}\nabla_{\nu}\psi -\overline{\psi}\gamma^0\overset{\leftarrow}{\nabla}_{\nu}e_a^{\nu}i\bold{e}_b^{\mu}\gamma^b\psi ]
\label{emt}
\end{align}
where we employed the relation
\begin{align}
0=\frac{\delta}{\delta e_{\nu}^c} \delta^a_b=\frac{\delta}{\delta e_{\nu}^c}(e_{\mu}^a e_b^{\mu}) \Leftrightarrow \frac{\delta e_b^{\nu}}{\delta e_{\mu}^a}=-e_a^{\nu} e_b^{\mu}.\nonumber 
\end{align}
The energy momentum tensor in Eq. (\ref{emt}) seems to be ambiguous or even inappropriate which can be seen from the rightmost equality. Had we chosen to rotate the $\Psi^+$ component instead of $\Psi^-$ and chosen $e_a^{\mu}=(e^-)_a^{\mu}$ this energy momentum tensor would have been different. It turns out that we may still stick with this definition of the energy momentum tensor, since
\begin{align}
T^{\mu}_{\,\,\,\,\nu}=T_a^{\mu}e^a_{\nu}=-\frac{1}{e}e^a_{\nu}\frac{\delta S}{\delta e^a_{\mu}(x)}=\frac{1}{4}\overline{\Psi}\gamma^0ie_b^{\mu}\gamma^bD_{\nu}\Psi -\frac{1}{4}\overline{\Psi}\gamma^0\overset{\leftarrow}{D}_{\nu}ie_b^{\mu}\gamma^b\Psi =\frac{1}{4}[\overline{\psi}\gamma^0i\bold{e}_b^{\mu}\gamma^b\nabla_{\nu}\psi -\overline{\psi}\gamma^0\overset{\leftarrow}{\nabla}_{\nu}i\bold{e}_b^{\mu}\gamma^b\psi ]
\end{align}
is unambiguous and coincides with the on-shell canonical energy momentum tensor derived from the Noether procedure.\par
%It coincides with the canonical energy momentum tensor obtained by
%\begin{align}
% (T_{can})_a^{\mu}=&(\frac{\partial\mathcal{L}}{\partial (\nabla_{\mu}\psi )}e_a^{\nu}\nabla_{\nu}\psi +[\nabla_{\nu}\overline{\psi}]e_a^{\nu}\frac{\partial\mathcal{L}}{\partial (\nabla_{\mu}\overline{\psi})})\\
%=&\sum_{C=L,R}(\frac{\partial\mathcal{L}}{\partial (\nabla_{\mu}\psi_C)}e_a^{\nu}\nabla_{\nu}\psi_C+[\nabla_{\nu}\overline{\psi}_C]e_a^{\nu}\frac{\partial\mathcal{L}}{\partial (\nabla_{\mu}\overline{\psi}_C)})\\
%=&\sum_{s=\pm ,C=L,R}(\frac{\partial\mathcal{L}}{\partial (\nabla_{\mu}\psi^s_C)}e_a^{\nu}\nabla_{\nu}\psi^s_C+[\nabla_{\nu}\overline{\psi}^s_C]e_a^{\nu}\frac{\partial\mathcal{L}}{\partial (\nabla_{\mu}\overline{\psi}^s_C)})
%\end{align}
%where we explicitly made use of the earlier finding $\mathcal{L}_C^s\rvert_{EOM}=0$. 
The energy momentum tensor is not conserved but obeys
\begin{align}
\nabla_{\mu}T^{\mu}_a=T_{ac}^{\rho}e^d_{\rho}e^c_{\mu}T^{\mu}_d-j_A^{\mu}e_a^{\nu}\mathcal{F}_{\mu\nu}
\label{tdivergence}
\end{align}
in the case $\nabla_{\mu}\bold{d}=0$. For inhomogeneous $\bold{d}$ we have to add the extra term
\begin{align}
G_a=\sum_{b=1,2}\Big[-\frac{1}{4}\overline{\psi}\gamma^0i\gamma^be_a^{\nu}(\boldsymbol{\sigma}\nabla_{\nu}\bold{d})e_b^{\mu}\nabla_{\mu}\psi +\frac{1}{4}\overline{\psi}\gamma^0\overset{\leftarrow}{\nabla}_{\mu}e_b^{\mu}i\gamma^be_a^{\nu}(\boldsymbol{\sigma}\nabla_{\nu}\bold{d})\psi \Big]
\label{tdivergenceextra}
\end{align}
to $\nabla_{\mu}T_a^{\mu}$. This term vanishes specifically for homogeneous $\bold{d}$ in both the $e_1$ and $e_2$ directions. $G_0=0$, if $\bold{d}$ is time independent and $G_i=0$, if $\bold{d}$ is homogeneous in space. It furthermore vanishes if only one of the eigenstates $s=\pm$ is present implying
\begin{align}
\nabla_{\mu}(T^{\pm})_a^{\mu}=T_{ac}^{\rho}e^d_{\rho}e^c_{\mu}(T^{\pm})^{\mu}_d-(j_A^{\pm})^{\mu}e_a^{\nu}\mathcal{F}_{\mu\nu} \,\,\text{for}\,\, S=\int d^4x e\mathcal{L}^{\pm}.
\end{align}
A derivation of this result is detailed in Appendix \ref{app2}. 
The change of the energy momentum tensor under translations is proportional to the curvature tensor (namely the torsion tensor) associated with the gauge field of translation, or vierbein, $e_a^{\mu}$ plus an extra term for inhomogeneous $\bold{d}$ that couples the eigenstates of $(\bold{d}\boldsymbol{\sigma})$.\par When the above steps in the calculation are repeated in the geometric formulation comprising the non-Abelian gauge field $\mathcal{B}_{\mu}$, the divergence of the energy momentum tensor may be expressed as
\begin{align}
\nabla_{\mu}T_a^{\mu}=&T_c^{\rho}e^b_{\rho}e^c_{\nu}T_{ab}^{\nu}-j_A^{\mu}e_a^{\nu}\mathcal{F}_{\mu\nu}+\frac{1}{2}\overline{\Psi}\gamma^0e_a^{\nu}e_b^{\mu}F^b_{\mu\nu}\Psi +\frac{1}{8}\overline{\Psi}\gamma^0(\nabla_{\mu}e_b^{\mu})\gamma^be_a^{\nu}\mathcal{B}_{\nu}\Psi +\frac{1}{8}\overline{\Psi}\gamma^0e_a^{\nu}\mathcal{B}_{\nu}(\nabla_{\mu}e_b^{\mu})\gamma^b\Psi \label{geononconservationofemt1}\\
%\overset{EOM}{=}&T_c^{\rho}e^b_{\rho}e^c_{\nu}T_{ab}^{\nu}+\frac{1}{2}\overline{\Psi}\gamma^0e_a^{\nu}e_b^{\mu}G^b_{\mu\nu}\Psi -\frac{1}{4}[\nabla_{\mu}\overline{\Psi}]\gamma^0e_b^{\mu}\gamma^be_a^{\nu}\mathcal{B}_{\nu}\Psi -\frac{1}{4}\overline{\Psi}\gamma^0e_a^{\nu}\mathcal{B}_{\nu}e_b^{\mu}\gamma^b\nabla_{\mu}\Psi \label{geononconservationofemt2}\\
=&T_c^{\rho}e^b_{\rho}e^c_{\nu}T_{ab}^{\nu}-j_A^{\mu}e_a^{\nu}\mathcal{F}_{\mu\nu}+G_a
\end{align}
with non-Abelian field strength tensor
\begin{align}
&2F^b_{\mu\nu}=\nabla_{\mu}\{\gamma^b,\mathcal{B}_{\nu}\}-\nabla_{\nu}\{\gamma^b,\mathcal{B}_{\mu}\}-\frac{i}{2}[\{\gamma^b,\mathcal{B}_{\mu}\},\mathcal{B}_{\nu}]\\
%&2G^b_{\mu\nu}=\nabla_{\mu}\{\gamma^b,\mathcal{B}_{\nu}\}-\nabla_{\nu}\{\gamma^b,\mathcal{B}_{\mu}\}-i[\{\gamma^b,\mathcal{B}_{\mu}\},\mathcal{B}_{\nu}].
\end{align}
The first two terms in Eq. (\ref{geononconservationofemt1}) indicate non-conservation of the energy momentum tensor proportional to the curvatures associated with the vierbein $e_a^{\mu}$ and the non-Abelian gauge field $\mathcal{B}_{\mu}$, respectively. The final two terms in Eq. (\ref{geononconservationofemt1}) would be absent, if the Euler-Lagrange equations had the form
\begin{align}
\frac{\partial\mathcal{L}}{\partial\Psi}-D_{a}\frac{\partial\mathcal{L}}{\partial (D_{a}\Psi)}=0=\frac{\partial\mathcal{L}}{\partial\overline{\Psi}}-D_{a}\frac{\partial\mathcal{L}}{\partial (D_{a}\overline{\Psi})}
\end{align}
instead of
\begin{align}
\frac{\partial\mathcal{L}}{\partial\Psi}-D_{\mu}\frac{\partial\mathcal{L}}{\partial (D_{\mu}\Psi)}=0=\frac{\partial\mathcal{L}}{\partial\overline{\Psi}}-D_{\mu}\frac{\partial\mathcal{L}}{\partial (D_{\mu}\overline{\Psi})}.
\end{align} 
The details of the derivation in the geometric formulation are exhibited in Appendix \ref{app2}. \par
The Lorentz transformation tensor may be written in terms of an orbital part and a spin part 
\begin{align}
M^{\mu}_{ab}=x^{\nu}e_{a\nu}T^{\mu}_b-x^{\nu}e_{b\nu}T^{\mu}_a+S^{\mu}_{ab}
\label{ltt}
\end{align}
with the spin tensor 
\begin{align}
S^{\mu}_{ab}(x)=\frac{1}{e}\frac{\delta S}{\delta \omega_{\mu}^{ab}(x)}\,\,\text{(with $D_{\mu}$ comprising the spin connection in the action $S$)}.
\label{spintensordefinition}
\end{align}
The orbital part is expressed in terms of the energy momentum tensor. We include it here but we will see in the next section that its direct presence is irrelevant within the Zubarev statistical operator due to a cancellation. Thus we may effectively set $M^{\mu}_{ab}\to S^{\mu}_{ab}$. Straightforward evaluation of Eq. (\ref{spintensordefinition}) yields
\begin{align}
S^{\mu}_{ab}=&\frac{i}{16}\overline{\Psi}\gamma^0e_c^{\mu}\{\gamma^c,[\gamma_a,\gamma_b]\}\Psi =\frac{1}{4}\epsilon_{abcd}\overline{\Psi}\gamma^0e^{c\mu}\gamma^5\gamma^d\Psi \label{spintensor1}\\
\neq&\frac{i}{16}\overline{\psi}\gamma^0\bold{e}_c^{\mu}\{\gamma^c,[\gamma_a,\gamma_b]\}\psi =\frac{1}{4}\epsilon_{abcd}\overline{\psi}\gamma^0\bold{e}^{c\mu}\gamma^5\gamma^d\psi \label{spintensor2}
\end{align}
The difference between Eqs. (\ref{spintensor1}) and (\ref{spintensor2}) resides in the additional phase factor multiplying $\Psi^-$ within $\Psi$ which does not commute with $[\gamma^a,\gamma^b]$ in general and may then lead to additional minus signs within some of the spin tensor components. Eq. (\ref{spintensor1}), when expressed entirely in spacetime indices, takes the form
\begin{align}
S^{\mu\nu\rho}=S^{\mu}_{ab}e^{a\nu}e^{b\rho}=\frac{i}{16}\overline{\Psi}\gamma^0e_a^{\nu}e_b^{\rho}e_c^{\mu}\{\gamma^c,[\gamma^a,\gamma^b]\}\Psi =\frac{i}{16}\overline{\psi}\gamma^0\bold{e}_a^{\nu}\bold{e}_b^{\rho}\bold{e}_c^{\mu}\{\gamma^c,[\gamma^a,\gamma^b]\}\psi .
\end{align}
The spin tensor in Eq. (\ref{spintensor1}), when expressed solely in either Lorentz or coordinate indices, is completely antisymmetric under permutations of (entirely covariant or contravariant) indices, while this is only true for the $s=+$-component of the expression in Eq. (\ref{spintensor2}). The complete antisymmetry implies duality to the axial current according to 
\begin{align}
S^{\mu}_{ab}=\frac{1}{4}\epsilon_{abcd}\overline{\Psi}\gamma^0e^{c\mu}\gamma^5\gamma^d\Psi =-\frac{1}{2}e_a^{\nu}e_b^{\rho}\epsilon^{\mu}_{\,\,\,\,\nu\rho\sigma}j_A^{\sigma}.\label{axialspinduality}
\end{align}
We will make use of the former form (meaning Eq. (\ref{spintensor1})) only. 

The divergence of the spin tensor is evaluated to be
\begin{align}
\nabla_{\mu}S^{\mu}_{ab}=e_{b\nu}T_a^{\nu}-e_{a\nu}T_b^{\nu}+P_{ab}
\label{sdivergence}
\end{align}
with
\begin{align}
P_{ab}=\frac{i}{16}\overline{\Psi}\gamma^0e_c^{\mu}\{\gamma^c,[[\gamma_a,\gamma_b],i\mathcal{B}_{\mu}]\}\Psi .
\label{sdivergenceextra}
\end{align}
The extra term $P_{ab}$ in Eq. (\ref{sdivergenceextra}) is present for inhomogeneous $\bold{d}$ (or nonzero $\mathcal{B}_{\mu}$). Notice that $P_{03}=P_{12}=0$, though. We furthermore find for the spatial Lorentz index combinations
\begin{align}
&P_{13}=\frac{i}{4}e_1^{\mu}(Re(\omega_{\mu 12})Q_A^1-Im(\omega_{\mu 12})Q_A^2),\,\,\,\,P_{23}=\frac{i}{4}e_2^{\mu}(Re(\omega_{\mu 12})Q_A^1-Im(\omega_{\mu 12})Q_A^2)\label{spatialPab}\\
&Q_A^1=-\frac{1}{2}\overline{\Psi}\gamma^5\sigma^1\Psi ,\,\,\,\,Q_A^2=-\frac{1}{2}\overline{\Psi}\gamma^5\sigma^2\Psi\nonumber
\end{align}
which are relevant for the discussion of disclinations and fractional vortices further below.\par
Finally, we obtain for the divergence of the Lorentz transformation tensor
\begin{align}
\nonumber \nabla_{\mu}M^{\mu}_{ab}=&e_{a\nu}T_b^{\nu}-e_{b\nu}T_a^{\nu}+x^{\nu}T_b^{\mu}\nabla_{\mu}e_{a\nu}-x^{\nu}T_a^{\mu}\nabla_{\mu}e_{b\nu}+x^{\nu}e_{a\nu}\nabla_{\mu}T_b^{\mu}-x^{\nu}e_{b\nu}\nabla_{\mu}T_a^{\mu}+\nabla_{\mu}S^{\mu}_{ab}\\
\nonumber =&e_{a\nu}T_b^{\nu}-e_{b\nu}T_a^{\nu}+x^{\nu}T_b^{\mu}\nabla_{\mu}e_{a\nu}-x^{\nu}T_a^{\mu}\nabla_{\mu}e_{b\nu}+x^{\nu}e_{a\nu}(T^{\rho}_{bc}e_{\rho}^de_{\lambda}^cT_d^{\lambda}-j_A^{\mu}e_b^{\rho}\mathcal{F}_{\mu\rho}+G_b)\\
\nonumber &-x^{\nu}e_{b\nu}(T^{\rho}_{ac}e_{\rho}^de^c_{\lambda}T_d^{\lambda}-j_A^{\mu}e_a^{\rho}\mathcal{F}_{\mu\rho}+G_a)+\nabla_{\mu}S^{\mu}_{ab}\\
\nonumber =&x^{\nu}T_b^{\mu}\nabla_{\mu}e_{a\nu}-x^{\nu}T_a^{\mu}\nabla_{\mu}e_{b\nu}+x^{\nu}e_{a\nu}(T^{\rho}_{bc}e_{\rho}^de_{\lambda}^cT_d^{\lambda}-j_A^{\mu}e_b^{\rho}\mathcal{F}_{\mu\rho}+G_b)\\
&-x^{\nu}e_{b\nu}(T^{\rho}_{ac}e_{\rho}^de^c_{\lambda}T_d^{\lambda}-j_A^{\mu}e_a^{\rho}\mathcal{F}_{\mu\rho}+G_a)+P_{ab}.
\end{align}
Notice that the Dirac spin current is a special case of generalized spin currents of the form
\begin{align}
S^{a\mu}_{cd}=\frac{i}{16}\overline{\Psi}\gamma^0e_b^{\mu}\{\gamma^b,[\gamma_c,\gamma_d]\}\sigma^a\Psi
\end{align}
which may be obtained by variation of the action with respect to fictitious gauge fields acting on both Dirac and internal spin spaces. Two such currents are obtained by variation of the action with respect to the real and imaginary parts of $\omega_{\mu 12}$ which are contained in $\mathcal{B}_{\mu}$. Another example is the ordinary spin current (or spin tensor). Their divergence is given by
\begin{align}
\nabla_{\mu}S^{a\mu}_{cd}=\frac{i}{16}\overline{\Psi}\gamma^0e_b^{\mu}\Big[\{\gamma^b,\mathcal{B}_{\mu}\},[\gamma_c,\gamma_d]\sigma^a\Big]\Psi +\frac{i}{4}\overline{\Psi}\gamma^0\overset{\leftarrow}{\nabla}_{\mu}(\gamma_ce_d^{\mu}-\gamma_de_c^{\mu})\sigma^a\Psi +\frac{i}{4}\overline{\Psi}\gamma^0(\gamma_de_c^{\mu}-\gamma_ce_d^{\mu})\sigma^a\nabla_{\mu}\Psi .
\end{align}
Since these currents are conserved neither for vanishing nor non-vanishing gauge field $B_{\mu}$ with a divergence not proportional to any other current, they will not be considered in the following section at all, an exception being the ordinary spin current.\par
To sum up we find that both energy momentum tensor and the orbital part of the Lorentz transformation tensor are conserved modulo terms comprising derivatives of the matrix-valued vierbein (scalar vierbein and non-Abelian gauge field). The spin tensor divergence contains the expected anti-symmetric part of the energy momentum tensor supplemented by a term comprising the non-Abelian gauge field. \par
The cases to distinguish are those of homogeneous and inhomogeneous $\bold{d}$ or equivalently vanishing and non-vanishing gauge field $\mathcal{B}_{\mu}$. As the two components $\Psi^{\pm}$ of $\Psi$ are only coupled for inhomogeneous $\bold{d}$ or non-vanishing gauge field $\mathcal{B}_{\mu}$ (more precisely non-vanishing $\omega_{\mu 12}$), the homogeneous case needs a weak interaction by other means in order to ensure that both components move together with the same macroscopic four velocity and share the same temperature. The same statement is also valid for the chiral spinor components. Otherwise individual four velocities and temperatures need to be introduced.

\subsubsection{Stationarity of the statistical operator: Assumption of global thermodynamic equilibrium}

The low energy considerations in superfluid $^3$He - A imply the condition $T \ll v_{\perp}k_F$ for the temperature in line with Eq. (\ref{timespaceconstraints}). It should be noted that in general the identification of conserved currents may be more involved. That is why it is not straightforward to obtain the correct form of the Zubarev statistical operator in global thermodynamic equilibrium for the case of $^3$He - A. We explicitly calculated the non-conservation of the energy momentum tensor field and spin tensor field. In addition the axial current of the Weyl fermions at the Fermi points is anomalous and further current non-conservation is implied by nonzero spin connection gauge field $\mathcal{B}_{\mu}$. The logarithm of the statistical operator ansatz we employ is expressed as
\begin{align}
\log\hat \rho =-\alpha -\int d\Sigma n_{\mu}(\hat{T}^{\mu}_aB^a-\frac{1}{2}\hat{M}^{\mu}_{ab}\Omega^{ab}-\sum_i\zeta_i\hat{j}_i^{\mu}).
\label{Zub}
\end{align}
$\hat{T}^{\mu}_a$ is the canonical stress energy tensor operator and $\hat{M}^{\mu}_{ab}$ is the canonical Lorentz transformation tensor operator. The currents of internal symmetries are represented by $\hat{j}_i^{\mu}$ and labelled by the subscript $i$. The Lorentz vector field $B^a$ describes translational motion, while the Lorentz tensor field $\Omega^{ab}$ describes vorticity which comprises spatial rotations and boosts. The scalars $\zeta_i$ indicate the strength with which internal charges contribute.

The spacetime considered here is flat rescaled Minkowski spacetime, the spacetime intrinsic to $^3$He - A.

The conserved currents are given by $j_V^{\mu}$ and $j_{CS}^{\mu}$, respectively, which are supplemented by the further vector and axial currents $j^{a\mu}_V$ and $j^{a\mu}_A$ of Eq. (\ref{vectorandaxialcurrents}) in the case where the corresponding algebraic condition in Eq. (\ref{algebraiccondition}) is fulfilled. We include a sum over currents into the Zubarev statistical operator indexed by $i$ which is meant to comprise only conserved currents. The conserved currents, except for $j_{CS}$, may be treated identically to $j_V$ within the following considerations. We assume that the current operators (including the energy momentum tensor operator) vanish at spacelike infinity.

The stationarity condition \(\frac{d\hat\rho}{d\sigma} =0\) implies global thermodynamic equilibrium, and requires the integrand to be divergence-free \cite{Zubarev1979}. In order to have equilibrium we should require that the expression of Eq. (\ref{Zub}) does not depend on the form of $\Sigma = \Sigma_{\sigma}$. In the case of vanishing Nieh - Yan form and axial gauge field anomaly the axial current $\hat{j}^{\mu}_A$ is conserved and may enter the Zubarev statistical operator in the same fashion as $\hat{j}^{\mu}_V$. For nonzero Nieh - Yan form  or axial gauge field anomaly the current operator $\hat j_A^{\mu}$ is no longer conserved and is to be replaced by $\hat j_{CS}^{\mu}$. Notice that the inclusion of a term for the current operator $\hat{j}^{\mu}_A$ into the Zubarev statistical operator for non-vanishing anomalous contributions will not result in an extra independent case which is why we omit it. The requirement of global thermodynamic equilibrium is equivalent to 
\begin{align}
\nonumber 0=&\nabla_{\mu}(\hat{T}^{\mu}_aB^a-\frac{1}{2}\hat{M}^{\mu}_{ab}\Omega^{ab}-\sum_i\zeta_i\hat{j}_i^{\mu}-\zeta_{CS}\hat{j}_{CS}^{\mu})\\
\nonumber =&(\nabla_{\mu}\hat{T}^{\mu}_a)B^a-\frac{1}{2}(\nabla_{\mu}\hat{M}^{\mu}_{ab})\Omega^{ab}+\hat{T}^{\mu}_a\nabla_{\mu}B^a-\frac{1}{2}\hat{M}^{\mu}_{ab}\nabla_{\mu}\Omega^{ab}\\
\nonumber &-\sum_i\zeta_i(\nabla_{\mu}\hat{j}_i^{\mu})-\zeta_{CS}(\nabla_{\mu}\hat{j}_{CS}^{\mu})-\sum_i(\nabla_{\mu}\zeta_i)\hat{j}_i^{\mu}-(\nabla_{\mu}\zeta_{CS})\hat{j}_{CS}^{\mu}\\
\nonumber =&(\nabla_{\mu}B^a+x^{\nu}e_{b\nu}\nabla_{\mu}\Omega^{ab}+T^{\rho}_{dc}e_{\rho}^ae_{\mu}^cB^d+x^{\nu}(\nabla_{\mu}e_{b\nu})\Omega^{ab}+x^{\nu}e_{b\nu}T^{\rho}_{dc}e_{\rho}^ae_{\mu}^c\Omega^{db})\hat T_a^{\mu}+(B^a-x^{\mu}e_{b\mu}\Omega^{ba})\hat{G}_a\\
&+(\frac{1}{4}e_a^{\nu}e_b^{\rho}\epsilon^{\sigma}_{\,\,\,\,\nu\rho\mu}(\nabla_{\sigma}\Omega^{ab})-e_a^{\nu}\mathcal{F}_{\mu\nu}B^a+x^{\nu}\mathcal{F}_{\mu\rho}\Omega_{\nu}^{\,\,\,\,\rho})\hat{j}_A^{\mu}-\frac{1}{2}\Omega^{ab}\hat{P}_{ab}-\sum_i(\nabla_{\mu}\zeta_i)\hat j_i^{\mu}-(\nabla_{\mu}\zeta_{CS})\hat j_{CS}^{\mu}.
\end{align}
The final equality has been obtained using Eqs. (\ref{tdivergence}), (\ref{tdivergenceextra}) and (\ref{sdivergence}) for the current divergences. Furthermore we made use of the duality of the spin current and the axial current for fermions expressed in Eq. (\ref{axialspinduality}).
Global thermodynamic equilibrium requires the coefficients of all the operators to vanish. With the definition
\begin{align}
B^a=x^{\mu}e_{b\mu}\Omega^{ba}+\beta^a
\label{Bintermsofbeta}
\end{align}
we may bring the constraint equation arising from the equilibrium condition into the following form
\begin{align}
\nonumber 0=&(\nabla_{\mu}\beta^a+T^{\rho}_{dc}e_{\rho}^ae_{\mu}^c\beta^d+e_{b\mu}\Omega^{ba})\hat T_a^{\mu}+\beta^a\hat{G}_a\\
&+(\frac{1}{4}e_a^{\nu}e_b^{\rho}\epsilon^{\sigma}_{\,\,\,\,\nu\rho\mu}(\nabla_{\sigma}\Omega^{ab})-\mathcal{F}_{\mu\nu}\beta^{\nu})\hat{j}_A^{\mu}-\frac{1}{2}\Omega^{ab}\hat{P}_{ab}-\sum_i(\nabla_{\mu}\zeta_i)\hat j_i^{\mu}-(\nabla_{\mu}\zeta_{CS})\hat j_{CS}^{\mu}.
\label{stationary1}
\end{align}
We assume now that the Nieh - Yan form and the axial gauge field anomaly vanish. We then find as a necessary and sufficient criterion for global thermodynamic equilibrium with $\zeta_{CS}=\zeta_A$ and $j_{CS}^{\mu}=j_A^{\mu}$
\begin{align}
\nonumber 0=&(\nabla_{\mu}B^a+x^{\nu}e_{b\nu}\nabla_{\mu}\Omega^{ab}+T^{\rho}_{dc}e_{\rho}^ae_{\mu}^cB^d+x^{\nu}(\nabla_{\mu}e_{b\nu})\Omega^{ab}+x^{\nu}e_{b\nu}T^{\rho}_{dc}e_{\rho}^ae_{\mu}^c\Omega^{db})\hat T_a^{\mu}+(B^a-x^{\mu}e_{b\mu}\Omega^{ba})\hat{G}_a\\
&+(\frac{1}{4}e_a^{\nu}e_b^{\rho}\epsilon^{\sigma}_{\,\,\,\,\nu\rho\mu}(\nabla_{\sigma}\Omega^{ab})-e_a^{\nu}\mathcal{F}_{\mu\nu}B^a+x^{\nu}\mathcal{F}_{\mu\rho}\Omega_{\nu}^{\,\,\,\,\rho}-\nabla_{\mu}\zeta_A)\hat{j}_A^{\mu}-\frac{1}{2}\Omega^{ab}\hat{P}_{ab}-\sum_i(\nabla_{\mu}\zeta_i)\hat j_i^{\mu}.
\end{align}
We again employ Eq. (\ref{Bintermsofbeta}) to find
\begin{align}
\nonumber 0=&(\nabla_{\mu}\beta^a+T^{\rho}_{dc}e_{\rho}^ae_{\mu}^c\beta^d+e_{b\mu}\Omega^{ba})\hat T_a^{\mu}+\beta^a\hat{G}_a\\
&+(\frac{1}{4}e_a^{\nu}e_b^{\rho}\epsilon^{\sigma}_{\,\,\,\,\nu\rho\mu}(\nabla_{\sigma}\Omega^{ab})-\mathcal{F}_{\mu\nu}\beta^{\nu}-\nabla_{\mu}\zeta_A)\hat{j}_A^{\mu}-\frac{1}{2}\Omega^{ab}\hat{P}_{ab}-\sum_i(\nabla_{\mu}\zeta_i)\hat j_i^{\mu}.
\label{stationary2}
\end{align}
Furthermore we have
\begin{align}
\hat{T}^{\mu}_aB^a-\frac{1}{2}\hat{M}^{\mu}_{ab}\Omega^{ab}=\hat{T}^{\mu}_a\beta^a-\frac{1}{2}\hat{S}^{\mu}_{ab}\Omega^{ab}.
\end{align}
This justifies the statement made in the last section regarding the orbital part of the Lorentz transformation tensor. It may be disregarded such that $\hat{M}^{\mu}_{ab}\to\hat{S}^{\mu}_{ab}$ which is accompanied here by the replacement $B^a\to\beta^a$. The orbital part of the Lorentz transformation tensor $M^{\mu}_{ab}$ plays no direct role within the Zubarev statistical operator method. The residual appearance of $\Omega^{ab}$ within the coefficient of the energy momentum tensor operator is due to the divergence of the spin tensor.\par

The coefficient of the energy momentum tensor operator (and that of the operator $\hat{G}_a$) is identical for the cases of both vanishing and non-vanishing anomalous contributions. It leads to the constraint equation
\begin{align}
\nonumber 0=&\nabla_{\mu}\beta^a+T_{dc}^{\rho}e_{\rho}^ae_{\mu}^c\beta^d+e_{b\mu}\Omega^{ba}\\
\nonumber &\Leftrightarrow\\
\nonumber 0=&e_{a\nu}(\nabla_{\mu}\beta^a+T_{dc}^{\rho}e_{\rho}^ae_{\mu}^c\beta^d+e_{b\mu}\Omega^{ba})\\
\nonumber =&e_{a\nu}\nabla_{\mu}e^a_{\rho}\beta^{\rho}-(e_d^{\sigma}\nabla_{\sigma}e_c^{\rho}-e_c^{\sigma}\nabla_{\sigma}e_d^{\rho})g_{\nu\rho}e_{\mu}^c\beta^d+\Omega_{\mu\nu}\\
\nonumber =&e_{a\nu}(\nabla_{\mu}e^a_{\rho})\beta^{\rho}+\nabla_{\mu}\beta_{\nu}-e_{\mu}^c(\nabla_{\sigma}e_{c\nu})\beta^{\sigma}+e^d_{\rho}(\nabla_{\mu}e_{d\nu})\beta^{\rho}+\Omega_{\mu\nu}\\
\nonumber =&\nabla_{\mu}\beta_{\nu}-e_{\mu}^c(\nabla_{\sigma}e_{c\nu})\beta^{\sigma}+\Omega_{\mu\nu}+(\nabla_{\mu}g_{\nu\rho})\beta^{\rho}\\
\nonumber =&\nabla_{\mu}\beta_{\nu}-e_{\mu}^c(\nabla_{\sigma}e_{c\nu})\beta^{\sigma}+\Omega_{\mu\nu}\\
\nonumber &\Leftrightarrow\\
&\nabla_{\mu}\beta_{\nu}=-\Omega_{\mu\nu}+e^a_{\mu}(\nabla_{\lambda}e_{a\nu})\beta^{\lambda}
\label{changeofbeta}
\end{align}
for $\hat{G}_a=0$. The right-hand side of Eq. (\ref{changeofbeta}) comprises only antisymmetric tensors in the uncontracted indices. This implies that $\beta^{\mu}$ has to fulfill the Killing equation
\begin{align}
\nabla_{\mu}\beta_{\nu}+\nabla_{\nu}\beta_{\mu}=0
\end{align}
for a flat spacetime (the symmetrized equation) which is known to have ten solutions given by
\begin{align}
\beta^{\mu}=b^{\mu}+\omega^{\mu}_{\,\,\,\,\nu}x^{\nu}
\end{align}
with constant coefficients $b^{\mu}$ and $\omega^{\mu}_{\,\,\,\,\nu}$, whereby $\omega_{\mu\nu}=-\omega_{\nu\mu}$. The spin tensor operator coefficient $\Omega^{ab}$ is fully determined by the antisymmetric part of Eq. (\ref{changeofbeta})
\begin{align}
\Omega_{\mu\nu}=\omega_{\mu\nu}+e^a_{\mu}(\nabla_{\lambda}e_{a\nu})\beta^{\lambda}=e_{\mu}^a[\beta^{\lambda}\nabla_{\lambda}e_{a\nu}+e_{a\lambda}\nabla_{\nu}\beta^{\lambda}].
\label{equationofomega}
\end{align}
The piece in angular brackets is the Lie derivative of the vierbein along the Killing vector field $\beta$. 

If $\hat{G}_a\neq 0$ the vector space of Killing vector solutions shrinks due to the additional conditions
\begin{align}
\beta^a=0 \,\,\text{for} \,\, \hat{G}_a\neq 0.
\label{furtherbetarestriction}
\end{align}
Remember that $\hat{G}_a$ is only nonzero if both $(s=\pm)$-components are present and the non-Abelian gauge field $\mathcal{B}$ is non-vanishing which is equivalent to $\nabla_{\mu}\bold{d}\neq 0$.\par
The constraint equation arising from the axial current operator coefficient translates into a constraint equation for the vierbein for a given Killing vector field $\beta$. With the definition
\begin{align}
T_{\mu\nu\rho}=e_{a\mu}T^a_{\nu\rho}
\end{align}
the vierbein has to fulfill
\begin{align}
\epsilon^{\alpha\beta\gamma\delta}[\beta^{\lambda}\nabla_{\lambda}T_{\beta\gamma\delta}+T_{\lambda\gamma\delta}\nabla_{\beta}\beta^{\lambda}+T_{\beta\lambda\delta}\nabla_{\gamma}\beta^{\lambda}+T_{\beta\gamma\lambda}\nabla_{\delta}\beta^{\lambda}]+4\mathcal{F}^{\alpha}_{\,\,\,\,\mu}\beta^{\mu}=0
\label{anomalyconstraint}
\end{align}
for nonzero Nieh - Yan form or axial gauge field anomaly. The expression in angular brackets is the Lie derivative of the torsion tensor field along the Killing vector field $\beta$. We may further derive a condition for the Nieh - Yan form. If we contract Eq. (\ref{anomalyconstraint}) with $\nabla_{\alpha}$ we obtain
\begin{align}
\beta^{\lambda}\nabla_{\lambda}a-\frac{1}{2\pi^2}\nabla_{\alpha}(\mathcal{F}^{\alpha}_{\,\,\,\,\mu}\beta^{\mu})=0.
\end{align}
Since we consider the case with nonzero Nieh - Yan form, Eq. (\ref{anomalyconstraint}) has to be supplemented by $a\not\equiv 0$. \par
For vanishing anomalous terms we obtain for $\zeta_A$ the condition
\begin{align}
\nabla_{\mu}\zeta_A=-\frac{1}{4}\epsilon^{\,\,\,\,\,\,\,\,\,\nu}_{ab\mu}\nabla_{\nu}\Omega^{ab}-\mathcal{F}_{\mu\nu}\beta^{\nu}
\label{equationforzetaa}
\end{align}
which implies a slight relaxation of the constraints for the vierbein in global thermodynamic equilibrium as compared to the case with nonzero anomalous terms. This equation has the solution
\begin{align}
\zeta_A=\zeta_A(x_0)-\int_{x_0}^x(\frac{1}{4}\epsilon^{\,\,\,\,\,\,\,\,\,\nu}_{ab\mu}\nabla_{\nu}\Omega^{ab}+\mathcal{F}_{\mu\nu}\beta^{\nu})dx^{\mu}
\end{align}
whereby the vierbein has to fulfill the corresponding integrability constraint
\begin{align}
\nonumber 0=\epsilon^{\mu\nu\rho\sigma}\{&\epsilon_{\sigma}^{\,\,\,\beta\gamma\delta}[\beta^{\lambda}\nabla_{\lambda}(\nabla_{\rho}T_{\beta\gamma\delta})+(\nabla_{\lambda}T_{\beta\gamma\delta})\nabla_{\rho}\beta^{\lambda}+(\nabla_{\rho}T_{\lambda\gamma\delta})\nabla_{\beta}\beta^{\lambda}\\
&+(\nabla_{\rho}T_{\beta\lambda\delta})\nabla_{\gamma}\beta^{\lambda}+(\nabla_{\rho}T_{\beta\gamma\lambda})\nabla_{\delta}\beta^{\lambda}]+4\nabla_{\rho}(\mathcal{F}_{\sigma\lambda}\beta^{\lambda})\}
\label{integrabilityconstraint}
\end{align}
derived from
\begin{align}
\epsilon^{\alpha\beta\gamma\delta}[\beta^{\lambda}\nabla_{\lambda}T_{\beta\gamma\delta}+T_{\lambda\gamma\delta}\nabla_{\beta}\beta^{\lambda}+T_{\beta\lambda\delta}\nabla_{\gamma}\beta^{\lambda}+T_{\beta\gamma\lambda}\nabla_{\delta}\beta^{\lambda}]+4\nabla^{\alpha}\zeta_A+4\mathcal{F}^{\alpha}_{\,\,\,\,\mu}\beta^{\mu}=0.
\label{constraintforzetaA}
\end{align}
The expression in angular brackets indicates again the Lie derivative along the Killing vector field $\beta$. The second constraint arises from the vanishing axial gauge field anomaly and Nieh - Yan form
\begin{align}
\nabla_{\mu}K^{\mu}_{\mathcal{A}}=0,\,\,\,\, a=\nabla_{\mu}K^{\mu}=0
\end{align}
with the Chern - Simons currents given in Eqs. (\ref{axialchernsimonscurrent}) and (\ref{chernsimonscurrent}), respectively. If we contract Eq. (\ref{constraintforzetaA}) with $\nabla_{\alpha}$ and use that $a=0$, it follows that $\zeta_A$ is a harmonic function modulo a source term
\begin{align}
\Box\zeta_A=-\nabla_{\alpha}(\mathcal{F}^{\alpha}_{\,\,\,\,\mu}\beta^{\mu}).
\end{align}
If we assume that $\zeta_A$ is bound to be finite with finite boundary conditions at infinity, the only solution in the absence of the source term is $\zeta_A=constant$ such that Eq. (\ref{constraintforzetaA}) reduces to Eq. (\ref{anomalyconstraint}). Then both for vanishing and non-vanishing anomalous terms (and vanishing axial source term in the former case) the torsion tensor field in equilibrium is constrained by the condition given in Eq. (\ref{anomalyconstraint}).\par
We provide now the general solution for the condition
\begin{align}
\nabla_{\mu}\zeta_A=\mathcal{F}_{\nu\mu}\beta^{\nu},
\label{zetaacondition}
\end{align}
relevant in the absence of anomalies and vanishing torsion term in Eq. (\ref{constraintforzetaA}), explicitly following section 4.3 of \cite{Buzzegoli2020ycf}. A necessary condition for global thermodynamic equilibrium is a vanishing Lie derivative of the axial field strength tensor along the frigidity vector field (apply a further derivative to Eq. (\ref{zetaacondition}) and use that here two consecutive derivatives commute)
\begin{align}
0=\beta^{\lambda}\nabla_{\lambda}\mathcal{F}_{\mu\nu}+(\nabla_{\mu}\beta^{\lambda})\mathcal{F}_{\lambda\nu}+(\nabla_{\nu}\beta^{\lambda})\mathcal{F}_{\mu\lambda}.
\end{align}
This condition follows from the axial vector field condition
\begin{align}
\beta^{\lambda}\nabla_{\lambda}\mathcal{A}_{\mu}+(\nabla_{\mu}\beta^{\lambda})\mathcal{A}_{\lambda}=\nabla_{\mu}\Phi ,\,\,\,\, \mathcal{A}_{\mu}\to\mathcal{A}_{\mu} +\nabla_{\mu}\phi\,\,\,\,\Phi\to \Phi +\beta^{\lambda}\nabla_{\lambda}\phi 
\end{align}
where we exhibit the necessary gauge transformation properties of the space and time dependent function $\Phi$. This leads to the general searched for solution 
\begin{align}
\zeta_A=\zeta_A^0-\beta^{\mu}\mathcal{A}_{\mu}+\Phi ,\,\,\,\, \zeta_A^0=constant.
\end{align}
The solutions for $\Omega^{ab}$ in both cases are further restricted by the conditions
\begin{align}
\Omega^{ab}=0 \,\,\text{for}\,\, \hat{P}_{ab} \neq 0.
\label{omegarestriction}
\end{align}
Since $\hat{P}_{03}=\hat{P}_{12}=0$, the constraint refers at most to the components $\Omega^{01}$, $\Omega^{02}$, $\Omega^{13}$ and $\Omega^{23}$ (with Lorentz indices). The contracted spin tensor $n_{\mu}S^{\mu}_{ab}$ has furthermore only nonzero spatial components. The conditions in Eq. (\ref{omegarestriction}) may be circumgone in the presence of the charges $Q_A^1$ and $Q_A^2$, respectively, following Eq. (\ref{spatialPab}). It amounts to (assuming that $\nabla_{\mu}\hat{j}_A^{1\mu}=\nabla_{\mu}\hat{j}_A^{2\mu}=0$)
\begin{align}
0=&-\Omega^{13}\hat{P}_{13}-\Omega^{23}\hat{P}_{23}-(\nabla_{\mu}\zeta_A^1)\hat{j}_A^{1\mu}-(\nabla_{\mu}\zeta_A^2)\hat{j}_A^{2\mu}
\end{align}
which is fulfilled for space and time independent charge coefficients by the conditions
\begin{align}
\zeta_A^1=-\frac{i}{4}\Omega^{i3}e_i^{\mu}Re(\omega_{\mu 12})t,\,\,\,\,\zeta_A^2=\frac{i}{4}\Omega^{i3}e_i^{\mu}Im(\omega_{\mu 12})t.
\label{relevantforvortices}
\end{align}
These configurations may be relevant, e. g., in the presence of vortices (if in addition $j_A^{1\phi}=j_A^{2\phi}=0$ where $\phi$ is the azimuth angle in cylindrical coordinates). The linear time dependence has been observed before in the case of accelerated motion as a solution of the Killing equation for the frigidity vector field $\beta^{\mu}$ corresponding to Lorentz boost symmetry.\par 
The remaining constraint equations are trivially solved by 
\begin{align}
\zeta_i=constant
\end{align}
where $i$ labels a conserved current, supplemented by $\zeta_{CS}=constant$ in the presence of nonzero anomalous contributions and (under the above mentioned conditions) $\zeta_A=constant$ in their absence.\par

Our original set of currents $(T_a^{\mu}\beta^a,S^{\mu}_{ab}\Omega^{ab},\zeta_{A/CS}j_{A/CS}^{\mu},\zeta_ij_i^{\mu})$ (omitting the generalized spin currents beyond the ordinary spin current) forms a basis in the space of currents of the theory for $^3$He-A with omission of the spin-orbit interaction. Not all of them are conserved separately. More specifically, the subspace spanned by $T_a^{\mu}\beta^a,S^{\mu}_{ab}\Omega^{ab}$ does not fulfill conservation. In fact the linear combination $T_a^{\mu}\beta^a+S^{\mu}_{ab}\Omega^{ab}$ with $\beta^a$ and $\Omega^{ab}$ fulfilling the conditions outlined in Eqs.(\ref{changeofbeta}), (\ref{furtherbetarestriction}) and (\ref{omegarestriction}) (the constraints coincide up to $a\neq 0$ in the presence of the nonzero Nieh - Yan form and $a=0$ in its absence), respectively, is a conserved current. Similar considerations are valid for the other operator coefficients, if additional currents are not conserved (especially in the case of inhomogeneous $\bold{d}$ or equivalently non-vanishing $\mathcal{B}_{\mu}$). We may thus think of composing the Zubarev statistical operator by a set of exclusively conserved currents by retaining only this linear combination in the mentioned subspace (and similarly for further non-conserved currents). The space of conserved currents is then smaller than that of the basis of currents.

\section{Effective Lagrangian in the presence of macroscopic motion within the path integral formulation}
\label{pathintegralformulation}

\mz{We proceed to formulate a macroscopic motion Lagrangian for $^3$He - A using the results established in previous sections. Throughout this section we follow closely the procedure proposed earlier in \cite{AZZ2023} within the path integral formulation of quantum field theory. This requires us to employ a few properties of the canonically quantized theory of $^3$He - A which was our original starting point. We collect some details on canonical quantization in appendices \ref{app1} and \ref{app1_5}, respectively.}\par
\mz{We note that the central idea is to identify a macroscopic motion Hamiltonian in global thermodynamic equilibrium (GTE) as follows
\begin{align}
\hat{\rho}_{GTE}=\frac{1}{Z_{GTE}}exp\Big(-\int d\Sigma\beta\mathcal{H}_{mm}\Big),\,\,\,\,\beta\mathcal{H}_{mm}=n_{\mu}\Big(\hat{T}^{\mu}_aB^a-\frac{1}{2}\hat{M}^{\mu}_{ab}\Omega^{ab}-\zeta_A\hat{j}_A^{\mu}-\sum_i\zeta_i\hat{j}_i^{\mu}\Big)
\end{align}
which is subsequently converted into a macroscopic motion Lagrangian $\mathcal{L}_{mm}$ via path integral methods. We will omit writing the subscript mm.}

We are from now on employing notation corresponding to the case of vanishing Nieh - Yan form as well as axial gauge field anomaly. This does not pose a restriction on generality, since the analogous expressions for the statistical operator in the case of nonzero anomalies are obtained by the trivial replacements $\zeta_A\to \zeta_{CS}$ as well as $\hat{Q}_A\to\hat{Q}_{CS}$.
The Zubarev statistical operator takes the two equivalent forms
\begin{align}
\nonumber \log\hat \rho =&-\alpha -\int d\Sigma  n_{\mu}\Big(\hat{T}^{\mu}_aB^a-\frac{1}{2}\hat{M}^{\mu}_{ab}\Omega^{ab}-\zeta_A\hat{j}_A^{\mu}-\sum_i\zeta_i\hat{j}_i^{\mu}\Big)\\
%\nonumber =&-\alpha -\int d\Sigma  \Big(n_{\mu}(\hat{T}^{\mu}_ae^a_{\nu}b^{\nu}-\frac{1}{2}\hat{M}^{\mu}_{ab}e^a_{\rho}e^b_{\nu}\omega^{\rho\nu}-\zeta_A(x_0)\hat{j}_A^{\mu}-\sum_i\zeta_i\hat{j}_i^{\mu})-\tilde{\Delta}_{\mu}\hat{j}_A^{\mu}\Big)\\
%\nonumber =&-\alpha -\int d\Sigma  \Big(n_{\mu}(\hat{T}_{BR}^{\mu\nu}\beta_{\nu}-\zeta_A(x_0)\hat{j}_A^{\mu}-\sum_i\zeta_i\hat{j}_i^{\mu})-\tilde{\Delta}_{\mu}\hat{j}_A^{\mu}\Big)\\
%\nonumber =&-\alpha -\int d\Sigma  \Big(n_{\mu}(\hat{T}_{sym}^{\mu\nu}\beta_{\nu}-\zeta_A(x_0)\hat{j}_A^{\mu}-\sum_i\zeta_i\hat{j}_i^{\mu})-\Delta_{\mu}\hat{j}_A^{\mu}\Big)\\
=&-\alpha -\int d\Sigma  n_{\mu}\Big(\hat{T}^{\mu}_a\beta^a-\frac{1}{2}\hat{S}^{\mu}_{ab}\Omega^{ab}-\zeta_A\hat{j}_A^{\mu}-\sum_i\zeta_i\hat{j}_i^{\mu}\Big)
\label{Zubinsert}
\end{align}
whereby we will stick with the latter version from now on.

We define $\beta^{\mu}=\beta (x)u^{\mu}$ with $u^{\mu}u_{\mu}=1$ as well as $\zeta_A(x)=\beta (x)\mu_A(x)$ and $\zeta_V(x)=\beta (x)\mu_V(x)$ with chemical potentials $\mu_A$ and $\mu_V$, respectively. The function $\beta(x)$ may be interpreted as inverse temperature depending on coordinates. The vector field $\beta^\mu(x)$ is called the frigidity vector field. We will not introduce conserved currents additional to the vector current $j^{\mu}_V$ and the axial current $j^{\mu}_A$ as these may be incorporated into the formalism in an identical manner. \par
In terms of the Poincar\'e and internal charges we obtain:
\begin{align}
\hat\rho = \frac{1}{Z}e^{\int d\Sigma (-b^{\mu} \hat P_{\mu} + \frac{1}{2}\omega^{\mu\nu} \hat M_{\mu\nu}+\frac{1}{2}\Omega^{\mu\nu}\hat S_{\mu\nu}+\zeta_A\hat Q_A +\zeta_V \hat Q_V)}
\label{rho00}
\end{align}
with normalization factor $Z$ such that $Tr(\rho )=1$, momentum, orbital and spin angular momentum and charge density operators
\begin{align}
&\hat{P}^{\mu} =  n_{\rho}\hat{T}^{\rho}_ae^{a\mu},\\
&\hat M^{\mu\nu}=n_{\rho} (x^{\mu}e^{a\nu}\hat T^{\rho}_a-x^{\nu}e^{a\mu}\hat T^{\rho}_a),\\
&\hat S_{\mu\nu}=n_{\rho} \hat S^{\rho}_{ab}e^a_{\mu}e^b_{\nu},\\
&\hat Q_k= n_{\rho} \hat j^{\rho}_k.
\end{align}
The tensor $\hat M^{\mu\nu}$ may be decomposed as
\begin{align}
\hat M_{\mu\nu} = \epsilon_{\mu\nu\alpha\beta}\hat J^\alpha u^\beta - \hat K_\mu u_\nu + \hat K_\nu u_\mu.
\end{align}
Here $\hat K_\mu$ is the generator of boosts, while $\hat J_\nu$ is the generator of rotations. 
In terms of these generators and with velocity $v^{\mu}=\frac{1}{\beta} b^{\mu}$, acceleration $a^{\mu}=-\frac{1}{\beta}\omega^{\mu\nu}u_{\nu}$, orbital vorticity \\
$\omega^{\mu}=-\frac{1}{\beta}\frac{1}{2}\epsilon^{\mu\nu\rho\sigma}\omega_{\nu\rho}u_{\sigma}$ and spin vorticity $\kappa^{\mu\nu}=\frac{1}{\beta}\Omega^{\mu\nu}$ we obtain the following expression for the statistical operator:
\begin{align}
 \hat	\rho = \frac{1}{Z}e^{\int d\Sigma \beta (-v^{\mu} \hat P_{\mu} - a^{\mu} \hat K_{\mu} +{\omega}^{\mu} \hat J_{\mu}+\frac{1}{2}\kappa^{\mu\nu}\hat{S}_{\mu\nu} +\mu_A\hat Q_A +\mu_V \hat Q_V)}.
\label{rho0}
\end{align}
We are now going to manipulate our results for the Zubarev statistcial operator obtained in Eq. (\ref{rho0}) to bring it into a more explicit form and introduce an effective Lagrangian of massless Weyl fermions in the presence of their macroscopic motion. We will focus on a lattice theory formulation in line with our previous work \cite{AZZ2023}.
We will further consider only the particular case, when the hypersurface $\Sigma$ is the hyperplane $t = 0$ in the inertial (laboratory) reference frame. Then $n^{\mu}=(1,0,0,0)$ in Cartesian coordinates. 
%The considerations can easily be extended to our geometrical formulation in terms of $\Psi$ by the simple replacements $\psi_L\to \Psi_L$, $\psi_R\to\Psi_R$, $\nabla_{\mu}\to D_{\mu}$ and $\bold{e}_a^{\mu}\to e_a^{\mu}$, respectively.
Let us introduce the notation 
\begin{align}
&{\mathcal R}[\beta(x),v(x),a(x),\omega (x),\kappa_{\mu\nu}(x),\mu_V(x),\mu_A(x),\hat\Psi_L(x),\hat\Psi_R(x),\hat{\overline{\Psi}}_L(x),\hat{\overline{\Psi}}_R(x)] \equiv 	-{\rm ln}\, \hat \rho -\alpha \nonumber\\   &=\int d\Sigma \beta (v^\mu \hat P_\mu +a^\mu \hat K_\mu -{\omega}^{\mu} \hat J_{\mu}-\frac{1}{2}\kappa^{\mu\nu}\hat{S}_{\mu\nu} -\mu_A\hat Q_A -\mu_V \hat Q_V)
\end{align}
with the explicit forms of the operators given by
\begin{align}
\hat P_i=&\frac{1}{4}[\hat{\overline{\Psi}}iD_i\hat\Psi -\hat{\overline{\Psi}}i\overset{\leftarrow}{D}_i\hat\Psi ]=\frac{1}{4}[\hat{\overline{\Psi}}_LiD_i\hat\Psi_L -\hat{\overline{\Psi}}_Li\overset{\leftarrow}{D}_i\hat\Psi_L +\hat{\overline{\Psi}}_RiD_i\hat\Psi_R -\hat{\overline{\Psi}}_Ri\overset{\leftarrow}{D}_i\hat\Psi_R ],\\
\nonumber \hat P_0=&-\frac{1}{8}\hat{\overline{\Psi}}[i\gamma^0\{\gamma^be_b^j,D_j\}-i\gamma^0\{\gamma^be_b^j,\overset{\leftarrow}{D}_j\}]\hat\Psi \\
=&-\frac{1}{8}\hat{\overline{\Psi}}_L[i\{\overline{\tau}^be_b^j,D_j\}-i\{\overline{\tau}^be_b^j,\overset{\leftarrow}{D}_j\}]\hat\Psi_L-\frac{1}{8}\hat{\overline{\Psi}}_R[i\{\tau^be_b^j,D_j\}-i\{\tau^be_b^j,\overset{\leftarrow}{D}_j\}]\hat\Psi_R ,\\
\hat{K}_{\mu}=&-\hat{M}_{\mu\nu}u^{\nu}=-(x_{\mu}\hat P_{\nu}-x_{\nu}\hat P_{\mu})u^{\mu},\\
\hat{J}_{\mu}=&-\frac{1}{2}\epsilon^{\,\,\,\nu\rho}_{\mu\,\,\,\,\,\,\sigma}\hat{M}_{\nu\rho}u^{\sigma}=-\frac{1}{2}\epsilon^{\,\,\,\nu\rho}_{\mu\,\,\,\,\,\,\sigma}(x_{\nu}\hat{P}_{\rho}-x_{\rho}\hat{P}_{\nu})u^{\sigma},\\
\hat{S}_{ij}=&\frac{1}{4}n^{\mu}\epsilon_{\mu ijc}\hat{\overline{\Psi}}\gamma^0\gamma^5\gamma^c\hat\Psi =\frac{1}{4}n^{\mu}\epsilon_{\mu ijc}[\hat{\overline{\Psi}}_L\overline{\tau}^c\hat\Psi_L-\hat{\overline{\Psi}}_R\tau^c\hat\Psi_R],\\
\hat{Q}_A=&-\frac{1}{2}\hat{\overline{\Psi}}\gamma^5\hat\Psi =\frac{1}{2}(\hat{\overline{\Psi}}_L\hat\Psi_L-\hat{\overline{\Psi}}_R\hat\Psi_R),\\
\hat{Q}_V=&\frac{1}{2}\hat{\overline{\Psi}}\hat{\Psi}=\frac{1}{2}(\hat{\overline{\Psi}}_L\hat\Psi_L+\hat{\overline{\Psi}}_R\hat\Psi_R).
\end{align}
The Heisenberg equation of motion (\ref{heisenbergeom}) has been employed in the case of $\hat{P}_0$ in order to remove explicit time derivatives of field operators. Notice that due to the Majorana condition of Eq. (\ref{MajoranaCond})  $\hat{Q}_V$ unlike $\hat{Q}_A$ vanishes identically. We, however, include both these operators here for completeness.  

We can introduce the notion of a coherent state associated with the Grassmann-valued fields ${a}_\pm(p), \bar{a}_\pm(p)$ entering Eq. (\ref{3heaction}). We assume that the hypersurface $\Sigma$ is the hyperplane $t = const$, so that $n_\mu = (1,0,0,0)$. We use the standard definition
\begin{align}
	|\psi\rangle = e^{\sum_{\bold{k},\pm} \hat{a}^\dagger_\pm(t,\bold{k}) \psi_\pm(t,\bold{k}) } |\Omega\rangle ,\,\,\,\, \langle \overline\psi | =\bra{\Omega} e^{\sum_{\bold{k},\pm} \psi_\pm(t,\bold{k})\hat{a}_\pm(t,\bold{k}) } . 
\end{align}	
The "vacuum state" $|\Omega\rangle$ is annihilated by the operators $\hat{a}_\pm(t,\bold{k})$ for all $\bold{k}$. We define the configuration space operators 
\begin{align}
a_{\pm}(x)=\int \frac{d^4p}{(2\pi )^4}a_{\pm}(p)e^{ipx}.
\end{align}
In the regime of relativistic invariance the Fourier components are confined to the vicinities of the two Fermi points. This implies a splitting of the momentum space integration into the neighborhoods of $K_{\pm}$, respectively.\par
Recall that the operators $\hat{\Psi}_{R/L}$, $\hat{\bar{\Psi}}_{R/L}$ and ${a}_\pm ,\, \bar{a}_\pm$ are related by Eqs. (\ref{psiLplus}) - (\ref{psiRminus}) in the vicinity of the Fermi points. This relation is linear, and we can express it as
\begin{equation}
	a_\pm(x) = \Theta[\bold{d}(x)] \hat{\Psi}^1_R(x), \quad \bar{a}_\pm(x) = \bar{\Theta}[\bold{d}(x)] \hat{\Psi}^2_R(x)
\end{equation}
with $2\times 2$-matrices $\Theta[\bold{d}(x)]$ and $\bar{\Theta}[\bold{d}(x)]$ which fulfill ${\rm det}\,\Theta[\bold{d}(x)] =-{\rm det}\,\bar{\Theta}[\bold{d}(x)] = e^{i {\rm Arg}\, (id_1+d_2)}$. 
The coherent states obey the following properties:
\begin{enumerate}
\item 
\begin{align}
%&\hat\Psi (x) |\Psi\rangle = \hat\Psi (x) e^{\int d\Sigma n_{\mu} \hat{\overline{\Psi}}e_a^{\mu}\gamma^a \Psi} |\Omega\rangle = \Psi (x) |\Psi\rangle \\
&a_\pm(x) |\psi\rangle =   \psi_\pm(x) |\psi\rangle
\end{align}
\item
\begin{align}
%\langle \Phi |\Psi\rangle = e^{\int d\Sigma n_{\mu} \overline{\Phi} e_a^{\mu}\gamma^a \Psi},\,\,\,\,
\langle \phi |\psi\rangle = e^{\sum_{\vec{k},\pm} \bar{\psi}_\pm(t,\vec{k}) \psi_\pm(t,\vec{k})}
\end{align}
\item
\begin{align}
%&\mathbb{1}= \int D\overline{\Psi} D\Psi e^{-\int d\Sigma n_{\mu} \overline{\Psi}e_a^{\mu}\gamma^a \Psi} |\Psi\rangle\langle \Psi |=\mathbb{1}_L+\mathbb{1}_R \\
&\mathbb{1} = \int D\overline{\psi} D\psi e^{-\sum_{\vec{k},\pm} \bar{\psi}_\pm(t,\vec{k}) \psi_\pm(t,\vec{k})} |\psi\rangle\langle \psi |\label{1}
\end{align}	
We define Grassmann - valued fields $\Psi$ as related to $\psi$ by the same expression that relates $\hat{\Psi}$ and $a$. 
One can check easily that $D\overline{\psi} D\psi = D\overline{\Psi}_R D\Psi_R$ (or equivalently for the left-handed fields). 
\end{enumerate}
We will employ the notation $D\overline{\Psi}_{L/R} D\Psi_{L/R}$ in the path integral in order to indicate that only fields of one handedness are being integrated over, while the fields of the opposite handedness will be eliminated by the Majorana condition. Notice that the path integral construction by coherent states is valid in the case of a normal ordered Lagrangian in terms of the fermion fields. The Lagrangian we consider does not fulfill normal ordering, but as it is only quadratic in the fermion fields the procedure goes through nonetheless.

We fix the surface $\Sigma$ as the hypersurface $t = 0$ in rescaled Minkowski spacetime and represent the Zubarev statistical operator defined on $\Sigma$ as 
\begin{align}
\hat \rho = e^{-\alpha} \lim_{N\to \infty}\Pi_{s=0,1,\ldots,N-1}e^{-{\mathcal R}[\beta(x),v(x),a(x),\omega (x),\kappa (x),\mu_V(x),\mu_A(x),\hat\Psi_L(x),\hat\Psi_R(x),\hat{\overline{\Psi}}_L(x),\hat{\overline{\Psi}}_R(x)]\delta(N)},\,\,\,\, \delta(N)=1/N.
\end{align}
Next, we insert unity from Eq. (\ref{1}) between each two multipliers in the above product, and arrive at the expression for the partition function 
\begin{align}
Z[n(x),\beta(x),v(x),a(x),\omega (x),\kappa (x),\mu_V(x),\mu_A(x)]= e^{\alpha} = \int D\overline{\Psi}_{L/R}D\Psi_{L/R} \, e^{\int_0^1 d \tau \int d\Sigma  {L}(\overline{\Psi}_L,\Psi_L,\overline{\Psi}_R,\Psi_R )}.
\end{align}
Now $\Psi_R(x)$ and $\overline{\Psi}_R(x)$ (or $\Psi_L(x)$ and $\overline{\Psi}_L(x)$) are independent Grassmann-valued fields depending on points $x=(\tau ,\bold{x})$ of rescaled Minkowski spacetime with $\bold{x}$ situated on $\Sigma$, and on the parameter $\tau$. $\Psi_L(x)$ and $\overline{\Psi}_L(x)$ are to be expressed through $\Psi_R(x)$ and $\overline{\Psi}_R(x)$ via Eqs. (\ref{MajoranaCond}) or vice versa. The "Lagrangian" is given by 
 \begin{align}
\nonumber  L(\overline{\Psi}_L,\Psi_L,\overline{\Psi}_R,\Psi_R)=&-\frac{1}{2}\bar{a}_\pm \overset{\leftrightarrow}{D}_\tau a_\pm+\beta(0,\bold{x})\Bigl(-v^{\mu}(0,\bold{x}) P_\mu -a^{\mu}(0,\bold{x}) K_\mu \\
&+{\omega}^{\mu}(0,\bold{x}) J_{\mu}+\frac{1}{2}\kappa^{\mu\nu}(0,\bold{x})S_{\mu\nu} +\mu_A(0,\bold{x}) Q_A +\mu_V(0,\bold{x}) Q_V) \Bigr)\\
=&-\frac{1}{4}\bar{\Psi}_R\overset{\leftrightarrow}{D}_\tau \Psi_R-\frac{1}{4}\bar{\Psi}_L\overset{\leftrightarrow}{D}_\tau \Psi_L+\beta(0,\bold{x})\Bigl(-v^{\mu}(0,\bold{x}) P_\mu -a^{\mu}(0,\bold{x}) K_\mu\nonumber \\
&+{\omega}^{\mu}(0,\bold{x}) J_{\mu}+\frac{1}{2}\kappa^{\mu\nu}(0,\bold{x})S_{\mu\nu} +\mu_A(0,\bold{x}) Q_A +\mu_V(0,\bold{x}) Q_V) \Bigr).
\label{Z1u} 
 \end{align}
The scalar field parameters, collectively denoted by $X(0,\vec{x})$, entering the above expression coincide with $X(x)$ at $x \in \Sigma$. Here the initial moment in time is set to $0$: the surface $\Sigma$ was initially taken as the hyperplane $t = 0$. The same refers to the vector field (and also tensor field) parameters, collectively denoted by $X^{\mu}(0,\vec{x})$, - they coincide with $X^{\mu}(x)$ at $x \in \Sigma$. In both cases the fields do not depend on $\tau$. 

One can represent 
\begin{align}
Z[n(x),\beta(x),v(x),a(x),\omega (x),\kappa (x),\mu_V(x),\mu_A(x)] = \mathcal{Z}[n(x),\beta(x),v(x),a(x),\omega (x),\kappa (x),\mu_V(x),\mu_A(x), - i],
\end{align}
where
\begin{align}
\nonumber &\mathcal{Z}[n(x),\beta(x),v(x),a(x),\omega (x),\kappa (x),\mu_V(x),\mu_A(x), h]\\
\nonumber &=\,{\rm Tr}\Big(\,{\rm exp}\,(-i h \mathcal{R}[\beta(x),v(x),a(x),\omega (x),\kappa (x),\mu_V(x),\mu_A(x),\hat\Psi_L(x),\hat\Psi_R(x),\hat{\overline{\Psi}}_L(x),\hat{\overline{\Psi}}_R(x)])\Big)\nonumber\\
&=\int D\overline{\Psi}_{L/R}D\Psi_{L/R} \, e^{i\int d \Sigma \int_0^h d w \, \mathcal{L}(\overline{\Psi}_L,\Psi_L,\overline{\Psi}_R,\Psi_R )}.
\end{align}
Integration in the exponent of the above expression is over the piece of $\Sigma \otimes R$ that consists of points $(\bold{x},w)$ with $w \in (0,h)$. The fields $\Psi_L(\bold{x},w)$, $\overline{\Psi}_L(\bold{x},w)$, $\Psi_R(\bold{x},w)$ and $\overline{\Psi}_R(\bold{x},w)$ are now functions of $\bold{x}\in \Sigma$ and $w \in R$. The new Lagrangian is given by
 \begin{align}
\nonumber \mathcal{L}(\overline{\Psi}_L,\Psi_L,\overline{\Psi}_R,\Psi_R)=&\frac{i}{4}\overline{\Psi}_L\overset{\leftrightarrow}{D}_w\Psi_L+\frac{i}{4}\overline{\Psi}_R\overset{\leftrightarrow}{D}_w\Psi_R+\beta(0,\bold{x})\Bigl(-v^{\mu}(0,\bold{x}) P_\mu -a^{\mu}(0,\bold{x}) K_\mu \\
&+{\omega}^{\mu}(0,\bold{x}) J_{\mu}+\frac{1}{2}\kappa^{\mu\nu}(0,\bold{x})S_{\mu\nu} +\mu_A(0,\bold{x}) Q_A +\mu_V(0,\bold{x}) Q_V \Bigr)	.
\label{Z2u} 
\end{align}
Now instead of $\Sigma \otimes R$ we restore Minkowski spacetime with the time variable related to $w$ via rescaling
\begin{equation}
	t = w\,{\mathfrak B}(\bold{x}) \label{rescaling}
\end{equation}
with a certain scaling function $\mathfrak B$ of spatial coordinates to be specified below. The new fields  $\Psi^{\prime}_L(t,\bold{x})$, $\overline{\Psi}^{\prime}_L(t,\bold{x})$, $\Psi^{\prime}_R(t,\bold{x})$ and $\overline{\Psi}^{\prime}_R(t,\bold{x})$ are defined as 
\begin{align}
\nonumber &\Psi^{\prime}_L(t,\bold{x}) = \Psi_L(t/{\mathfrak B}(\bold{x}),\bold{x}), \,\,\,\, \overline{\Psi}^{\prime}_L(t,\bold{x}) =  \overline{\Psi}_L(t/{\mathfrak B}(\bold{x}),\bold{x}),\\
&\Psi^{\prime}_R(t,\bold{x}) = \Psi_R(t/{\mathfrak B}(\bold{x}),\bold{x}), \,\,\,\, \overline{\Psi}^{\prime}_R(t,\bold{x}) =  \overline{\Psi}_R(t/{\mathfrak B}(\bold{x}),\bold{x})
\end{align}
where $\bold{x}\in \Sigma$. We will work with these new fields in the following and drop the prime at the same time. We then have 
\begin{align}
\mathcal{Z}[n(x),\beta(x),v(x),a(x),\omega (x),\kappa (x),\mu_V(x),\mu_A(x), h]=\int D\overline{\Psi}_{L/R}D\Psi_{L/R} \, e^{i\int d^4 x   \,\mathcal{L}(\overline{\Psi}_L,\Psi_L,\overline{\Psi}_R,\Psi_R)}.
\label{calZ}
\end{align}
The integration in the exponent here is along a $4D$ shell with the hyperplane $\Sigma$ as one of its boundaries, its second boundary is a (in general, curved) hypersurface depending on the function $\mathfrak B$ introduced above: \\
$\Sigma_h = \{(h {\mathfrak B}(\bold{x}),\bold{x})|\bold{x} \in \Sigma\}$. At the same time the Lagrangian is given by:
\begin{align}
\nonumber \mathcal{L}(\overline{\Psi}_L,\Psi_L,\overline{\Psi}_R,\Psi_R)=&\frac{i}{4}\overline{\Psi}_L\overset{\leftrightarrow}{D}_0\Psi_L+\frac{i}{4}\overline{\Psi}_R\overset{\leftrightarrow}{D}_0\Psi_R-\mathfrak{v}^{\mu}(\bold{x})  P_{\mu} -\mathfrak{a}^\mu(\bold{x}) K_{\mu} \\
&+\upomega^{\mu}(\bold{x}) J_{\mu}+\frac{1}{2}\upkappa^{\mu\nu}(\bold{x})S_{\mu\nu} +\upmu_A(\bold{x}) Q_A +\upmu_V(\bold{x}) Q_V.
\label{Z1u} 
\end{align}
In this expression we introduce the scalar and vector field (as well as tensor field) parameters with collective notation
\begin{align}
\mathfrak{X}(\bold{x}) = \frac{\beta(0,\bold{x})}{{\mathfrak B}(\bold{x})} X(0, \bold{x}),\,\,\,\, \mathfrak{X}^{\mu}(\bold{x}) = \frac{\beta(0,\bold{x})}{{\mathfrak B}(\bold{x})} X^{\mu}(0, \bold{x}).
\label{mU}
\end{align}
Anti-periodic boundary conditions are implied: 
\begin{align}
\nonumber &\Psi_L(  {\mathfrak B}(\bold{x}) h, \bold{x}) = - \Psi_L(0,\bold{x}), \,\,\,\, \overline{\Psi}_L(  {\mathfrak B}(\bold{x}) h, \bold{x}) = - \overline{\Psi}_L(0,\bold{x}),\\
&\Psi_R(  {\mathfrak B}(\bold{x}) h, \bold{x}) = - \Psi_R(0,\bold{x}), \,\,\,\, \overline{\Psi}_R(  {\mathfrak B}(\bold{x}) h, \bold{x}) = - \overline{\Psi}_R(0,\bold{x}).
\end{align}

Eq. (\ref{Z1u}) is the effective Lagrangian of the system that remains in global thermodynamic equilibrium for motion with four-velocity field $u$. {\it Notice that, although we define here the partition function in rescaled Minkowski spacetime, the functions $\mathfrak{X}$ and $\mathfrak{X}^{\mu}$ entering the effective Lagrangian remain functions of the spatial components of $x$ only.} The types of fields $X^{\mu}(0,\bold{x})$ forming part of the frigidity vector field $\beta^{\mu} (0,\bold{x})$ that are allowed for global thermodynamic equilibrium include motion with constant velocity (which is reduced to the system at rest in the corresponding boosted reference frame, the frame of the thermal bath), the rigid rotation and accelerated motion.

The choice of the function ${\mathfrak B}(\bold{x})$ is free. We may choose $\mathfrak{B}(\bold{x}) = \beta(0,\bold{x})$. In this case 
\begin{align}
\mathfrak{X}(\bold{x}) =  X(0, \bold{x}),\,\,\,\, \mathfrak{X}^{\mu}(\bold{x}) =  X^{\mu}(0, \bold{x}).
\label{mU1}
\end{align}
Then the scalar and vector field (as well as tensor field) parameters $\mathfrak{X}$ and $\mathfrak{X}^{\mu}$ may be interpreted as the corresponding scalar and vector parameter distributions at the initial moment.

\begin{figure}
\begin{center}
\includegraphics[scale=0.48]{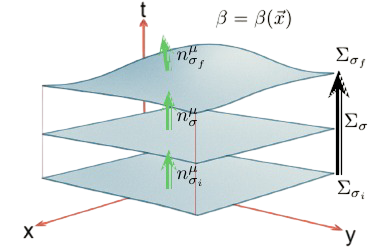}
\end{center}
\caption{Illustration of our path integral procedure. We parametrize our foliation of spacetime by hypersurfaces via the parameter $\sigma$ and embed it into flat Minkowski spacetime. The inverse temperature is allowed to vary as a function of spacelike hypersurface coordinates (illustrated here by Cartesian $x$- and $y$-coordinates in horizontal direction). The vertical direction is the direction of time $t$. A sequence of spacelike hypersurfaces is shown with the uppermost hypersurface being curved and a representative normal vector for each hypersurface. Due to the formulation being restricted to flat Minkowski spacetime, all but the uppermost hypersurfaces are actually hyperplanes. The uppermost hypersurface is curved due to the coordinate dependence of the inverse temperature $\beta$ which sets the scale for the Minkowski spacetime temporal extension via the scaling function $\mathfrak{B}$.}
\label{pathintegralillus}
\end{figure}

\mz{We illustrate our path integral formulation in Fig. (\ref{pathintegralillus}). The inverse temperature is allowed to vary as a function of spacelike hypersurface coordinates (illustrated here by Cartesian $x$- and $y$-coordinates spanning the horizontal directions). The vertical direction is the direction of time $t$. The sequence of hypersurfaces is parametrized by the hypersurface foliation parameter $\sigma$. We show a sequence of spacelike hypersurfaces with the uppermost hypersurface being curved and a representative normal vector for each hypersurface. Since we consider only flat Minkowski spacetime, all but the uppermost hypersurfaces are actually hyperplanes. The uppermost hypersurface is curved due to the coordinate dependence of the inverse temperature $\beta$ which sets the scale for the Minkowski spacetime temporal extension via the scaling function $\mathfrak{B}$.}\par
The constraint $\overline{\Psi}=\Psi^T\hat{U}\hat{d}^{\ast}$ (see Eq. (\ref{majoranaconstraint2}) or equivalently its chiral representation in Eq. (\ref{constraintchiralcomponents})) will reduce the number of integration variables in the path integral by half. In terms of chiral fields we may integrate over either left- or right-handed fields after eliminating the opposite chiral field by the constraint. The integration variables may thus be chosen to be either the pair $(\overline{\Psi}_L,\Psi_L)$ or $(\overline{\Psi}_R ,\Psi_R)$, respectively. In terms of the Dirac field the constraint implies to integrate only over the field configurations of $\Psi$ (or only over those of $\overline{\Psi}$).
The final result of our Lagrangian for $^3$He-A comprising macroscopic motion with $\mathfrak{U}(\bold{x}) = \frac{\beta(0,\bold{x})}{{\mathfrak B}(\bold{x})} u(0, \bold{x})$ reads
\begin{align}
\nonumber \mathcal{L}(\overline{\Psi},\Psi )=&\frac{i}{4}\overline{\Psi}\overset{\leftrightarrow}{D}_0\Psi +\mathfrak{U}^0\frac{1}{8}\overline{\Psi}[i\gamma^0\{\gamma^be_b^j,D_j\}-i\gamma^0\{\gamma^be_b^j,\overset{\leftarrow}{D}_j\}]\Psi -\mathfrak{U}^i\frac{1}{4}[\overline{\Psi}iD_i\Psi -\overline{\Psi}i\overset{\leftarrow}{D}_i\Psi ]+\frac{1}{8}\mathfrak{U}^{\rho}n^{\mu}\epsilon_{\mu\nu\rho\sigma}\nabla^{\nu}\overline{\Psi}\gamma^0\gamma^5e_c^{\sigma}\gamma^c\Psi \\
&+\frac{1}{8}\mathfrak{U}^{\lambda}n^{\mu}\epsilon_{\mu\nu\rho\sigma}e^{a\nu}(\nabla_{\lambda}e_a^{\rho})\overline{\Psi}\gamma^0\gamma^5e_c^{\sigma}\gamma^c\Psi +\frac{\upmu_V}{2}n_{\mu}\overline{\Psi}\gamma^0\gamma^{\mu}\Psi +\frac{\upmu_A}{2}n_{\mu}\overline{\Psi}\gamma^0\gamma^5\gamma^{\mu}\Psi .
\end{align}
The particular choice ${\mathfrak B}(\bold{x})=\beta (0,\bold{x}) u^0(0, \bold{x})$ leads to $\mathfrak{U}^0(\bold{x})\equiv 1$ and the simplification
\begin{align}
\nonumber \mathcal{L}(\overline{\Psi},\Psi )=&\frac{1}{4}\overline{\Psi}\gamma^0ie_b^{\mu}\gamma^bD_{\mu}\Psi -\frac{1}{4}\overline{\Psi}\gamma^0\overset{\leftarrow}{D}_{\mu}ie_b^{\mu}\gamma^b\Psi -\mathfrak{U}^i\frac{1}{4}[\overline{\Psi}iD_i\Psi -\overline{\Psi}i\overset{\leftarrow}{D}_i\Psi ]+\frac{1}{8}\mathfrak{U}^{i}n^{\mu}\epsilon_{\mu\nu i\sigma}\nabla^{\nu}\overline{\Psi}\gamma^0\gamma^5e_c^{\sigma}\gamma^c\Psi \\
&+\frac{1}{8}\mathfrak{U}^{\lambda}n^{\mu}\epsilon_{\mu\nu\rho\sigma}e^{a\nu}(\nabla_{\lambda}e_a^{\rho})\overline{\Psi}\gamma^0\gamma^5e_c^{\sigma}\gamma^c\Psi +\frac{\upmu_V}{2}n_{\mu}\overline{\Psi}\gamma^0\gamma^{\mu}\Psi +\frac{\upmu_A}{2}n_{\mu}\overline{\Psi}\gamma^0\gamma^5\gamma^{\mu}\Psi 
\label{completelagrangian}
\end{align}
where $i$ is a spacelike coordinate index orthogonal to the normal vector $n$. Enforcing $\overline{\Psi}=\Psi^T\hat{U}\hat{d}^{\ast}$ for this Lagrangian implies a vanishing vector charge, while the axial charge persists. More precisely, the currents $j_V^{0\mu}$, $j_V^{1\mu}$, $j_V^{2\mu}$ and $j_A^{3\mu}$ introduced in Eqs. (\ref{generalizedvectorcurrents}) and (\ref{vectorandaxialcurrents}), respectively, vanish on the constrained variable space. This may be proven in a straightforward way. The quantum operators follow this pattern and read explicitly as follows (the vector field ${\bf d}(x)$ below depends on $x$ which we will leave implicit):
\begin{align}
\nonumber\hat{Q}_A^0(x)=\hat{j}_A^{00}(x)=&\int\frac{d^3\delta\bold{q}}{(2\pi )^3}\frac{d^3\delta\bold{p}}{(2\pi )^3}\Bigl[[\overline{a}_+(K_++\delta\bold{q})a_+(K_++\delta\bold{p})+\overline{a}_-(K_++\delta\bold{q})a_-(K_++\delta\bold{p})]e^{i(\delta\bold{p}-\delta\bold{q})x}\\
&-[\overline{a}_+(K_--\delta\bold{q})a_+(K_--\delta\bold{p})+\overline{a}_-(K_--\delta \bold{q})a_-(K_--\delta\bold{p})]e^{-i(\delta\bold{p}-\delta\bold{q})x}\Bigr]\\
\nonumber \hat{Q}_A^1(x)=\hat{j}_A^{10}(x)=&\int\frac{d^3\delta\bold{q}}{(2\pi )^3}\frac{d^3\delta\bold{p}}{(2\pi )^3}\frac{1}{\sqrt{d_1^2+d_2^2}}\Bigl[[(-id_1-d_2)\overline{a}_+(K_++\delta\bold{q})a_-(K_++\delta\bold{p})\\
\nonumber &+(id_1-d_2)\overline{a}_-(K_++\delta\bold{q})a_+(K_++\delta\bold{p})]e^{i(\delta\bold{p}-\delta\bold{q})x}+[(-id_1+d_2)\overline{a}_-(K_--\delta\bold{q})a_+(K_--\delta\bold{p})\\
&+(id_1+d_2)\overline{a}_+(K_--\delta \bold{q})a_-(K_--\delta\bold{p})]e^{-i(\delta\bold{p}-\delta\bold{q})x}\Bigr]\\
\nonumber \hat{Q}_A^2(x)=\hat{j}_A^{20}(x)=&\int\frac{d^3\delta\bold{q}}{(2\pi )^3}\frac{d^3\delta\bold{p}}{(2\pi )^3}\Bigl[\frac{1}{\sqrt{d_1^2+d_2^2}}\{[(-id_1-d_2)(-id_3)\overline{a}_+(K_++\delta\bold{q})a_-(K_++\delta\bold{p})+(id_1-d_2)(id_3)\cdot\\
\nonumber &\,\,\overline{a}_-(K_++\delta\bold{q})a_+(K_++\delta\bold{p})]e^{i(\delta\bold{p}-\delta\bold{q})x}+[(-id_1+d_2)(id_3)\overline{a}_-(K_--\delta\bold{q})a_+(K_--\delta\bold{p})\\
\nonumber &+(id_1+d_2)(-id_3)\overline{a}_+(K_--\delta \bold{q})a_-(K_--\delta\bold{p})]e^{-i(\delta\bold{p}-\delta\bold{q})x}\}+[\overline{a}_+(K_++\delta\bold{q})a_+(K_++\delta\bold{p})\\
\nonumber &-\overline{a}_-(K_++\delta\bold{q})a_-(K_++\delta\bold{p})]e^{i(\delta\bold{p}-\delta\bold{q})x}+[-\overline{a}_+(K_--\delta\bold{q})a_+(K_--\delta\bold{p})\\
&+\overline{a}_-(K_--\delta\bold{q})a_-(K_--\delta\bold{p})]e^{-i(\delta\bold{p}-\delta\bold{q})x}\Bigr]\\
\nonumber \hat{Q}_V^3(x)=\hat{j}_V^{30}(x)=&\int\frac{d^3\delta\bold{q}}{(2\pi )^3}\frac{d^3\delta\bold{p}}{(2\pi )^3}\Bigl[[(d_1-id_2)\overline{a}_+(K_++\delta\bold{q})a_-(K_++\delta\bold{p})\\
\nonumber &+(d_1+id_2)\overline{a}_-(K_++\delta\bold{q})a_+(K_++\delta\bold{p})]e^{i(\delta\bold{p}-\delta\bold{q})x}+[(d_1+id_2)\overline{a}_-(K_--\delta\bold{q})a_+(K_--\delta\bold{p})\\
\nonumber &+(d_1-id_2)\overline{a}_+(K_--\delta \bold{q})a_-(K_--\delta\bold{p})]e^{-i(\delta\bold{p}-\delta\bold{q})x}+[d_3\overline{a}_+(K_++\delta\bold{q})a_+(K_++\delta\bold{p})\\
\nonumber &-d_3\overline{a}_-(K_++\delta\bold{q})a_-(K_++\delta\bold{p})]e^{i(\delta\bold{p}-\delta\bold{q})x}+[d_3\overline{a}_+(K_--\delta\bold{q})a_+(K_--\delta\bold{p})\\
&-d_3\overline{a}_-(K_--\delta\bold{q})a_-(K_--\delta\bold{p})]e^{-i(\delta\bold{p}-\delta\bold{q})x}\Bigr].
\end{align}

\section{Thermodynamic equilibrium solutions for the vierbein}
\label{thermoequisolutions}

In the previous subsections we considered the conditions for thermodynamic equilibrium with and without Nieh Yan form (and axial anomaly) and formulated an effective Lagrangian for macroscopic motion in thermodynamic equilibrium. We postponed there to provide explicit solutions for the vierbein (and therefore as well for $\Omega^{ab}$) and consider them subsequently.
Two obvious types of solutions for the case of vanishing Nieh Yan form and axial anomaly (which is the only case we cover) exist in conjunction with $\beta^{\mu}=b^{\mu}+\omega^{\mu}_{\,\,\,\,\nu}x^{\nu}$ (supplemented by Eq. (\ref{furtherbetarestriction})). These are
\begin{itemize}
\item[1)] $\Omega^{ab}=0$. This case is equivalent to a vanishing Lie derivative of the vierbein along the Killing vector field $\beta^{\mu}$. 
\item[2)] $\nabla_{\mu}e_{\nu}^a=0$ (supplemented by Eq. (\ref{omegarestriction}) in the case of inhomogeneous $\bold{d}$ or non-vanishing $\mathcal{B}_{\mu}$). This case implies homogeneity of the vierbein in space and time and moreover that $\Omega^{ab}=e^{a\mu}e^{b\nu}\omega_{\mu\nu}$. 
\end{itemize}
We would now like to discuss several topological solutions of $^3$He-A in the context of global thermodynamic equilibrium in the presence of macroscopic motion. These comprise pure mass vortices, radial and tangential disclinations and fractional (or spin mass) vortices. Each solution is accompanied by a general discussion of the space of solutions compatible for thermodynamic equilibrium. We will provide considerable details only in the case of pure mass vortices. We start by providing expressions for the components of $\Omega_{\mu}^{\,\,\,\nu}$, which appear in the constraints implied by global thermodynamic equilibrium, in terms of the triad $(\bold{m},\bold{n},\bold{l})$
\begin{align}
\Omega_0^{\,\,\,i}=\omega_0^{\,\,\,i},\,\,\,\, \Omega_i^{\,\,\,j}=(\beta^{\lambda}\nabla_{\lambda}m^j+m^k\omega_k^{\,\,\,j})m^i+(\beta^{\lambda}\nabla_{\lambda}n^j+n^k\omega_k^{\,\,\,j})n^i+(\beta^{\lambda}\nabla_{\lambda}l^j+l^k\omega_k^{\,\,\,j})l^i\label{Omegavector}.
\end{align}
In the following we present explicit topological solutions. Real $^3$He - A features a spin-orbit interaction term with interaction energy $E_{SO}\propto -(\bold{d}\cdot\bold{l})^2$. This is why the natural, energy minimizing configuration is given by $\bold{d}=\pm\bold{l}$. This configuration will be termed dipole locked in the following. In practice this configuration may be obstructed by a strong enough magnetic field in order to achieve misalignment of $\bold{d}$ from $\bold{l}$.\par
The so-called vacuum manifold of $^3$He-A is the factor space
\begin{align}
R_A=G/H_A=(SO(3)\times S^2)/\mathbb{Z}_2.
\end{align} 
The $2$-sphere $S^2$ is spanned by the spin vector field $\bold{d}$, while $SO(3)$ implies rotation of the vector fields $\bold{m}$ and $\bold{n}$ (and therefore also $\bold{l}$). The discrete group $\mathbb{Z}_2=P$ plays a non-trivial role in the classification of topological defects. It gives rise to fractional vortices. This follows from the homotopy group of linear defects $\pi_1(R_A)=\mathbb{Z}_4$. The homotopy group comprises the elements $n_1=0,\pm\frac{1}{2},1$. $n_1=0$ corresponds to the case without topological defect. Deformations may be unwound smoothly. The pure mass vortex and the disclinations are members with $n_1=1$. They may be continuously deformed into each other, although they seem to be different. Indeed, the energy minimizing, interpolating defect of class $n_1=1$ in real superfluid $^3$He-A is a pure mass vortex asymptotically in the direction transverse to the singular defect line whch becomes a vortex texture in a soft core region before transforming to a disclination in the hard core. The $n_1=\pm\frac{1}{2}$ cases cover the fractional vortices. We have $n_1=n_1 \,mod \,2$ with addition among the elements as the homotopy group multiplication. Details beyond our treatment may be found in \cite{Volovik2003}. 

\begin{itemize}
\item[1)] Pure (integer) mass vortices:
We consider a stationary setup with vanishing translational velocity $v^{\mu}$, acceleration $a^{\mu}$ but nonzero rigid rotation around the $z$-axis with $\beta\omega =-\omega^1_{\,\,\,2}=\omega^2_{\,\,\,1}\neq 0$ with angular velocity $\omega$ and inverse temperature $\beta =\frac{1}{T}$. It is convenient to introduce cylindrical coordinates expressed through Cartesian coordinates by 
\begin{align}
x=\rho cos(\phi ),\,\,\,\, y=\rho sin(\phi ),\,\,\,\, z=z.
\end{align}
We denote Cartesian unit vectors by $\hat{x}$, $\hat{y}$ and $\hat{z}$ and those of the cylindrical reference frame by $\hat{\rho}$, $\hat{\phi}$ and $\hat{z}$, respectively.
Then the pure mass vortices are given by 
\begin{align}
\bold{m}+i\bold{n}=e^{-in_1\phi}(\hat{x}+i\hat{y})\, (\, \overset{n_1=1}{\Leftrightarrow} \bold{m}=\hat{\rho},\, \bold{n}=\hat{\phi}),\,\,\, \bold{l}=\hat{z},\,\,\,\, \bold{d}=\pm \hat{z}
\end{align}
with $n_1\in\mathbb{Z}$. The vector $\bold{d}$ is dipole locked with $\bold{l}$. This configuration implies $\bold{E}=\bold{B}=0$, $2M_3\bold{v}_s=\frac{n_1}{\rho}\hat{\phi}$ and 
\begin{align}
&\Omega_0^{\,\,\,i}=0,\,\,\,\, \Omega_i^{\,\,\,j}=[\omega^2_{\,\,\,1}n_1n^j+m^k\omega_k^{\,\,\,l}n^ln^j]m^i+[\omega^2_{\,\,\,1}(-n_1)m^j+n^k\omega_k^{\,\,\,l}m^lm^j]n^i,\\
&\Omega_i^{\,\,\,j}m^in^j=\omega^2_{\,\,\,1}(n_1-1)=\omega (n_1-1),\\
&T^1_{\,\,\,12}=v_s^1=-\frac{n_1}{\rho}sin(n_1\phi ),\,\,\,\, T^2_{\,\,\,12}=v_s^2=\frac{n_1}{\rho}cos(n_1\phi ),\,\,\,\,T^{\mu}_{\,\,\,\nu\rho}=0\,\,\text{otherwise}\,\,\Rightarrow a=\nabla_{\mu}K^{\mu}=0.
\end{align}
We also have $G_a=P_{ab}=0$. The pure mass vortex is a global thermodynamic equilibrium solution in the presence of both $(s=+)$- and $(s=-)$-components, as they are compatible with Eqs. (\ref{furtherbetarestriction}) and (\ref{omegarestriction}). 
Consider now a slight misalignment of the rotation axis of macroscopic motion from that of the vortex axis $\omega_{\mu}^{\,\,\,\nu}\to \omega_{\mu}^{\,\,\,\nu} +\delta\omega_{\mu}^{\,\,\,\nu}$. We have $\Omega_3^{\,\,\,j}=\delta\omega_3^{\,\,\,j}\neq 0$ which is compatible with Eq. (\ref{omegarestriction}), since $P_{ab}=0$ and also with Eq. (\ref{anomalyconstraint}). This means that global thermodynamic equilibrium is achieved for any relative orientation of vortex and macroscopic motion rotation axis. This is a consequence of the fact that not the Lie derivative of the torsion tensor field along the frigidity vector field is supposed to vanish in global thermodynamic equilibrium but only its antisymmetrization. If $\bold{d}$ gets dipole unlocked with $\bold{l}$ such that $P_{j3}\neq 0$, global thermodynamic equilibrium is achievable if and only if the vortex axis is aligned with the rotation axis of macroscopic motion.\par
We consider now the Lagrangian and the equations of motion in the case of alignment of the vortex axis and the rotation axis of macroscopic motion with dipole locked $\bold{d}$. It suffices to treat the $(s=+)$-component, since both components are not coupled for homogeneous $\bold{d}$. We further restrict to the left-handed fermions. The case of right-handed fermions may be obtained by setting $v_{\parallel},v_{\perp}\to -v_{\parallel},-v_{\perp}$ implying a helicity flip. The Lagrangian of Eq. (\ref{Z1u}) for our case is given by (with $n^{\mu}=(1,0,0,0)$)
\begin{align}
\mathcal{L}^+_L=\frac{i}{4}\overline{\Psi}_L^+\overset{\leftrightarrow}{\nabla}_0\Psi_L^+-\mathcal{H},\,\,\, \mathcal{H}=\mathcal{H}_0+\mathcal{H}_{\omega}
\end{align}
with Hamiltonian $\mathcal{H}$ which splits into the contribution $\mathcal{H}_0$ without macroscopic motion
\begin{align}
\nonumber \mathcal{H}_0\equiv &(P^{L,+})_0=-\frac{1}{2}\overline{\Psi}_L^+ie_b^j\overline{\tau}^b\nabla_j\Psi_L^+-\frac{1}{4}\overline{\Psi}_L^+i(\nabla_je_b^j)\overline{\tau}^b\Psi_L^+\\
\nonumber =&\frac{1}{2}\overline{\Psi}_L^+(\sigma^1v_{\perp}im^j\nabla_j+\sigma^2v_{\perp}in^j\nabla_j+v_{\parallel}\sigma^3i\nabla_z+\frac{1}{2}(\nabla_je_k^j)\sigma^k)\Psi_L^+\\
=&\frac{1}{2}\overline{\Psi}_L^+\begin{pmatrix}v_{\parallel}i\nabla_z & iv_{\perp}e^{i(n_1-1)\phi}(\nabla_{\rho}-\frac{1}{\rho}i\nabla_{\phi}+\frac{n_1}{2}\frac{1}{\rho})\\  
iv_{\perp}e^{i(1-n_1)\phi}(\nabla_{\rho}+\frac{1}{\rho}i\nabla_{\phi}+\frac{n_1}{2}\frac{1}{\rho}) & -v_{\parallel}i\nabla_z\end{pmatrix}\Psi_L^+,
\end{align}
with macroscopic motion contribution $\mathcal{H}_{\omega}$ given by
\begin{align}
\nonumber \beta\mathcal{H}_{\omega}=&-\overline{\Psi}_L^+(\frac{1}{2}\omega^{ij}(x_i(P^{L,+})_j-x_j(P^{L,+})_i)+\frac{1}{2}\Omega^{ij}(S^{L,+})_{ij})\Psi_L^+\\
=&-\overline{\Psi}_L^+(\frac{1}{2}\omega^{ij}x_ii\nabla_j+\frac{1}{8}\Omega^{ij}n^{\mu}\epsilon_{\mu ijc}\sigma^c)\Psi_L^+=\beta\overline{\Psi}_L^+(\frac{1}{2}\omega i\nabla_{\phi}+\frac{1}{4}\omega (n_1-1)\sigma^3)\Psi_L^+.
\end{align}
%and misaligned macroscopic motion contribution $\mathcal{H}_{\delta\omega}$ which we treat perturbatively and which takes the form
%\begin{align}
%\beta\mathcal{H}_{\delta\omega}=&-\overline{\Psi}_L^+(\frac{1}{2}\delta\omega^{ij}(x_i(P^{L,+})_j-x_j(P^{L,+})_i)+\frac{1}{2}\delta\Omega^{ij}(S^{L,+})_{ij})\Psi_L^+\\
%=&-\overline{\Psi}_L^+(\frac{1}{2}\delta\omega^{ij}x_ii\nabla_j+\frac{1}{8}\delta\Omega^{ij}n^{\mu}\epsilon_{\mu ijc}\sigma^c)\Psi_L^+=-\overline{\Psi}_L^+(\frac{1}{2}\delta\omega_3^{\,\,\,j}zi\nabla_j+\frac{1}{2}\delta\omega_i^{\,\,\,3}ix^i\nabla_z+\frac{1}{4}\delta\omega^{i3}\epsilon_{0i3j}e^j_c\sigma^c)\Psi_L^+
%\end{align}
where we made use of $\epsilon_{0123}=1$ (with Lorentz indices). The equations of motion $\frac{\delta\mathcal{L}_L^+}{\delta \overline{\Psi}_L^+}$ can be easily read from $\mathcal{L}_L^+$. Due to the stationarity of the pure mass vortex, its rotational and translational symmetries we may proceed to find the eigenvalues $E$ and eigenfunctions of $\mathcal{H}_0+\mathcal{H}_{\omega}$ by employing the spinorial ansatz
\begin{align}
\Psi_L^+=\begin{pmatrix} B_1(\rho )e^{im_1\phi} \\
 B_2(\rho )e^{im_2\phi}\end{pmatrix}e^{ip^zz}e^{-iEt}.
\end{align}
with $m_1,m_2\in\mathbb{Z}$. We obtain
\begin{align}
0\overset{!}{=}&(-\frac{i}{2}\nabla_0\Psi^+_L+\frac{\delta\mathcal{H}_0}{\delta\overline{\Psi}_L^+}+\frac{\delta\mathcal{H}_{\omega}}{\delta\overline{\Psi}_L^+})\\
=&\begin{pmatrix} (-\frac{E}{2}-\frac{v_{\parallel}}{2}p^z-(\frac{m_1}{2}-\frac{1}{4}(n_1-1))\omega)B_1(\rho )e^{im_1\phi}+i\frac{v_{\perp}}{2}e^{i(n_1-1)\phi}(\frac{B_2^{\prime}}{B_2}+\frac{\frac{n_1}{2}+m_2}{\rho})B_2(\rho )e^{im_2\phi} \\ i\frac{v_{\perp}}{2}e^{i(1-n_1)\phi}(\frac{B_1^{\prime}}{B_1}+\frac{\frac{n_1}{2}-m_1}{\rho})B_1(\rho )e^{im_1\phi}+(-\frac{E}{2}+\frac{v_{\parallel}}{2}p^z-(\frac{m_2}{2}+\frac{1}{4}(n_1-1))\omega)B_2(\rho )e^{im_2\phi}\end{pmatrix}e^{ip^zz}e^{-iEt}.
\label{eigenvalue1}
\end{align}
The pure mass vortices we are considering have an infinitely thin core. We are interested in the normalizable bound states. 
The four terms in Eq. (\ref{eigenvalue1}) can not vanish independently, unless either $B_1=0$ or $B_2=0$. We therefore require $m_1=n_1-1+m_2$ (, unless either $B_1=0$ or $B_2=0$, but this is not compatible with normalizability). 

%Let us start assuming that $A_1=0$. This implies for a non-trivial solution
%\begin{align}
%A_2(\rho )=A\rho^{-m_2},\,\,A\in\mathbb{C},\,\,m_2>1,\,\,\,\, E=E_+=v_{\parallel}p^z+(m_2+\frac{1}{2}(n_1-1))\Omega.
%\end{align}
%Similarly for $A_2=0$ we obtain as the non-trivial solution
%\begin{align}
%A_1(\rho )=A\rho^{m_1},\,\,A\in\mathbb{C},\,\,m_1<-1,\,\,\,\, E=E_-=-v_{\parallel}p^z+(m_1-\frac{1}{2}(n_1-1))\Omega .
%\end{align}
In the case where both $B_1\neq 0$ and $B_2\neq 0$ %and $p^z\neq 0$ 
the eigenvalue problem with $m\equiv m_1=n_1-1+m_2$ becomes  
\begin{align}
0\overset{!}{=}\begin{pmatrix} (-E-v_{\parallel}p^z-(m-\frac{1}{2}(n_1-1))\omega)B_1(\rho )+iv_{\perp}(B_2^{\prime}(\rho )+\frac{m+1-\frac{n_1}{2}}{\rho}B_2(\rho )) \\ iv_{\perp}(B_1^{\prime}(\rho )+\frac{\frac{n_1}{2}-m}{\rho}B_1(\rho ))+(-E+v_{\parallel}p^z-(m-\frac{1}{2}(n_1-1))\omega )B_2(\rho )\end{pmatrix}.
\label{eigenvalue2}
\end{align}
%in a first step. Assume now that $E=E_+$. This implies 
%\begin{align}
%A_1(\rho )=A\rho^{m},\,\,\,\, A_2(\rho )=-iA\frac{2}{2(m+1)-n_1}\frac{v_{\parallel}}{v_{\perp}}p^z\rho^{m+1},\,\,\,\,A\in\mathbb{C},\,\,m<-2,\,\,m\neq \frac{n_1}{2}-1.
%\end{align}
%The case $E=E_-$ leads to
%\begin{align}
%A_1(\rho )=-iA\frac{2}{-2m+n_1}\frac{v_{\parallel}}{v_{\perp}}p^z\rho^{-m+n_1},\,\,\,\, A_2(\rho )=A\rho^{-m+n_1-1},\,\,m>n_1+1,\,\,m\neq \frac{n_1}{2}.
%\end{align}
%The special case $p^z=0$ implies $E_+=E_-=m-\frac{1}{2}(n_1-1)$ and results with $E=E_+=E_-$ in 
%\begin{align}
%A_1(\rho )=A_1\rho^{m},\,\,\,\, A_2(\rho )=A_2\rho^{-m+n_1-1},\,\, A_1,A_2\in\mathbb{C},\,\, m<-1,\,\, m>n_1.
%\end{align}
%For $E\neq E_{\pm}$ 
Define $m=n+\frac{n_1}{2}$ such that $n\in \mathbb{Z}$ for even $n_1$ and $n\in \mathbb{Z}+\frac{1}{2}$ for odd $n_1$.
Eq. (\ref{eigenvalue2}) then becomes
\begin{align}
0\overset{!}{=}\begin{pmatrix} (-E-v_{\parallel}p^z-(n+\frac{1}{2})\omega)B_1(\rho )+iv_{\perp}(B_2^{\prime}(\rho )+\frac{n+1}{\rho}B_2(\rho )) \\ iv_{\perp}(B_1^{\prime}(\rho )-\frac{n}{\rho}B_1(\rho ))+(-E+v_{\parallel}p^z-(n+\frac{1}{2})\omega)B_2(\rho )\end{pmatrix}
\label{eigenvalue3}
\end{align}
Notice that we reduced the problem to the case without vortex by this index shift. We will proceed by assuming a solution of the form 
\begin{align}
B_2^{\prime}(\rho )+\frac{n+1}{\rho}B_2(\rho )=qB_1,\,\,\,\, B_1^{\prime}(\rho )-\frac{n}{\rho}B_1(\rho )=rB_2,\,\,\,\,q,r\in\mathbb{C}.
\end{align}
This leads to the new decoupled differential equations
\begin{align}
&\rho^2B_1^{\prime\prime}(\rho )+\rho B_1^{\prime}(\rho )+(-qr\rho^2-n^2)B_1=0,\,\,\,\,\rho^2B_2^{\prime\prime}(\rho )+\rho B_2^{\prime}(\rho )+(-qr\rho^2-(n+1)^2)B_2=0.
\end{align}
We require $-qr>0$ and define $\rho_0^{-2}=-qr$. This requirement is necessary in order to interpret the eigenfunctions as bound states. Together with the coordinate redefinition $\rho\to\frac{\rho}{\rho_0}$ solutions to these equations are Bessel functions of the first kind with $B_1(\frac{\rho}{\rho_0} )=B_1J_n(\frac{\rho}{\rho_0})$ as well as $B_2(\frac{\rho}{\rho_0} )=B_2J_{n+1}(\frac{\rho}{\rho_0})$ and $B_1,B_2\in\mathbb{C}$. 
Eq. (\ref{eigenvalue3}) then becomes fully algebraic. Employing the relations
\begin{align}
\hat{P}_{\pm}(e^{in\phi}J_n(\frac{\rho}{\rho_0}))=\mp \frac{i}{\rho_0}e^{i(n\pm 1)\phi}J_{n\pm 1}(\frac{\rho}{\rho_0}),\,\,\,\, \hat{P}_{\pm}=ie^{\pm i\phi}(\nabla_{\rho}\pm\frac{i}{\rho}\nabla_{\phi})
\end{align}
we find
\begin{align}
0\overset{!}{=}\begin{pmatrix} -E-v_{\parallel}p^z-(n+\frac{1}{2})\omega & i\frac{v_{\perp}}{\rho_0} \\ -i\frac{v_{\perp}}{\rho_0} & -E+v_{\parallel}p^z-(n+\frac{1}{2})\omega \end{pmatrix}
\begin{pmatrix} B_1 \\  B_2 \end{pmatrix}.
\label{eigenvalue4}
\end{align}
The eigenvalues and amplitudes are finally given by
\begin{align}
&E_{\pm}=-(n+\frac{1}{2})\omega\pm \sqrt{(v_{\parallel}p^z)^2+(\frac{v_{\perp}}{\rho_0})^2},\\
&B^{\pm}_2=i\frac{(\mp \sqrt{(v_{\parallel}p^z)^2+(\frac{v_{\perp}}{\rho_0})^2}-v_{\parallel}p^z)\rho_0}{v_{\perp}}B^{\pm}_1\equiv iN^{\pm}(v_{\parallel},v_{\perp})B^{\pm}_1\equiv iN^{\pm}(v_{\parallel},v_{\perp})B^{\pm},
\label{energyeigenvaluesandamplitudes}
\end{align}
respectively. A basis of normalizable solutions involves only those functions with $n\geq 0$. We furthermore choose $E=E_+$ with $n\geq 0$ (and $B=B^+$, $N=N^+$) in order to get an energy spectrum which is bounded from below. Our eigenspinors are then given by
\begin{align}
\Psi^+_L=Be^{-iEt}e^{ip^zz}\begin{pmatrix} e^{i(n+\frac{n_1}{2})\phi}J_n(\frac{\rho}{\rho_0}) \\  iN(v_{\parallel},v_{\perp})e^{i(n+1-\frac{n_1}{2})\phi}J_{n+1}(\frac{\rho}{\rho_0}) \end{pmatrix}.
\end{align}
Since only $2n\in\mathbb{Z}$ in general, we introduce $l=m-n_1=n-\frac{n_1}{2}$ with $l\in\mathbb{Z}$ such that $l\geq -\frac{n_1}{2}$.
We obtain the solutions of the right-handed fermions by the replacement $N(v_{\parallel},v_{\perp})\to N(-v_{\parallel},-v_{\perp})$ which changes the eigenstates, while the energy levels remain invariant. We thus have a degeneracy of the four distinct degrees of freedom spanned by the helicity (or chirality) eigenstates with eigenvalues $\frac{\bold{p}\boldsymbol{\tau}}{|\bold{p}|}=\pm 1$ and the eigenstates with intrinsic spin eigenvalues $\bold{d}\boldsymbol{\sigma}=\pm 1$. The original number of degrees of freedom are restricted from eight to four due to the constraint Eq. (\ref{majoranaconstraint}) (see also Eqs. (\ref{majoranaconstraint2}) and (\ref{constraintchiralcomponents}) in Appendix \ref{app1}).\par
\begin{figure}
\begin{center}
\includegraphics[scale=0.48]{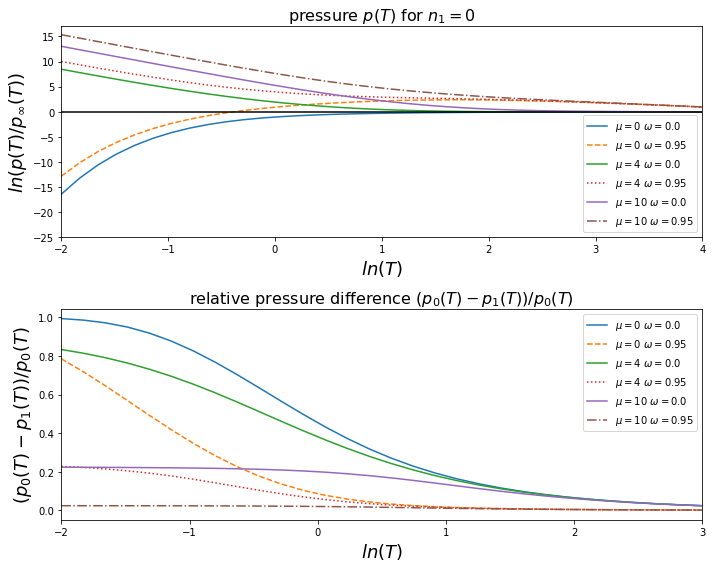}
\end{center}
\caption{Thermodynamic equilibrium pressure $p$ in units of $p_0$ (see Table \ref{tableunits}) as a function of temperature $T$ (in the units of $\omega_0$) for four fermionic particle species ($(L/\overline{L},\pm )$ or equivalently $(R/\overline{R},\pm )$) confined to a cylinder with finite transverse size but infinite longitudinal size subject to MIT bag boundary conditions without fermion doubling. The pressure is compared to its high temperature expression $p_{\infty}$ (upper plot) as well as among the two topologically different vortex configurations with topological indices $n_1=0$ (no vortex) and $n_1=1$ (vortex) (lower plot). The vortex is located at the transverse center featuring an infinitely thin core. We employ our natural units $\hbar =k_B=(v_{\parallel}v_{\perp}^2)^{\frac{1}{3}}=R/v_{\perp}=1$. Therefore, both $\mu$ and $\omega$ are measured in the units of $\omega_0$ (see Table \ref{tableunits}).}
\label{pressure}
\end{figure}
We proceed to calculate the grand canonical potential $\Omega$ for our system of fermions at finite temperature $T$, chemical potential $\mu$ and angular velocity $\omega$. The chemical potential is due to the axial charge, since the vector charge of the Lagrangian in Eq. (\ref{completelagrangian}) vanishes after restriction to the constrained variables. The system is necessarily bound to be finite by the causality constraint $\omega R\leq v_{\perp}$ with $R$ the transverse radius. We will then need to introduce boundary conditions for the spinors. We choose MIT bag boundary conditions 
\begin{align}
(i\gamma^{\mu}n_{\mu}-1)\Psi^{\pm}\Big\rvert_{\rho =R}=0\,\,\Rightarrow \,\, j^{\mu}_{\pm}n_{\mu}=0,\,\,\,\, j^{\mu}_{\pm}=\frac{1}{2}\overline{\Psi}^{\pm}\gamma^0\gamma^{\mu}\Psi^{\pm}.
\end{align}
These boundary conditions imply a mixing of the chiral components according to
\begin{align}
\Psi^{\pm}=C^{\pm}_L\Psi^{\pm}_L+C^{\pm}_R\Psi^{\pm}_R,\,\,\,\, \frac{C^{\pm}_R}{C^{\pm}_L}=N(-v_{\parallel},-v_{\perp})\frac{J_{l+\frac{n_1}{2}}(\frac{R}{\rho_0})}{J_{l+1+\frac{n_1}{2}}(\frac{R}{\rho_0})}\equiv \pm N(-v_{\parallel},-v_{\perp}).
\label{MITbag}
\end{align}
The final equality within Eq. (\ref{MITbag}) implies the quantization of $\frac{1}{\rho_0}$. We will call its quantized values $q_{l,s}$. Since the MIT bag boundary conditions imply a relation between the chiral components, we will restore a factor of two in the number of degrees of freedom counting. The (vacuum subtracted and therefore renormalized) pressure (or the negative grand canonical potential) reads
\begin{align}
p(T,\mu ,\omega )=&-\Omega (T,\mu ,\omega )=lim_{L\to\infty}\frac{1}{\pi R^2L}Tln(Z)=\frac{4}{\pi R^2}\int \frac{dp^z}{2\pi}\sum_{l\geq -\frac{n_1}{2},s}\sum_{\pm}Tln(1+e^{-\beta (E_{l,s}(p^z)\pm\mu )})
\label{pressureformula}
\end{align}
with inverse temperature $\beta =\frac{1}{T}$ and longitudinal size $L$. The Euclidean partition function is denoted by $Z$. We have written the energy eigenvalues in terms of their discrete ($l,s$) as well as continuous ($p^z$) variables. Notice that we chose units such that $\hbar = k_B=1$ and $(v_{\parallel}v_{\perp}^2)^{\frac{1}{3}}=1$. We will further restrict ourselves to dimensionless units by setting $\frac{R}{v_{\perp}}=1$. These units are further discussed in Appendix \ref{app3}. Causality then demands $\omega <1$. We do not consider here the instability occuring due to the violation of the inequalities of Eq. (\ref{velocityinequalities}). The limit $T\gg 1$ implies that the impact of the boundary to the pressure and dependent quanitites is negligibly small, while the opposite regime $T\lesssim 1$ is significantly dependent on $R$. With the definition $Q_{l,s}=v_{\perp}q_{l,s}$ we may finally write (in these dimensionless units):
\begin{figure}
\begin{center}
\includegraphics[scale=0.48]{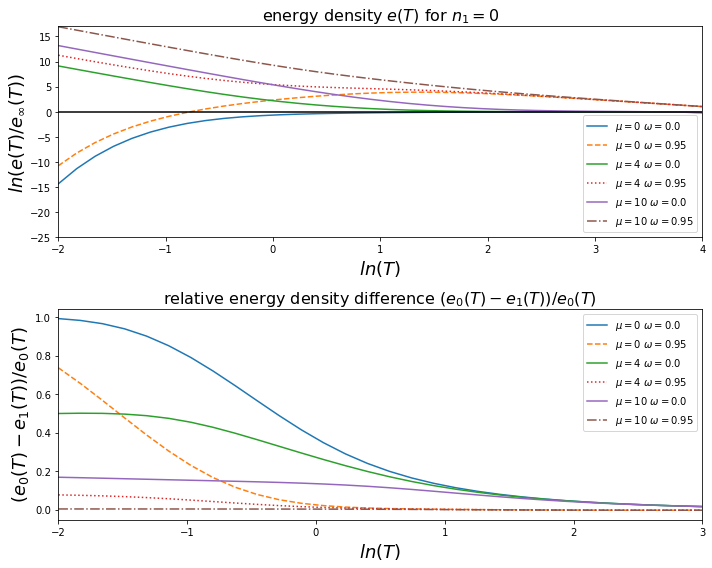}
\end{center}
\caption{Thermodynamic equilibrium energy density $e$ (in the units of $p_0$) as a function of temperature $T$ (in the units of $\omega_0$) for four fermionic particle species ($(L/\overline{L},\pm )$ or equivalently $(R/\overline{R},\pm )$) confined to a cylinder with finite transverse size but infinite longitudinal size subject to MIT bag boundary conditions without fermion doubling. The energy density is compared to its high temperature expression $e_{\infty}$ (upper plot) as well as among the two topologically different vortex configurations with topological indices $n_1=0$ (no vortex) and $n_1=1$ (vortex) (lower plot). The vortex is located at the transverse center featuring an infinitely thin core. We employ our natural units $\hbar =k_B=(v_{\parallel}v_{\perp}^2)^{\frac{1}{3}}=R/v_{\perp}=1$. Therefore, both $\mu$ and $\omega$ are measured in the units of $\omega_0$ (see Table \ref{tableunits}).}
\label{energy}
\end{figure}
\begin{align}
&p(T,\mu ,\omega )=\frac{2T}{\pi^2}\int_{-\infty}^{\infty}dx\sum_{l\geq -\frac{n_1}{2},s}\sum_{\pm}ln\Big(1+e^{-\frac{E_{l,s}(x)\pm\mu}{T}}\Big)=\frac{2}{\pi^2}\int_{-\infty}^{\infty}dx\sum_{l\geq -\frac{n_1}{2},s}\sum_{\pm}\frac{x^2}{\sqrt{x^2+Q_{l,s}^2}}n_F(T,\pm\mu ,\omega ),\nonumber\\
\nonumber
& n_F(T,\mu ,\omega )=\frac{1}{1+e^{\frac{E_{l,s}(x)-\mu}{T}}},\,\,\,\, E_{l,s}(x)=-(l+\frac{n_1}{2}+\frac{1}{2})\omega +\sqrt{x^2+Q_{l,s}^2},\,\,\,\, \frac{J_{l+\frac{n_1}{2}}(Q_{l,s})}{J_{l+1+\frac{n_1}{2}}(Q_{l,s})}=\pm 1.
\end{align}
In ordinary relativistic units we have:
	\begin{align}
		&p(T,\mu ,\omega )=p_0\frac{2}{\pi^2}\int_{-\infty}^{\infty}dx\sum_{l\geq -\frac{n_1}{2},s}\sum_{\pm}\frac{x^2}{\sqrt{x^2+Q_{l,s}^2}}n_F(T,\pm\mu ,\omega ),\label{pressureformula2}\\ & p_0 = \frac{d^3}{f(R)^4}(eV)^4, \quad d=\frac{c}{(v_{\parallel}v_{\perp}^2)^{\frac{1}{3}}}, \quad f(R)=\frac{(eV)\cdot R}{v_{\perp}} 
		\nonumber\\
		& n_F(T,\mu ,\omega )=\frac{1}{1+e^{\frac{E_{l,s}(x)-\mu}{T}}},\,\,\,\, E_{l,s}(x)=-(l+\frac{n_1}{2}+\frac{1}{2})\omega +\omega_0\sqrt{x^2+Q_{l,s}^2},\,\,\,\, \frac{J_{l+\frac{n_1}{2}}(Q_{l,s})}{J_{l+1+\frac{n_1}{2}}(Q_{l,s})}=\pm 1, \\& \omega_0 = \frac{1}{f(R)}eV \nonumber
	\end{align}

In the following we use the units of Table \ref{tableunits}.
	\begin{table}
\begin{center}
	\begin{tabular}{lclclclcl}
		energy  & $\omega_0\cong \frac{1}{f(R)}eV$ & \,\,\,\,\,\,\,\,\,\,\, temperature & $\omega_0\cong \frac{1}{f(R)}eV$ \\
		momentum & ${\cal P}_0\cong\frac{d}{f(R)}eV$ & \,\,\,\,\,\,\,\,\,\,\, pressure & $p_0\cong\frac{d^3}{f(R)^4}(eV)^4$ \\
		mass & $m_0\cong\frac{d^2}{f(R)}eV$ & \,\,\,\,\,\,\,\,\,\,\, entropy density & $n_0\cong\frac{d^3}{f(R)^3}(eV)^3$ \\
		time & $t_0\cong f(R) (eV)^{-1}$ & \,\,\,\,\,\,\,\,\,\,\, particle number density & $n_0\cong\frac{d^3}{f(R)^3}(eV)^3$ \\
		position & $x_0\cong\frac{f(R)}{d}(eV)^{-1}$ & \,\,\,\,\,\,\,\,\,\,\, angular momentum density & $n_0\cong\frac{d^3}{f(R)^3}(eV)^3$
	\end{tabular}
\caption{Units of physical quantities represented in the main text
 ($d=\frac{c}{(v_{\parallel}v_{\perp}^2)^{\frac{1}{3}}}, \quad f(R)=\frac{(eV)\cdot R}{v_{\perp}}$). \label{tableunits}}
\end{center}
\end{table}

We plot the pressure $p$ (in the units of $p_0$) as well as the particle number density 
\begin{align}
n=\frac{\partial p}{\partial \mu}=n_0\frac{2}{\pi^2}\int_{-\infty}^{\infty}dx\sum_{l\geq -\frac{n_1}{2},s}\sum_{\pm}n_F(T,\pm\mu ,\omega ),
\end{align}
the angular momentum density 
\begin{align}
j=\frac{\partial p}{\partial \omega}=n_0\frac{2}{\pi^2}\int_{-\infty}^{\infty}dx\sum_{l\geq -\frac{n_1}{2},s}\sum_{\pm}(l+\frac{n_1}{2}+\frac{1}{2})n_F(T,\pm\mu ,\omega ),
\end{align}
the entropy density 
\begin{align}
s=\frac{\partial p}{\partial T}=n_0\frac{p}{T}+n_0\frac{2}{\pi^2T}\int_{-\infty}^{\infty}dx\sum_{l\geq -\frac{n_1}{2},s}\sum_{\pm}(E_{l,s}(x)\mp\mu )n_F(T,\pm\mu ,\omega )
\label{entropyformula}
\end{align}
and the energy density 
\begin{align}
e=-p+sT+n\mu +j\omega =p_0\frac{2}{\pi^2}\int_{-\infty}^{\infty}dx\sum_{l\geq -\frac{n_1}{2},s}\sum_{\pm}\sqrt{x^2+Q_{l,s}^2}n_F(T,\pm\mu ,\omega )
\label{energyformula}
\end{align} 
as functions of temperature $T$ for different values of the the chemical potential $\mu$ and the angular velocity $\omega$ for the choices $n_1=0,1$. Inspection of the formula in Eq. (\ref{pressureformula}) reveals that the two choices of the topological index already exhaust all cases. This observation is in line with all mass vortices modulo two in the topological index being topologically equivalent (see, e. g., \cite{Volovik2003} for more details). \par
\begin{figure}
\begin{center}
\includegraphics[scale=0.48]{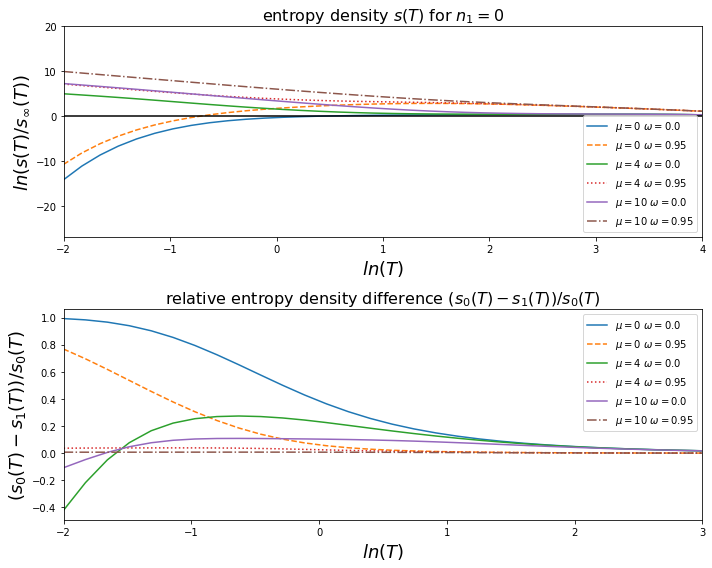}
\end{center}
\caption{Thermodynamic equilibrium entropy density $s$ (in the units of $n_0$) as a function of temperature $T$ (in the units of $\omega_0$) for four fermionic particle species ($(L/\overline{L},\pm )$ or equivalently $(R/\overline{R},\pm )$) confined to a cylinder with finite transverse size but infinite longitudinal size subject to MIT bag boundary conditions without fermion doubling. The entropy density is compared to its high temperature expression $s_{\infty}$ (upper plot) as well as among the two topologically different vortex configurations with topological indices $n_1=0$ (no vortex) and $n_1=1$ (vortex) (lower plot). The vortex is located at the transverse center featuring an infinitely thin core. We employ our natural units $\hbar =k_B=(v_{\parallel}v_{\perp}^2)^{\frac{1}{3}}=R/v_{\perp}=1$. Therefore, both $\mu$ and $\omega$ are measured in the units of $\omega_0$ (see Table \ref{tableunits}).}
\label{entropy}
\end{figure}
A numerical evaluation of the pressure, energy density and entropy density over temperature ranges of five and six $e$-folds yields the results presented in Figs. (\ref{pressure}), (\ref{energy}) and (\ref{entropy}), respectively. \par
In the upper plots we compare the calculated values with the expected asymptotic high temperature limits
\begin{align}
\frac{p_{\infty}(T)}{p_0}=4\cdot\frac{7}{8}\frac{1}{90}\pi^2\Bigl(\frac{T}{\omega_0}\Bigr)^4=\frac{7\pi^2}{180}\Bigl(\frac{T}{\omega_0}\Bigr)^4,\,\,\,\, \frac{e_{\infty}(T)}{p_0}=3p_{\infty}=\frac{7\pi^2}{60}\Bigl(\frac{T}{\omega_0}\Bigr)^4,\,\,\, \frac{s_{\infty}(T)}{n_0}=\frac{7\pi^2}{45}\Bigl(\frac{T}{\omega_0}\Bigr)^3
\end{align}
via their ratios. At high temperatures $T\gg \omega_0$ we expect that $lim_{T\to\infty}ln(X(T)/X_{\infty}(T))=0$ where $X=p,e,s$ which is in line with our calculations, though it can be seen in the plots that $ln(T/\omega_0)=4$ still deviates visibly from the asymptotic limit, which is represented by the black horizontal line, for the case of large angular velocity $\omega$ (in distinction to relatively large chemical potential where convergence is much faster). Towards lower temperatures, the aforementioned ratio begins to deviate significantly from the asymptotic behaviour. This applies not only for different finite choices of the chemical potential $\mu$ and angular velocity $\omega$ but also for the case $\mu =\omega =0$ and is a consequence of the finite system size. Finite size effects become significant for $T\lesssim \frac{v_{\perp}}{R}$. Not all modes are allowed at finite system size but only those compatible with the boundary conditions. This amounts to a considerable drop in pressure as a function of temperature as compared to the ''infinite volume'' limit in the transverse direction. The pressure is enhanced in the presence of particles ($\mu >0$) as well as rotation ($\omega >0$). The same applies to the energy and entropy densities. The pressure, energy density and entropy density increase with chemical potential and angular momentum for fixed temperature. \par
\begin{figure}
\begin{center}
\includegraphics[scale=0.48]{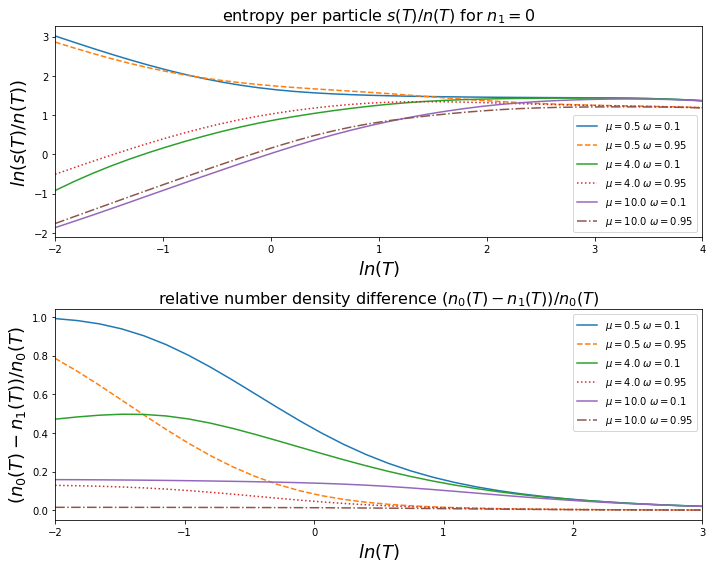}
\end{center}
\caption{Thermodynamic equilibrium particle number density $n$ (in the units of $n_0$) as a function of temperature $T$ (in the units of $\omega_0$) for four fermionic particle species ($(L/\overline{L},\pm )$ or equivalently $(R/\overline{R},\pm )$) confined to a cylinder with finite transverse size but infinite longitudinal size subject to MIT bag boundary conditions without fermion doubling. The particle number density is compared to the entropy density via the entropy per particle number ratio $s/n$ (upper plot) as well as among the two topologically different vortex configurations with topological indices $n_1=0$ (no vortex) and $n_1=1$ (vortex) (lower plot). The vortex is located at the transverse center featuring an infinitely thin core. We employ our natural units $\hbar =k_B=(v_{\parallel}v_{\perp}^2)^{\frac{1}{3}}=R/v_{\perp}=1$. Therefore, both $\mu$ and $\omega$ are measured in the units of $\omega_0$ (see Table \ref{tableunits}).}
\label{entropyperparticle}
\end{figure}
\begin{figure}
\begin{center}
\includegraphics[scale=0.48]{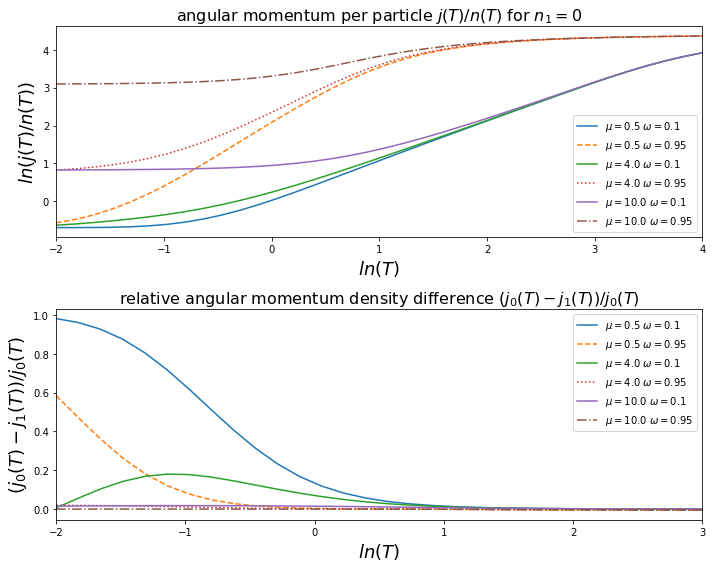}
\end{center}
\caption{Thermodynamic equilibrium angular momentum density $j$ (in the units of $n_0$) as a function of temperature $T$ (in the units of $\omega_0$) for four fermionic particle species ($(L/\overline{L},\pm )$ or equivalently $(R/\overline{R},\pm )$) confined to a cylinder with finite transverse size but infinite longitudinal size subject to MIT bag boundary conditions without fermion doubling. The angular momentum density is compared to the particle number density via the angular momentum per particle ratio $j/n$ (upper plot) as well as among the two topologically different vortex configurations with topological indices $n_1=0$ (no vortex) and $n_1=1$ (vortex) (lower plot). The vortex is located at the transverse center featuring an infinitely thin core. We employ our natural units $\hbar =k_B=(v_{\parallel}v_{\perp}^2)^{\frac{1}{3}}=R/v_{\perp}=1$. Therefore, both $\mu$ and $\omega$ are measured in the units of $\omega_0$ (see Table \ref{tableunits}).}
\label{angularmomentumperparticle}
\end{figure}
In the lower plots we compare the topologically distinct cases $n_1=0$ and $n_1=1$ by considering the relative difference of the two cases for pressure, energy density and entropy density, respectively. While at high temperatures the presence of a vortex has only a minor effect on the thermodynamic quantities, the low temperature limit exposes the effect of a vortex quite visibly, since at low temperatures only a small number of discrete modes is thermodynamically accessible. An increase in temperature enhances the number of discrete modes participating in thermodynamic fluctuations. The pressure, energy density and entropy density get significantly reduced in the presence of a vortex. The relative suppression decreases with chemical potential and angular velocity. If both chemical potential and angular velocity are large, the presence of a vortex is basically irrelevant to the thermodynamics. For low angular velocities and intermediate to large chemical potentials, the entropy density even exhibits an enhancement in the presence of a vortex at low temperatures, which is observed neither for the pressure nor the entropy densities, however.\par
A numerical evaluation of the entropy per particle as well as the angular momentum per particle over a temperature range of six $e$-folds is presented in the upper plots of Figs. (\ref{entropyperparticle}) and (\ref{angularmomentumperparticle}), respectively. We furthermore show, in line with the previous plots of the pressure, energy density and entropy density, the relative difference of particle number densities and angular momentum densities for the two topologically distinct cases $n_1=0$ and $n_1=1$ over a range of five e-folds in temperature. The latter (lower) plots exhibit the same patterns as outlined before in the cases of the pressure and energy density, so we proceed to discuss the upper plots.\par
At high temperatures the entropy per particle $s/n$ converges to a finite value which is larger for lower angular velocity but not dependent on the chemical potential. Towards lower temperatures the entropy per particle develops a strong dependence on the chemical potential. For very small particle density ($\mu\lesssim 1$) it increases for decreasing temperature, while it increases for very large particle number density ($\mu \gg 1$). It is monotonously decreasing with chemical potential. In constrast the entropy per particle depends only weakly on the angular velocity without any specific properties regarding monotony. 
At high temperatures the angular momentum per particle $j/n$ converges to a finite value as well. The asymptotic region implied by the convergence is reached at smaller temperatures for large angular velocity. For fixed temperatures the angular momentum per particle increases both for increasing angular velocity and chemical potential. Towards low temperatures a strong dependence of the angular momentum per particle both on chemical potential and angular velocity arises. It decreases monotonously with temperature.\par
The number of degrees of freedom consideration mentioned further above is mostly irrelevant for the discussed plots, as the considered ratios of thermodynamic quantities are independent of the number of degrees of freedom. 

\item[2)] Disclinations: 
We again assume stationarity with vanishing translational velocity $v^{\mu}$ and acceleration $a^{\mu}$. We would like to consider once more the situation with nonzero rigid rotation of macroscopic motion. A radial (tangential) disclination in $^3$He-A is defined by
\begin{align}
\bold{m}=\hat{\phi},\,\,\,\, \bold{n}=\hat{z},\,\,\,\, \bold{l}=\hat{\rho},\,\,\,\, \bold{d}=\pm \hat{\rho}\,\,\,\,(\bold{m}=-\hat{\rho},\,\,\,\, \bold{n}=\hat{z},\,\,\,\, \bold{l}=\hat{\phi},\,\,\,\, \bold{d}=\pm \hat{\phi}).
\end{align}
The vector $\bold{d}$ is again dipole locked with $\bold{l}$. We find $\bold{E}=\bold{B}=2M_3\bold{v}_s=0$ ($\bold{E}=2M_3\bold{v}_s=0$, $\bold{B}=-\frac{k_F}{\rho}\hat{z}$) and 
\begin{align}
T^1_{\,\,\,\,12}=-\frac{n_1}{\rho}sin(n_1\phi ),\,\,\,\, T^2_{\,\,\,\,12}=\frac{n_1}{\rho}cos(n_1\phi ),\,\,\,\, T^{\mu}_{\,\,\,\,\nu\rho}=0\,\,\text{otherwise}\,\,\Rightarrow a=\nabla_{\mu}K^{\mu}=0.
\end{align}
Now $G_{\phi}$ is not naturally vanishing for disclinations which implies for the Lorentz indexed vector fields
\begin{align}
0\neq G_1=\frac{1}{8\rho}\overline{\Psi}\gamma^0\gamma^3\Psi -i\frac{\pi}{8\rho}Q_A^1\,\,\,\, (0\neq G_3=-\frac{1}{8\rho}\overline{\Psi}\gamma^0\gamma^1\Psi ) 
\end{align}
and consequently $\beta^1=0$ (with Lorentz index) ($\beta^3=0$ (with Lorentz index)), or more succinctly $\beta^{\phi}=0$, and consequently $\omega_{12}=0$ (with spacetime Cartesian coordinate indices). A direct further implication of $\beta^1=0$ ($\beta^3=0$) and Eq. (\ref{equationofomega}) is $\Omega_{\mu\nu}=\omega_{\mu\nu}$. Furthermore $b_{\phi}^s=\frac{1}{2}$ and $\omega_{\phi 12}=-\pi$ which implies $F^b_{\mu\nu}=0$ such that only the final two terms in Eq. (\ref{geononconservationofemt1}) contribute to $G_1$ ($\nabla_{\mu}e_{1,2}^{\mu}=0,\,\,\nabla_{\mu}e_3^{\mu}\neq 0$) ($G_3$ ($\nabla_{\mu}e_{2,3}^{\mu}=0,\,\,\nabla_{\mu}e_1^{\mu}\neq 0$)). Moreover $\nabla_{\mu}j_A^{1\mu}=0$ and
\begin{align}
P_{13}=\frac{i}{4}Re(\omega_{\phi 12})Q^1_A,\,\,\,\,Q^1_A=-\frac{1}{2}\overline{\Psi}\gamma^5\sigma^1\Psi ,\,\,\,\,P_{23}=0\,\,\,\,(P_{13}=P_{23}=0)
\end{align}
which in the presence of excitations with charge $Q_A^1$ means $P_{13}\neq 0$ and then $\Omega_{13}=\omega_{13}=0$ (with Lorentz indices). The previous constraint is not posing any further restrictions and may even be eliminated for $j_A^{1\phi}=0$ for $\zeta_A^1$ fulfilling Eq. (\ref{relevantforvortices}). In the presence of both $(s=+)$- and $(s=-)$-components Eq. (\ref{omegarestriction}) for $\Omega^{ab}$ together with $\beta^1=0$ can be fulfilled only for orientations of the rigid rotation axis orthogonal to the $z$-axis, as $\omega_{13},\omega_{23}\neq 0$ (with spacetime indices) is possible. The vierbein constraint of Eq. (\ref{anomalyconstraint}) is trivial here, as is the case for pure mass vortices. Thus global thermodynamic equilibrium is possible in this case for all orientations of the rotation axis orthogonal to the $z$-axis.

\item[3)] Fractional vortices:
We consider fractional (or spin mass) vortices corresponding to the situation outlined above in the discussion of pure mass vortices, but now with $2n_1\in\mathbb{Z}$ such that $n_1\not\in\mathbb{Z}$. In order for the order parameter of the $^3$He-A-phase to be single-valued, $\bold{d}$ can no longer be dipole locked with $\bold{l}$ but instead fulfills 
\begin{align}
\bold{d}=cos(n_1\phi)\hat{x}+sin(n_1\phi )\hat{y}.
\end{align}
Therefore $G_2$ (or equivalently $G_{\phi}$) is naturally nonzero 
\begin{align}
G_2=\frac{1}{8\rho}\overline{\Psi}\gamma^0\gamma^1\Psi
\end{align}
which implies that correspondingly $\beta^2$ (or equivalently $\beta^{\phi}$) has to be zero and comprises exactly the $\omega_{12}$ component (in spacetime Cartesian coordinate indices). We again find $\Omega_{\mu\nu}=\omega_{\mu\nu}$ from Eq. (\ref{equationofomega}). Furthermore $b_{\phi}^s=\frac{1}{2}n_1$ and $\omega_{\phi 12}=-\pi n_1$ which implies $F^b_{\mu\nu}=0$. Both $P_{13}$ and $P_{23}$ are generically nonzero 
\begin{align}
P_{13}=\frac{i}{4}Re(\omega_{\phi 12})sin((n_1-1)\phi )Q^1_A,\,\,\,\,P_{23}=\frac{i}{4}Re(\omega_{\phi 12})cos((n_1-1)\phi )Q^1_A
\end{align}
in the presence of excitations with charge $Q^1_A\neq 0$ thereby forcing $\Omega_{13}$ and $\Omega_{23}$ ($\omega_{13}$ and $\omega_{23}$) to vanish. We find $\nabla_{\mu}j_A^{1\mu}=0$. If we assume that $j_A^{1\phi}=0$ (which means that there is no macroscopic motion of this charge around the vortex axis), then $\zeta_A^1$ fulfilling Eq. (\ref{relevantforvortices}) lifts the constraints on $\Omega_{13}$ and $\Omega_{23}$. We come to the conclusion that fractional vortices in the presence of both the $(s=+)$- and the $(s=-)$-components may allow for a global thermodynamic equilibrium solution for orientations of the rotation axis orthogonal to the vortex axis either for $Q_A^1=0$ or $j_A^{1\phi}=0$ and $\zeta_A^1$ fulfilling Eq.(\ref{relevantforvortices}).
\end{itemize} 

\section{Discussion}
\label{SectDisc}
\mz{This work comprises two salient achievements. On the one hand we performed a reformulation of the emergent relativistic theory of Weyl fermions in a superfluid background component in terms of a uniform vierbein and a spin connection gauge field which combines the entire emergent spin dynamics. We combined the two Weyl fermions at the two momentum space Fermi points into a single Dirac fermion. Within superfluid $^3$He - A these fermions form a doublet due to a discrete $\mathbb{Z}_2$-symmetry. This doublet may be thought of as a spin up and a spin down component of the superfluid order parameter spin structure (supplemented by an orbital sign flip). The matrix-valued vierbein formulation suggested that each component of this doublet is subject to a different vierbein (the spin up and down states where distinguished by a sign $\pm$)
\begin{align}
\nonumber \bold{e}_a^{\mu}\Psi^{\pm}=(e^{\pm})_a^{\mu}\Psi^{\pm}.
\end{align}
The spin structure implicit within the matrix-valued vierbein is distinct from the Dirac spin structure in this formulation. The ordinary spin connection $\omega^{\mu}_{ab}$ was found to vanish identically (or more precisely to be gauge equivalent to zero). We performed a reparametrization of the theory which we regard as more natural for the two just mentioned reasons. Both components of the Dirac fermion doublet share the same vierbein. They answer universally to the ''emergent gravity'' implied by the superfluid component. In order to achieve this we ''rotated'' one of the two components (for definiteness we chose the lower ''$-$''-component) so as to obtain two “spin up” components coupled to the same vierbein field $e_a^{\mu}=(e^+)_a^{\mu}$. This operation implied a reformulation in the spin one gauge sector
\begin{align}
\nonumber \nabla_{\mu}=\partial_{\mu}-i\mathcal{A}_{\mu}\gamma^5\to D_{\mu}=\nabla_{\mu}-i\mathcal{B}_{\mu}.
\end{align}
The original unit normalized spin vector $\bold{d}$ becomes the source of two Abelian Berry connection gauge fields together with a non-Abelian spin connection gauge field which we called $\omega_{\mu 12}$. Together these gauge fields were labelled by $\mathcal{B}_{\mu}$, which we also name spin connection gauge field, acting in the eight dimensional vector bundle space of the ''spin up'' Dirac spinor doublet. It governs the entire spin dynamics associated with the fermionic normal component. Notice that not all of the gauge fields within $\mathcal{B}_{\mu}$ are independent of each other.}\par
\mz{On the other hand we analyzed our reformulated theory of superfluid $^3$He - A under simultaneous motion of the superfluid component and the macroscopic motion of the normal Weyl fermion component within the Zubarev statistical operator method. We started from the conventional Zubarev statistical operator defined for an arbitrary foliation of spacetime to spacelike surfaces $\Sigma (\sigma )$ with foliation (or time) parameter $\sigma$. We restricted ourselves to flat Minkowski spacetime. The macroscopic four velocity $u^{\mu}(x)$ appears in this approach naturally, and it appears that the following types of macroscopic motion are possible in thermodynamic equilibrium \cite{Zubarev1979}} under the simplified assumption of Poincar\'e symmetry 

\begin{enumerate}
	\item 
	
	Motion with constant four-velocity $u^{\mu}(x) = {\rm const}$. Correspondingly, in this case temperature and chemical potential are constant as functions of time.
	
	\item
	
	Rigid rotation with constant angular velocity $\omega$. In this case temperature becomes a function of spatial coordinates. The whole theory becomes ill-defined at the distances larger than $1/\omega$ from the rotation axis. This means that we can use the Zubarev statistical operator for the case when $R \omega < 1$, where $R$ is the size (transverse extension) of the considered system. This admits, in particular, the possibility of rotation with relativistic velocities. The inequality is valid by causality. Any physical system will desintegrate (or be ripped apart) before reaching the limit implied by causality.
	
	If rotation is along the $z$-axis, then we have 
	$$
	u^{\mu}(x) = \frac{1}{\sqrt{1-\omega^2(x^2 + y^2)}}(1,-y\omega,x\omega,0)
	$$
	and
	$$
	\beta(x) = \beta_0 \sqrt{1-\omega^2(x^2+y^2)}, \quad b = (\beta_0,0,0,0)
	$$ 
	with constant $\beta_0$ of dimension of inverse temperature.
	
	\item 
	
	Accelerated motion with constant acceleration vector $a$. The acceleration $a$ appears as the thermodynamically conjugated quantity to the boost operator. The interpretation of the theory in terms of the four velocity $u^{\mu}(x)$ becomes ill-defined at times $t>1/a$. This situation is avoided in a physical system, as maintaining the finite acceleration $a$ for times larger than $1/a$ implies an injection of an infinite amount of energy. However, the spatial size of the corresponding system is not limited. 
	
	In the case when acceleration is along the $x$-axis, we have: 
	$$
	u^{\mu}(x) = \frac{1}{\sqrt{(1+ax)^2 - a^2 t^2}}(1+ax,at,0,0)
	$$
	and
	$$
	\beta(x) = \beta_0 \sqrt{(1+ax)^2 -a^2 t^2}
	$$

	\item 
	
	The combination of the three types of motion explained above is also admitted for thermal equilibrium. 
	
\end{enumerate}

\mz{The statistical partition function of an equilibrium system of fermions interacting with the non-Abelian gauge field may be denoted as
$Z[n^{\mu}(x),u^{\mu}(x),\beta(x),\mu_i(x)]$. It is a function of velocity of the macroscopic motion $u^{\mu}(x)$, temperature $\beta(x)$ depending on spatial coordinates and varying chemical potentials $\mu_i(x)$ corresponding to the conserved charges of the system. We considered the case of constant vector $n$ orthogonal to the hypersurface $\Sigma$ such thatwe may assume $n^{\mu} = (1,0,0,0)$. We derived a representation of this partition function in the form of the Euclidean functional integral over fermionic fields, the gauge field (taken in temporal gauge), and the corresponding conjugate momentum. We represent it as an analytical continuation of the partition function for the effective quantum field theory in Minkowski spacetime. The latter effective theory seems to us especially instructive. We represent it via the relation}\newline   
\mz{$Z[n^{\mu}(x),u^{\mu}(x),\beta(x),\mu_i(x)] = {\cal Z}[n^{\mu}(x),u^{\mu}(x),\beta(x),\mu_i(x),-i]$ where the Minkowski space partition function depends on the parameter $h$. It is to be taken equal to $-i$ in order to arrive at the original statistical partition function
\begin{equation}
{\cal 	Z}[n^{\mu}(x),u^{\mu}(x),\beta(x),\mu_i(x),h]=\int D\overline{\psi} D\psi DA_\mu   \, e^{i\int  d^4 x \, \mathcal{L}(\overline{\psi},\psi ,A_{\mu})}.\label{ZZZ}
\end{equation} 
We denote $x = (t,\bold{x})$. }

\mz{The integral in the exponent of Eq. (\ref{ZZZ}) is to be taken along a piece of spacetime of extent ${\mathfrak B}(\bold{x}) h $ that starts from the given hyperplane $\Sigma$ corresponding to $t = t_0$ (usually we put $t_0=0$). Here $\mathfrak B$ is an arbitrarily chosen function of spatial coordinates (physical observables should not depend on this choice). The effective Lagrangian is not relativistically invariant. It depends on the macroscopic four velocity $u^{\mu}(t_0,\bold{x})$ and function ${\mathfrak B}(\bold{x})$ through the four vector field $\mathfrak U^{\mu}$ given by ${\mathfrak U}(\bold{x}) = (\beta(t_0,\bold{x})/{\mathfrak B}(\bold{x})) u(t_0, \bold{x})$. We have two convenient choices of the function $\mathfrak B$. On the one hand ${\mathfrak B}(\bold{x}) =\beta(t_0,\bold{x})$ such that ${\mathfrak U}^\mu(\bold{x}) =  u^\mu(t_0, \bold{x})$. The vector field $\mathfrak U^{\mu}$ may be interpreted as the four velocity distribution at the initial moment, while the function $\mathfrak B$ is the inverse temperature depending on spatial coordinates. On the other hand ${\mathfrak B}(\bold{x}) =\beta(t_0,\bold{x}) u^0(t_0,\bold{x})$. In this case ${\mathfrak U}^\mu(\bold{x}) = u^\mu(t_0, \bold{x})/u^0(t_0, \bold{x})$. The vector field $\mathfrak U^{\mu}$ can not be interpreted as a four velocity of macroscopic motion ($\mathfrak B$, though, may still be interpreted as inverse temperature). For the second choice of $\mathfrak U^{\mu}$ the effective Lagrangian is simplified. The macroscopic motions allowed in global thermodynamic equilibrium may be written in terms of $\mathfrak{U}^{\mu}$ for this second choice as follows.} 

\begin{enumerate}
	\item 
	Uniform linear motion along the $x$-direction with constant four velocity  
	$$
	u^{\mu}(x) = \gamma (v)(1,v,0,0).
	$$
	In this case $\mathfrak{U}^{\mu}(x)$ is constant as well
	$$
	\mathfrak{U}^{\mu}(x)=(1,v,0,0).
	$$
	\item	
	Rigid rotation with constant angular velocity $\omega$ around the $z$-axis
	$$
	u^{\mu}(x) = \frac{1}{\sqrt{1-\omega^2(x^2 + y^2)}}(1,-y\omega,x\omega,0).
	$$
	Then we have
	$$
	{\mathfrak U}^{\mu}(x) = (1,-y\omega,x\omega,0).
	$$	
	\item 
	The initially accelerated motion with acceleration $a$ along the $x$-axis
	$$
	u^{\mu}(x) = \frac{1}{\sqrt{(1+ax)^2 - a^2 t_0^2}}(1+ax,at_0,0,0).
	$$
	For the choice $t_0=0$ we have 
	$$
	{\mathfrak U}^{\mu}(x) = (1,0,0,0).
	$$
One can see that in this case (especially for accelerated motion) the effective Lagrangian is especially simple. For accelerated motion it is reduced to the Lagrangian of the system remaining at rest. The only effect of acceleration is manifest through the temperature depending on spatial coordinates.
\end{enumerate}

The considerations up to now apply to a superfluid component which is homogeneous. In this case the canonical energy momentum tensor of Eq. (\ref{emt}) and the canonical Lorentz transformation tensor of Eq. (\ref{ltt}) are conserved. This case is basically covered in the introduction of section \ref{zubarevformalism}. This is the case in which only the normal component is moving, while the superfluid “vacuum” component is at rest. The novelty in our work is the consideration of macroscopic motion of the normal component while the superfluid component is moving in simultaneity. This more involved case was studied in the main body of section \ref{zubarevformalism} and analyzed in the presence of vortices in section \ref{thermoequisolutions}. These analyses are specific to $^3$He - A for whose details we specifically refer to these sections. Section \ref{pathintegralformulation} aimed at deriving a macroscopic motion Lagrangian which comprises both motion of the superfluid and normal component Wick rotated to Minkowski spacetime.

\mz{The Lagrangian comprising motion of the superfluid component as well as the normal component has been found to be
\begin{align}
\nonumber \mathcal{L}(\overline{\Psi},\Psi )=&\frac{1}{4}\overline{\Psi}\gamma^0ie_b^{\mu}\gamma^bD_{\mu}\Psi -\frac{1}{4}\overline{\Psi}\gamma^0\overset{\leftarrow}{D}_{\mu}ie_b^{\mu}\gamma^b\Psi -\mathfrak{U}^i\frac{1}{4}[\overline{\Psi}iD_i\Psi -\overline{\Psi}i\overset{\leftarrow}{D}_i\Psi ]+\frac{1}{8}\mathfrak{U}^{i}n^{\mu}\epsilon_{\mu\nu i\sigma}\nabla^{\nu}\overline{\Psi}\gamma^0\gamma^5e_c^{\sigma}\gamma^c\Psi \\
\nonumber &+\frac{1}{8}\mathfrak{U}^{\lambda}n^{\mu}\epsilon_{\mu\nu\rho\sigma}e^{a\nu}(\nabla_{\lambda}e_a^{\rho})\overline{\Psi}\gamma^0\gamma^5e_c^{\sigma}\gamma^c\Psi +\frac{\upmu_V}{2}n_{\mu}\overline{\Psi}\gamma^0\gamma^{\mu}\Psi +\frac{\upmu_A}{2}n_{\mu}\overline{\Psi}\gamma^0\gamma^5\gamma^{\mu}\Psi .
\end{align}}

\mz{The allowed types of macroscopic motion (admitted for global thermodynamic equilibrium) may be considered using the Zubarev statistical operator for any substance described by relativistic quantum field theory. We reached the final Lagrangian description for $^3$He - A via the functional integral technique and considered the simplest possible foliation of spacetime, in which hypersurfaces $\Sigma (\sigma)$ for any value of $\sigma$ are the hyperplanes $t = const$. It would be interesting to consider the extension of the presented formalism to an arbitrary form of $\Sigma (\sigma )$.}

\section{Conclusions}
\label{conclusions}

{We constructed the theory of the normal component of $^3$He - A in the presence of both the moving superfluid component and macroscopic motion of the normal component itself in the regime of emergent relativistic invariance. \mz{The dynamics of the superfluid has not been considered. Instead we treated the vierbein as an external background coupled to the fermionic normal component neglecting the backreaction of the normal component on the superfluid component.}}

{In this theory the moving superfluid component manifests itself via a space and time dependent (matrix - valued) emergent vierbein and space and time dependent emergent axial gauge field (which is not independent of the vierbein, though). Alternatively, we may represent the theory in terms of the ordinary real - valued vierbein, implying in turn the appearance of a nontrivial spin connection and two extra nontrivial emergent vector gauge fields. \mz{We present both formulations but advertise the latter formulation which is not common. The normal fermionic component is described by a Dirac spinor doublet in both cases. The doublet degeneracy originated in the additional internal spin space present within superfluid $^3$He - A. Though both formulations imply a slightly different interpretation of the doublet degeneracy expressed in the different emergent gauge and gravitational couplings.}}

{In order to take into account the macroscopic motion of the normal component we apply the Zubarev statistical operator approach within a path integral formulation. The effective action for the fermions remains in the form of the action of Dirac fermions in the presence of background fields. The motion of the superfluid component is represented by a vierbein, an axial gauge field and a spin connection gauge field, while the macroscopic motion of the normal component is represented by the frigidity vector field. Finite particle densities further imply nonzero chemical potentials. In the presence of thermodynamic equilibrium the dynamics of the normal component is severely restricted, admiting only a small number of types of macroscopic motion (linear uniform, rotated and uniformly accelerated motion). Motion of the superfluid component of $^3$He - A differs, in general, from the motion of the normal component.}

\mzc{There exist alternative ways to describe the macroscopic motion of $^3$He. In particular, it is possible to take into account this macroscopic motion at the level of the original theory of Sect. \ref{SectIIA}. However, we choose another way. We consider the superfluid $ ^3$He - A as consisting of the two components: the superfluid component, and the normal component. Technically this separation appears as the separation between the dynamics of the order parameter field (the auxiliary field, which appears during the Hubbard – Stratonovich transformation) and the normal component (the fermionic quasiparticles that are excitations above the superfluid component). This separation is important for the interpretation of the system as simulating high energy physics in the laboratory. Namely, the superfluid component simulates vacuum (which by itself may be in motion with superfluid velocity), while the normal component simulates matter. For more details about this separation see in \cite{Volovik2003}. Therefore, we consider “matter” simulated by the normal component in the presence of macroscopic motion on the background “vacuum” moving with superfluid velocity. Motion of the “vacuum” is considered in our paper as given, while the developed theory is responsible for the description of the normal component in the presence of macroscopic motion. 	
The advantage of the formalism based on the Zubarev statistical operator is that it allows us to identify rather easily the macroscopic motion of the system with emergent relativistic invariance, and derive the effective action for the equilibrium system in the presence of such a motion. In our previous publication \cite{AZZ2023} we applied this method to quark – gluon plasma. Now we extend it to $^3$He - A. The emergent relativistic invariance is an important feature of the latter, which allows us to use the Zubarev statistical operator.} 

{We analyse the interrelation of the two types of motion by providing three explicit examples assuming global thermodynamic equilibrium. We demonstrate the developed theory by applying it to the description of the normal component of $^3$He - A in global thermodynamic equilibrium in the presence of the pure mass vortex, a disclination and fractional vortices. We calculate several thermodynamic quantities of this system in the presence of macroscopic rotation around the axis of the integer mass vortex. \mz{We found, in particular, that in the presence of pure mass vortices the normal component's rotation axis is not necessarily aligned with that of the vortex. Angular velocities of the two rotations may be different. We regard this outcome as a shortcoming of our treatment and expect perfect aligned as a mandatory requirement as soon as the vierbein is allowed to become dynamical. Similar findings apply to the disclinations and the fractional vortices.}}
\mzc{We do not consider in the present paper the dynamics of vortices. However, the obtained results for the thermodynamical quantities of the normal component in the presence of vortices may be used as a building block for the description of this dynamics. Namely, these quantities may describe the influence of the normal component on the dynamics of vortices. However, we expect that this effect is subleading and should be taken into account as a correction to the main sources of vortex dynamics (see, e. g., \cite{Volovik2003}.)}

\mzc{As we already mentioned above, we do not consider the dynamics of the order parameter field (the dynamics of superlfuid component, i.e. the dynamics of “vacuum”). The motion of the "vacuum" is assumed to be known. And it gives the background for the motion of the normal component considered in the present paper. 	
	There are several reasons why we do not consider the dynamics of the “vacuum” and focus on the dynamics of “matter”. First of all, the dynamics of the “vacuum” unfortunately does not exhibit emergent relativistic invariance. Therefore, it is not of interested for the simulation of relativistic physics in the laboratory. Second, this rather complicated description has been given in sufficient details in several textbooks including the mentioned above \cite{Volovik2003} as well as the older book by the same author \cite{volovik1993exotic}, and the seminal textbook \cite{VollhardtWolfle1990} by Dieter Vollhardt, Peter Wolfle. 
	The application of the formalism developed in our present paper to the description of the normal component in the presence of vortices allows us to represent the framework in which our technique may, in principle, be verified experimentally. 	
}

{\mz{We identify three future research directions of this work. Firstly, we may indeed generalize our treatment to a dynamical superfluid component. This allows for a full understanding of the superfluid $^3$He - A phase in global thermodynamic equilibrium. This has not been considered within this work, as we primarily intended to apply our path integral formalism in line with our previous findings within quantum chromodynamics \cite{AZZ2023}. Secondly, we may provide a bridge to the known literature on $^3$He - A. Our formalism allows for the inclusion of a host of transport properties within superfluid $^3$He - A simultaneously, while at the same time being formulated concisely in terms of our final results. These are manifest in our macroscopic motion Lagrangian in Eq. (\ref{completelagrangian}). A recent study of effective field theory dynamics of the superfluid $^3$He $A$- and $B$-phases identified the dissipationless Hall viscosity \cite{fujii2018low,furusawa2021hall} which might be worthwhile to discuss within our formalism. The relation of the Hall viscosity to the angular momentum density of a substance makes it worthwhile to consider the role of the Hall viscosity in the presence of vortices in $^3$He - A. This justifies our focus on rotation and vortices in the presence of normal component macroscopic motion. Thirdly, we may go beyond the consideration of global thermodynamic equilibrium and employ the Keldysh path integral technique to describe the full dynamics of the superlfuid $A$-phase of $^3$He out of equilibrium.}}

{The authors are grateful to G.E. Volovik and L.Melnikovsky for useful comments on the content of the paper.}

\appendix

\section{Equations of motion and canonical formalism for Nambu-Gorkov spinors without doubling of degrees of freedom}\label{app1}

Consider again the constraints of Eq. (\ref{majoranaconstraint})
\begin{align}
\psi_R(p)=i\tau^1\sigma^2\psi^{\ast}_L(-p),\,\,\,\,\psi_L(p)=-i\tau^1\sigma^2\psi^{\ast}_R(-p)
\end{align}
for the spinors $\psi_{L/R}$ together with their relation to the spinors $\Psi^{\pm}_{L/R}$ which, in components, may be written as 
\begin{equation}
	\Psi^{\pm}_{L/R, \alpha} \eta^\pm_i = \frac{\delta_{ij}\pm({\bf d} \boldsymbol{\sigma}_{ij})}{2} \psi_{L/R,\alpha j}  
\end{equation}
where the index $i$ corresponds to spin, while $\alpha$ corresponds to Nambu spin. In terms of $\Psi^{\pm}$ defined in coordinate space the constraint reads 
\begin{align}
\Psi^{s}_{R, \alpha}(x) =(\tau^1)_{\alpha}^{\beta}(H^{su})^{\ast}(x)(\Psi^u)^{\ast}_{L,\beta}(x), \quad \Psi^{s}_{L, \alpha}(x) =-(\tau^1)_{\alpha }^{\beta}(H^{su})^{\ast}(x)(\Psi^u)^{\ast}_{R,\beta}(x)
\label{majoranaconstraint2}
\end{align}
with
\begin{equation}
H^{su} = {\eta}^s_i\epsilon_{ij}{\eta}^u_{j} = \frac{i}{\sqrt{d_1^2+d_2^2}} (d_1 - i d_2) (\sigma^2)^{su}. 
\end{equation}
The $(\pm )$-components finally need to be subjected to the phase rotation or field redefinition of Eq. (\ref{phaserotation}). Then the $8$ - component spinor $\Psi$ obeys
\begin{align}
\overline{\Psi} = \Psi^T \hat{U} \hat{d}^{\ast},\,\,\,\,\hat{U}  =\gamma^2\otimes\sigma^1,\,\,\,\,\hat{d}=\frac{d_1 +i d_2}{\sqrt{d_1^2+d_2^2}}.
\label{majoranaconstraint2}
\end{align}
The explicit constraints for the chiral components read
\begin{align}
\overline{\Psi}_L^{\pm}=\hat{d}^{\ast}\tau^2\Psi_R^{\mp},\,\,\,\, \overline{\Psi}_R^{\pm}=\hat{d}^{\ast}\overline{\tau}^2\Psi_L^{\mp}.
\label{constraintchiralcomponents}
\end{align}
The action may then be written as
\begin{align}
S_{eff}=\frac{1}{4}\int d^4xe[{\Psi}^T \hat{U}\hat{d}^{\ast} i\gamma^0\gamma^be_b^{\mu}D_{\mu}\Psi -[{\Psi}^T\hat{U}\hat{d}^{\ast}\gamma^0\overset{\leftarrow}{D}_{\mu}]i\gamma^be_b^{\mu}\Psi ].
\end{align}
In the following we will derive the equations of motion as well as the canonical quantization procedure for the above formulation of the low energy theory of $^3$He-A and employ the relations
\begin{align}
(\gamma^{0/2})^T=\gamma^{0/2},\,\,(\gamma^{1/3})^T=-\gamma^{1/3},\,\Rightarrow [\gamma^1,\gamma^2]^T=[\gamma^1,\gamma^2],\,\, (\sigma^{1/3})^T=\sigma^{1/3},\,\, (\sigma^2)^T=-\sigma^2,\,\Rightarrow \hat{U}^T=\hat{U}.
\label{transpositionrules}
\end{align}

\subsection{Equations of motion}

Stationarity of the action with respect to the Grassmann-valued spinor $\Psi$ implies 
\begin{align} 
\nonumber 0=\frac{\delta S}{\delta \Psi (x)}=&(\frac{1}{4}\Psi^T\hat{U}\hat{d}^{\ast}\gamma^0\overset{\leftarrow}{D}_{\mu}i\gamma^be_b^{\mu})^T+(\frac{1}{4}\Psi^T\hat{U}\hat{d}^{\ast}i\gamma^0\gamma^be_b^{\mu}\overset{\leftarrow}{D}_{\mu})^T \\
\nonumber &+\frac{1}{4}\hat{U}\hat{d}^{\ast}i\gamma^0\gamma^be_b^{\mu}D_{\mu}\Psi +\frac{1}{4}\hat{U}\hat{d}^{\ast}\gamma^0D_{\mu}i\gamma^be_b^{\mu}\Psi\\
\nonumber =&\frac{1}{4}i(\gamma^b)^Te_b^{\mu}(\nabla_{\mu}+i\mathcal{B}_{\mu}^T)(\gamma^0)^T\hat{U}^T\hat{d}^{\ast}\Psi +\frac{1}{4}(\nabla_{\mu}+i\mathcal{B}_{\mu}^T)i(\gamma^0\gamma^b)^Te_b^{\mu}\hat{U}^T\hat{d}^{\ast}\Psi \\
&+\frac{1}{4}\hat{U}\hat{d}^{\ast}i\gamma^0\gamma^be_b^{\mu}(\nabla_{\mu}-i\mathcal{B}_{\mu})\Psi +\frac{1}{4}\hat{U}\hat{d}^{\ast}\gamma^0(\nabla_{\mu}-i\mathcal{B}_{\mu})i\gamma^be_b^{\mu}\Psi .
\end{align}
Employing the relations in Eq. (\ref{transpositionrules}) together with the commutation relations of Dirac- and Pauli-matrices leads to the final form
\begin{align}
\nonumber 0=&i\gamma^0\gamma^be_b^{\mu}\nabla_{\mu}\Psi +\frac{1}{2}i\gamma^0\gamma^be_b^{\mu}(\hat{d}\nabla_{\mu}\hat{d}^{\ast})\Psi +\frac{1}{2}i\gamma^0\gamma^b(\nabla_{\mu}e_b^{\mu})\Psi +\frac{1}{2}\gamma^0\{\tilde{B}_{\mu},\gamma^b\} e_b^{\mu}\Psi\\
\nonumber =&i\gamma^0\gamma^be_b^{\mu}\nabla_{\mu}\Psi +\gamma^0\gamma^be_b^{\mu}(\frac{b_{\mu}^++b_{\mu}^-}{2})\Psi +\frac{1}{2}i\gamma^0\gamma^b(\nabla_{\mu}e_b^{\mu})\Psi +\frac{1}{2}\gamma^0\{\tilde{B}_{\mu},\gamma^b\} e_b^{\mu}\Psi\\
=&i\gamma^0\gamma^be_b^{\mu}\nabla_{\mu}\Psi +\frac{1}{2}i\gamma^0\gamma^b(\nabla_{\mu}e_b^{\mu})\Psi +\frac{1}{2}\gamma^0\{B_{\mu},\gamma^b\} e_b^{\mu}\Psi
\label{eomwithconstraint}
\end{align}
with
\begin{align}
\tilde{B}_{\mu}=\frac{1}{8}Re(\omega_{\mu 12})[\gamma^1\gamma^2]\sigma^1-\frac{1}{8}Im(\omega_{\mu 12})[\gamma^1,\gamma^2]\sigma^2+\frac{1}{2}(b_{\mu}^+-b_{\mu}^-)\mathbb{1}_D\sigma^3.
\end{align}
Due to the reality constraint $\overline{\Psi} = \Psi^T \hat{U} \hat{d}^*$ the phase term in the first line of Eq. (\ref{eomwithconstraint}) can not be removed by a field redefinition of $\Psi$. \par
We will subsequently proof explicitly the equivalence of Eqs. (\ref{eomwithoutconstraint2}) and (\ref{eomwithoutconstraint}) as well as their reduction to Eq. (\ref{eomwithconstraint}) after enforcement of the reality constraint. To achieve this we make use of the identity
\begin{align}
i\hat{d}\nabla_{\mu}\hat{d}^{\ast}=b^+_{\mu}+b^-_{\mu}
\label{dderivative}
\end{align}
which may be verified with the help of Eqs. (\ref{bplus}) and (\ref{bminus}). We may then manipulate Eq. (\ref{eomwithoutconstraint2}) as follows
\begin{align}
0=\nonumber&-2\overline{\Psi}\gamma^0\overset{\leftarrow}{\nabla}_{\mu}i\gamma^be_b^{\mu}+\overline{\Psi}\gamma^0\gamma^be_b^{\mu}\mathcal{B}_{\mu}+\overline{\Psi}\gamma^0\mathcal{B}_{\mu}\gamma^be_b^{\mu}-\overline{\Psi}i\gamma^0\gamma^b(\nabla_{\mu}e_b^{\mu})\\
=\nonumber &-2\Psi^T\hat{U}\hat{d}^{\ast}\gamma^0\overset{\leftarrow}{\nabla}_{\mu}i\gamma^be_b^{\mu}+\Psi^T\hat{U}\hat{d}^{\ast}\gamma^0\gamma^be_b^{\mu}\mathcal{B}_{\mu}+\Psi^T\hat{U}\hat{d}^{\ast}\gamma^0\mathcal{B}_{\mu}\gamma^be_b^{\mu}-\Psi^T\hat{U}\hat{d}^{\ast}i\gamma^0\gamma^b(\nabla_{\mu}e_b^{\mu})\\
=\nonumber &-2\Psi^T\overset{\leftarrow}{\nabla}_{\mu}\hat{U}\hat{d}^{\ast}i\gamma^0\gamma^be_b^{\mu}-2\Psi^T\hat{U}(\hat{d}^{\ast}\overset{\leftarrow}{\nabla}_{\mu})i\gamma^0\gamma^be_b^{\mu}+\Psi^T\hat{U}\hat{d}^{\ast}\gamma^0\gamma^be_b^{\mu}\mathcal{B}_{\mu}+\Psi^T\hat{U}\hat{d}^{\ast}\gamma^0\mathcal{B}_{\mu}\gamma^be_b^{\mu}-\Psi^T\hat{U}\hat{d}^{\ast}i\gamma^0\gamma^b(\nabla_{\mu}e_b^{\mu})\\
&\nonumber \Leftrightarrow\\
0=\nonumber &-2ie_b^{\mu}(\gamma^b)^T(\gamma^0)^T\hat{U}^T(\hat{d}\nabla_{\mu}\hat{d}^{\ast})\Psi -2ie_b^{\mu}(\gamma^b)^T(\gamma^0)^T\hat{U}^T\hat{d}^{\ast}\nabla_{\mu}\Psi \\
\nonumber &+\mathcal{B}_{\mu}^Te_b^{\mu}(\gamma^b)^T(\gamma^0)^T\hat{U}^T\hat{d}^{\ast}\Psi +e_b^{\mu}(\gamma^b)^T(\gamma^0)^T\mathcal{B}_{\mu}^T\hat{U}^T\hat{d}^{\ast}\Psi -i(\nabla_{\mu}e_b^{\mu})(\gamma^b)^T(\gamma^0)^T\hat{U}^T\hat{d}^{\ast}\Psi\\
=\nonumber &\hat{U}(-2i\gamma^0\gamma^be_b^{\mu}\nabla_{\mu}\Psi -2\gamma^0\gamma^be_b^{\mu}(b_{\mu}^++b_{\mu}^-)\Psi +\hat{\mathcal{B}}_{\mu}\gamma^0\gamma^be_b^{\mu}\Psi +\gamma^0\gamma^be_b^{\mu}\hat{\mathcal{B}}_{\mu}\Psi -i(\nabla_{\mu}e_b^{\mu})\gamma^0\gamma^b\Psi )\\
=\nonumber &\hat{U}(-2i\gamma^0\gamma^be_b^{\mu}\nabla_{\mu}\Psi +\mathcal{B}_{\mu}\gamma^0\gamma^be_b^{\mu}\Psi +\gamma^0\gamma^be_b^{\mu}\mathcal{B}_{\mu}\Psi -i(\nabla_{\mu}e_b^{\mu})\gamma^0\gamma^b\Psi )\\
&\nonumber \Leftrightarrow\\
0=\nonumber &2i\gamma^0\gamma^be_b^{\mu}\nabla_{\mu}\Psi -\mathcal{B}_{\mu}\gamma^0\gamma^be_b^{\mu}\Psi -\gamma^0\gamma^be_b^{\mu}\mathcal{B}_{\mu}\Psi +i(\nabla_{\mu}e_b^{\mu})\gamma^0\gamma^b\Psi .
\label{proofofequivalence}
\end{align}
In the fourth equality we applied a transposition and multiplied by $\hat{d}$. Subsequently we employed the identities in Eq. (\ref{transpositionrules}) as well as the anti-commutation relations of the $\gamma$- and Pauli-matrices. The auxiliary gauge field $\hat{\mathcal{B}}_{\mu}$ is obtained from $\mathcal{B}$ after the operation $\sigma^i\to -\sigma^i$ on the Pauli matrices. Finally we multiply the equation by another $\hat{U}^T$ after having moved the latter to the left.

\subsection{Canonical quantization}

The canonical conjugate momentum for the massless fermion field is given by (after applying partial integration)
\begin{align}
\Pi =\frac{\partial \mathcal{L}}{\partial (D_0\Psi )}=\frac{1}{2c}\Psi^T\hat{U}\hat{d}^{\ast}i.
\end{align}
In the quantum theory this implies the canonical anti-commutation relation
\begin{align}
\{\hat{\Pi}_{\alpha}^s(\bold{x}),\hat{\Psi}_{\beta}^r(\bold{y})\}=i\delta^3(\bold{x}-\bold{y})\delta_{\alpha\beta}\delta^{rs}\,\,\Leftrightarrow\,\,\{\hat{\Psi}_{\alpha}^s(\bold{x}),\hat{\Psi}_{\beta}^r(\bold{y})\}=-2\cdot c\cdot \hat{d}\delta^3(\bold{x}-\bold{y})\gamma^2_{\alpha\beta}\delta^{(-r)s}
\label{anticommutatorwithconstraint}
\end{align}
The Hamiltonian density operator is given by
\begin{align}
\hat{\mathcal{H}}=\hat{\Pi}\cdot c\cdot D_0\hat{\Psi}-\mathcal{L}=-\frac{1}{4}\hat{\Psi}^T\hat{U}\hat{d}^{\ast}i\gamma^0\{D_j,\gamma^be_b^j\}\hat{\Psi} .
\end{align}
With the Hamiltonian operator $\hat{H}=\int d\Sigma\hat{\mathcal{H}}$ we obtain the commutation relation
\begin{align}
[\hat{H},\hat{\Psi}_{\alpha}^s(\bold{x})]=c\cdot [\gamma^0(2i\gamma^be_b^j\nabla_j\hat{\Psi} (\bold{x})+i\gamma^b(\nabla_je_b^j)\hat{\Psi} (\bold{x})+i(\hat{d}\nabla_j\hat{d}^{\ast})\gamma^be_b^j\hat{\Psi} (\bold{x})+\{\tilde{\mathcal{B}}_j,\gamma^be_b^j\}\hat{\Psi} (\bold{x}))]_{\alpha}^s.
\end{align}
The numerical factor $c$ has the same meaning as in the canonical quantization procedure outlined in the main text where the physical constraint of Eq. (\ref{majoranaconstraint}) has not been enforced a priori.

\section{Canonical quantization and fermion doubling}\label{app1_5}

The canonically conjugate momenta for the massless (chiral) fermions are given by (after applying partial integration)
\begin{align}
&\Pi =\frac{\partial\mathcal{L}}{\partial (c\cdot D_0\Psi )}=\frac{1}{2\cdot c}\overline{\Psi}i,\,\,\,\, \overline{\Pi}=\frac{\partial\mathcal{L}}{\partial (c\cdot \overline{\Psi}\overset{\leftarrow}{D}_0)}=-\frac{1}{2\cdot c}i\Psi \\
&\Pi_L=\frac{\partial\mathcal{L}}{\partial (c\cdot D_0\Psi_L)}=\frac{1}{2\cdot c}\overline{\Psi}_Li,\,\,\,\, \overline{\Pi}_L=\frac{\partial\mathcal{L}}{\partial (c\cdot D_0\overline{\Psi}_L)}=-\frac{1}{2\cdot c}i\Psi_L,\,\,\,\,L\,\leftrightarrow R,\,\tau\,\leftrightarrow \overline{\tau}
\end{align}
In the quantum theory fields $O$ get replaced by corresponding operators $O\to \hat O$. The non-trivial, equal-time canonical anti-commutation relations for the corresponding operators of the elementary fields in the quantum theory are 
\begin{align}
\nonumber &\{ (\hat\Pi )_{\alpha}(t,\bold{x}),(\hat\Psi )_{\beta}(t,\bold{y})\}=i\delta^3(\bold{x}-\bold{y})\delta_{\alpha\beta}\,\,\Leftrightarrow \,\,\{ (\hat{\overline{\Psi}})_{\alpha}(t,\bold{x}),(\hat\Psi )_{\beta}(t,\bold{y})\}=2\cdot c\cdot \delta^3(\bold{x}-\bold{y})\delta_{\alpha\beta}\\
 &\{ (\hat\Pi_L)_{\alpha}(t,\bold{x}),(\hat\Psi_L)_{\beta}(t,\bold{y})\}=i\delta^3(\bold{x}-\bold{y})\delta_{\alpha\beta}\,\, \Leftrightarrow \,\,\{ (\hat{\overline{\Psi}}_L)_{\alpha}(t,\bold{x}),(\hat\Psi_L)_{\beta}(t,\bold{y})\}=2\cdot c\cdot \delta^3(\bold{x}-\bold{y})\delta_{\alpha\beta},\,\,\,\,L\,\leftrightarrow R \label{comm1}
\end{align}
If the constraint $\hat{\overline{\Psi}} = \hat{\Psi}^T \hat{U} \hat{d}^{\ast}$ is enforced on operators, these anti-commutation relations become identical to those of Eq. (\ref{anticommutatorwithconstraint}) in Appendix \ref{app1}.\par
Notice the parameter $c$ which we inserted into the canonical formalism. Naively we have $c=1$. The problem with this choice is that we know how we should quantize the theory, as the quantum description was our original starting point. The fundamental field is $\chi$ defined in Eq. (\ref{fundamental1}) (see also Eq. (\ref{fundamental2})) comprising the annihilation operators $a_{\pm}(p)$. We observe (following Eq. (\ref{comm1})) that the equal time commutation relations are
\begin{align}
\{a_{\pm}(t,\bold{p}),\overline{a}_{\pm}(t,\bold{q})\}=2\cdot c\cdot (2\pi )^3\delta^3(\bold{p}-\bold{q})\,\,\,\,\text{(and equivalently for $\bold{p}\to K_{\pm}+\delta \bold{p},\,\,\bold{q}\to K_{\pm}+\delta \bold{q}$)}
\end{align}
We therefore take $c=\frac{1}{2}$ to be consistent with the anti-commutation relations of the annihilation and creation operators $a_{\pm}(t,\bold{p})$ and $\overline{a}_{\pm}(t,\bold{q})$, respectively. This peculiarity is a consequence of the doubling of degrees of freedom. More precisely, we have (with $\hat{\Psi} =(\hat{\Psi}^+,\hat{\Psi}^-)$ and four-momenta as arguments)
\begin{align}
&\hat{\Psi}_L^+(p)\Big|_{p=K_-+\delta p}=\int \frac{d^4x d^4 q}{(2\pi)^4}\frac{1}{\sqrt{2(1-d_3(x))}}\begin{pmatrix}
[-(d_1(x)+id_2(x))a_+(q)+(d_3(x)-1)a_-(q)]\\
[(d_3(x)-1)\overline{a}_+(-q)+(d_1(x)+id_2(x))\overline{a}_-(-q)] 
\end{pmatrix}e^{ix (q-p)}\label{psiLplus}\\
&\hat{\Psi}_L^-(p)\Big|_{p=K_-+\delta p}=\int \frac{d^4x d^4 q}{(2\pi)^4}\frac{1}{\sqrt{2(1+d_3(x))}}\begin{pmatrix}
[(-id_1(x)+d_2(x))a_+(q)+(id_3(x)+i)a_-(q)]\\
[(-id_3(x)-i)\overline{a}_+(-q)+(-id_1(x)+d_2(x))\overline{a}_-(-q)]
\end{pmatrix}e^{ix (q-p)}\label{psiLminus}\\
&\hat{\Psi}_R^+(p)\Big|_{p=K_++\delta p}=\int \frac{d^4x d^4 q}{(2\pi)^4}\frac{1}{\sqrt{2(1-d_3(x))}}\begin{pmatrix}
[-(d_1(x)+id_2(x))a_+(q)+(d_3(x)-1)a_-(q)]\\
[(-d_3(x)+1)\overline{a}_+(-q)-(d_1(x)+id_2(x))\overline{a}_-(-q)]
\end{pmatrix}e^{ix (q-p)}\label{psiRplus}\\
&\hat{\Psi}_R^-(p)\Big|_{p=K_++\delta p}=\int \frac{d^4x d^4 q}{(2\pi)^4}\frac{1}{\sqrt{2(1+d_3(x))}}\begin{pmatrix}
[(-id_1(x)+d_2(x))a_+(q)+(id_3(x)+i)a_-(q)]\\
[(id_3(x)+i)\overline{a}_+(-q)+(id_1(x)-d_2(x))\overline{a}_-(-q)]
\end{pmatrix}e^{ix (q-p)}\label{psiRminus}
\end{align}
We may consider, e. g., at $\bold{d} = const$ 
\begin{align}
\{(\overline{\Psi}_L^+)_1(t,\bold{q}),(\Psi_L^+)_1(t,\bold{p})\}=&\frac{1}{2(1-d_3)}[(d_1^2+d_2^2)\{\overline{a}_+(t,K_-+\delta \bold{q}),a_+(t,K_-+\delta \bold{p})\}\\
&+(d_3^2+1-2d_3)\{\overline{a}_-(t,K_-+\delta \bold{q}),a_-(t,K_-+\delta \bold{p})\}]\\
=&2\cdot c\cdot (2\pi )^3\delta^3(\bold{p}-\bold{q}) 
\label{comm0}
\end{align}
where we made use of $\{\overline{a}_{\pm}(t,K_-+\delta \bold{q}),a_{\pm}(t,K_-+\delta \bold{p})\} =2\cdot c\cdot (2\pi )^3\delta^3(\bold{p}-\bold{q})$ and $|\bold{d}|=1$.
One can easily check, however, that the same anti-commutation relation follows from the anti-commutation relation for the fields $a,\bar{a}$ also for space dependent (but time independent) vector field $\bold{d}$.
\par
The action is of first order in derivatives. Therefore the canonical momenta are given in terms of the Dirac fields (or their chiral components) themselves which therefore span the phase space of the theory. 
The Hamiltonian operator is given by $\hat H=\int d\Sigma \hat{\mathcal{H}}$ with Hamiltonian density operator
\begin{align}
\nonumber \hat{\mathcal{H}}=&\hat\Pi \cdot c\cdot D_0\hat\Psi -\mathcal{L}\\
\cong&-\frac{1}{4}[\hat{\overline{\Psi}}\gamma^0i\gamma^be_b^{i}(x)D_{i}\hat\Psi -[\hat{\overline{\Psi}}\gamma^0\overset{\leftarrow}{D}_i]i\gamma^be_b^{i}(x)\hat\Psi ]\\
\cong&-\frac{1}{4}[\hat{\overline{\Psi}}_Lie_b^{i}(x)\overline{\tau}^bD_{i}\hat\Psi_L-[\hat{\overline{\Psi}}_L\overset{\leftarrow}{D}_i]ie_b^{i}(x)\overline{\tau}^b\hat\Psi_L+\hat{\overline{\Psi}}_Rie_b^{i}(x)\tau^bD_{i}\hat\Psi_R-[\hat{\overline{\Psi}}_R\overset{\leftarrow}{D}_i]ie_b^{i}(x)\tau^b\hat\Psi_R]\label{Hzero}
\end{align}
where $\cong$ is meant to imply equality up to partial integration. Notice that the value $c$ appears in the definition of the Hamiltonian due to the doubling of degrees of freedom. It may be checked that Eq. (\ref{Hzero}) simply subtracts from the Lagrangian the term containing the time derivative, and flips the sign. This is the correct procedure to yield the Hamiltonian from the Lagrangian if we take into account the latter as a function of the original fields $a_\pm, \bar{a}_\pm$. Operators $\hat O$ satisfy the Heisenberg equation of motion 
\begin{align}
i\cdot D_0\hat O=-[\hat H,\hat O]
\label{heisenbergeom}
\end{align}
where $\hat H$ is the Hamiltonian operator defined above with normal ordering implied. When the degrees of freedom are subject to doubling, Eq. (\ref{heisenbergeom}) comes with an additional factor of $\frac{1}{2}$ on the left-hand side. The commutation relations we need later on are those for $\hat O=\hat\Psi,\,\hat{\overline{\Psi}},\,\hat\Psi_L,\,\hat{\overline{\Psi}}_L,\,(L\leftrightarrow R)$, respectively. The relevant commutators read
\begin{align}
\nonumber & [\hat H, \hat{\overline{\Psi}}_{\alpha}(x)]=c\cdot \frac{1}{2}(\hat{\overline{\Psi}}(x)\gamma^0i\gamma^be_b^i\overset{\leftarrow}{D}_i+\hat{\overline{\Psi}}(x)\gamma^0\overset{\leftarrow}{D}_ii\gamma^be_b^i)_{\alpha}\\
\nonumber & [\hat H,\hat \Psi_{\alpha}(x)]=c\cdot \frac{1}{2}(i\gamma^0\gamma^be_b^iD_i\hat\Psi (x)+i\gamma^0D_i\gamma^be_b^i\hat\Psi (x))_{\alpha}
\end{align}
as well as
\begin{align}
\nonumber &[\hat H,(\hat{\overline{\Psi}}_L)_{\alpha}(x)]=c\cdot \frac{1}{2}(\hat{\overline{\Psi}}_L(x)ie_b^i(x)\overline{\tau}^b\overset{\leftarrow}{D}_i+\hat{\overline{\Psi}}_L(x)\overset{\leftarrow}{D}_iie_b^i(x)\overline{\tau}^b)_{\alpha},\,\,\,\,L \leftrightarrow R,\, \tau \leftrightarrow \overline{\tau}\\
\nonumber &[\hat H,(\hat{\Psi}_L)_{\alpha}(x)]=c\cdot \frac{1}{2}(ie_b^i(x)\overline{\tau}^bD_i\hat\Psi_L(x)+D_iie_b^i(x)\overline{\tau}^b\hat\Psi_L(x))_{\alpha}, \,\,\,\, L \leftrightarrow R,\, \tau \leftrightarrow \overline{\tau}.
\end{align}
In conclusion we may enforce the reality constraint both classically and quantum mechanically in a trivial way in conjunction with the choice $c=\frac{1}{2}$.

\section{Energy momentum tensor  and spin tensor non-conservation}\label{app2}

The calculation of the divergence of the energy momentum tensor in Eq. (\ref{emt}) in the formulation which keeps the spin vector $\bold{d}$ explicit proceeds as follows. Consider 
\begin{align}
\nabla_{\mu}(\frac{1}{4}\overline{\psi}i\bold{e}_b^{\mu}\gamma^0\gamma^be_a^{\nu}\nabla_{\nu}\psi )=&\frac{1}{4}\overline{\psi}\gamma^0\overset{\leftarrow}{\nabla}_{\mu}i\bold{e}_b^{\mu}\gamma^be_a^{\nu}\nabla_{\nu}\psi +\frac{1}{4}\overline{\psi}\gamma^0i(\nabla_{\mu}\bold{e}_b^{\mu})\gamma^be_a^{\nu}\nabla_{\nu}\psi \nonumber \\
&+\frac{1}{4}\overline{\psi}\gamma^0i\bold{e}_b^{\mu}\gamma^b(\nabla_{\mu}e_a^{\nu})\nabla_{\nu}\psi +\frac{1}{4}\overline{\psi}\gamma^0i\bold{e}e_b^{\mu}\gamma^be_a^{\nu}\nabla_{\mu}\nabla_{\nu}\psi \nonumber \\
=&-\frac{1}{8}\overline{\psi}\gamma^0i(\nabla_{\mu}\bold{e}_b^{\mu})\gamma^be_a^{\nu}\nabla_{\nu}\psi +\frac{1}{4}\overline{\psi}\gamma^0i(\nabla_{\mu}\bold{e}_b^{\mu})\gamma^be_a^{\nu}\nabla_{\nu}\psi \nonumber\\
&+\frac{1}{4}\overline{\psi}\gamma^0i\bold{e}_b^{\mu}\gamma^b(\nabla_{\mu}e_a^{\nu})\nabla_{\nu}\psi +\frac{1}{4}\overline{\psi}\gamma^0i\bold{e}_b^{\mu}\gamma^be_a^{\nu}\nabla_{\nu}\nabla_{\mu}\psi -\frac{1}{4}\overline{\psi}\gamma^0\gamma^5\bold{e}_b^{\mu}\gamma^b\psi e_a^{\nu}\mathcal{F}_{\mu\nu}\nonumber\\
=&\frac{1}{8}\overline{\psi}\gamma^0i(\nabla_{\mu}\bold{e}_b^{\mu})\gamma^be_a^{\nu}\nabla_{\nu}\psi +\frac{1}{4}\overline{\psi}\gamma^0i\bold{e}_b^{\mu}\gamma^b(\nabla_{\mu}e_a^{\nu})\nabla_{\nu}\psi\nonumber\\
&-\frac{1}{4}\overline{\psi}\gamma^0ie_a^{\nu}(\nabla_{\nu}\bold{e}_b^{\mu})\gamma^b\nabla_{\mu}\psi +\frac{1}{4}\overline{\psi}\gamma^0e_a^{\nu}\nabla_{\nu}i\bold{e}_b^{\mu}\gamma^b\nabla_{\mu}\psi -\frac{1}{4}\overline{\psi}\gamma^0\gamma^5\bold{e}_b^{\mu}\gamma^b\psi e_a^{\nu}\mathcal{F}_{\mu\nu}\nonumber\\
=&\frac{1}{8}\overline{\psi}\gamma^0i(\nabla_{\mu}\bold{e}_b^{\mu})\gamma^be_a^{\nu}\nabla_{\nu}\psi +\frac{1}{4}\overline{\psi}\gamma^0i\gamma^b(\bold{e}_b^{\nu}\nabla_{\nu}e_a^{\mu}-e_a^{\nu}\nabla_{\nu}\bold{e}_b^{\mu})\nabla_{\mu}\psi\nonumber\\
&-\frac{1}{8}\overline{\psi}\gamma^0e_a^{\nu}\nabla_{\nu}i(\nabla_{\mu}\bold{e}_b^{\mu})\gamma^b\psi -\frac{1}{4}\overline{\psi}\gamma^0\gamma^5\bold{e}_b^{\mu}\gamma^b\psi e_a^{\nu}\mathcal{F}_{\mu\nu}\nonumber\\
=&\frac{1}{8}\overline{\psi}\gamma^0i(\nabla_{\mu}\bold{e}_b^{\mu})\gamma^be_a^{\nu}\nabla_{\nu}\psi +\frac{1}{4}\overline{\psi}\gamma^0i\gamma^b(\bold{e}_b^{\nu}\nabla_{\nu}e_a^{\mu}-e_a^{\nu}\nabla_{\nu}\bold{e}_b^{\mu})\nabla_{\mu}\psi\nonumber\\
&-\frac{1}{8}\overline{\psi}\gamma^0i(\nabla_{\mu}\bold{e}_b^{\mu})\gamma^be_a^{\nu}\nabla_{\nu}\psi -\frac{1}{8}\overline{\psi}\gamma^0e_a^{\nu}i(\nabla_{\nu}\nabla_{\mu}\bold{e}_b^{\mu})\gamma^b\psi -\frac{1}{4}\overline{\psi}\gamma^0\gamma^5\bold{e}_b^{\mu}\gamma^b\psi e_a^{\nu}\mathcal{F}_{\mu\nu}\nonumber\\
=&\frac{1}{4}\overline{\psi}\gamma^0i\gamma^b(\bold{e}_b^{\nu}\nabla_{\nu}e_a^{\mu}-e_a^{\nu}\nabla_{\nu}\bold{e}_b^{\mu})\nabla_{\mu}\psi -\frac{1}{8}\overline{\psi}\gamma^0e_a^{\nu}i(\nabla_{\nu}\nabla_{\mu}\bold{e}_b^{\mu})\gamma^b\psi -\frac{1}{4}\overline{\psi}\gamma^0\gamma^5\bold{e}_b^{\mu}\gamma^b\psi e_a^{\nu}\mathcal{F}_{\mu\nu}\nonumber .
\end{align}
We employed the equations of motion as well as the product rule of differentiation and the commutativity of derivatives in the above calculation. The second term in the ultimate line is cancelled by an analogous term due to the second term of Eq. (\ref{emt}) (obtained by hermitian conjugation) whose divergence is to be taken. Another way to see that this term vanishes is its anti-hermiticity. Adding up this result and its hermitian conjugate then produces the divergence claimed in Eqs. (\ref{tdivergence}) and (\ref{tdivergenceextra}), respectively.\par
The divergence of the energy momentum tensor in terms of the geometric formulation featuring the gauge field $\mathcal{B}_{\mu}$ will be outlined subsequently. We write $4\nabla_{\mu}T_a^{\mu}=A_a+B_a$ with
\begin{align}
\nonumber A_a=&\overline{\Psi}\gamma^0\overset{\leftarrow}{\nabla}_{\mu}i\gamma^0\gamma^be_a^{\nu}e_b^{\mu}D_{\nu}\Psi +\overline{\Psi}i\gamma^0\gamma^b(\nabla_{\mu}e_a^{\nu})e_b^{\mu}D_{\nu}\Psi +\overline{\Psi}i\gamma^0\gamma^be_a^{\nu}(\nabla_{\mu}e_b^{\mu})D_{\nu}\Psi \\
\nonumber &+\overline{\Psi}i\gamma^0\gamma^be_a^{\nu}e_b^{\mu}\nabla_{\mu}\nabla_{\nu}\Psi -\overline{\Psi}i\gamma^0\gamma^be_a^{\nu}e_b^{\mu}i(\nabla_{\mu}\mathcal{B}_{\nu})\Psi \\
\nonumber =&-\frac{1}{2}\overline{\Psi}\gamma^0i\mathcal{B}_{\mu}i\gamma^be_b^{\mu}e_a^{\nu}D_{\nu}\Psi -\frac{1}{2}\overline{\Psi}i\gamma^0\gamma^be_b^{\mu}i\mathcal{B}_{\mu}e_a^{\nu}D_{\nu}\Psi -\frac{1}{2}\overline{\Psi}i\gamma^0\gamma^b(\nabla_{\mu}e_b^{\mu})e_a^{\nu}D_{\nu}\Psi +\overline{\Psi}i\gamma^0\gamma^b(\nabla_{\mu}e_a^{\nu})e_b^{\mu}D_{\nu}\Psi \\
\nonumber &+\overline{\Psi}i\gamma^0\gamma^be_a^{\nu}(\nabla_{\mu}e_b^{\mu})D_{\nu}\Psi +\overline{\Psi}i\gamma^0\gamma^be_a^{\nu}e_b^{\mu}\nabla_{\nu}\nabla_{\mu}\Psi -\overline{\Psi}\gamma^0\gamma^5e_b^{\mu}\gamma^b\Psi e_a^{\nu}\mathcal{F}_{\mu\nu}-\overline{\Psi}i\gamma^0\gamma^be_a^{\nu}e_b^{\mu}i(\nabla_{\mu}\mathcal{B}_{\nu})\Psi \\
\nonumber =&-\frac{1}{2}\overline{\Psi}\gamma^0i\mathcal{B}_{\mu}i\gamma^be_b^{\mu}e_a^{\nu}D_{\nu}\Psi -\frac{1}{2}\overline{\Psi}i\gamma^0\gamma^be_b^{\mu}i\mathcal{B}_{\mu}e_a^{\nu}D_{\nu}\Psi +\frac{1}{2}\overline{\Psi}i\gamma^0\gamma^be_a^{\nu}(\nabla_{\mu}e_b^{\mu})D_{\nu}\Psi +\overline{\Psi}i\gamma^0\gamma^b(\nabla_{\mu}e_a^{\nu})e_b^{\mu}D_{\nu}\Psi \\
\nonumber &+\overline{\Psi}i\gamma^0\gamma^be_a^{\nu}\nabla_{\nu}e_b^{\mu}\nabla_{\mu}\Psi -\overline{\Psi}i\gamma^0\gamma^be_a^{\nu}(\nabla_{\nu}e_b^{\mu})\nabla_{\mu}\Psi -\overline{\Psi}\gamma^0\gamma^5e_b^{\mu}\gamma^b\Psi e_a^{\nu}\mathcal{F}_{\mu\nu}-\overline{\Psi}i\gamma^0\gamma^be_a^{\nu}e_b^{\mu}i(\nabla_{\mu}\mathcal{B}_{\nu})\Psi \\
\nonumber =&-\frac{1}{2}\overline{\Psi}\gamma^0i\mathcal{B}_{\mu}i\gamma^be_b^{\mu}e_a^{\nu}D_{\nu}\Psi -\frac{1}{2}\overline{\Psi}i\gamma^0\gamma^be_b^{\mu}i\mathcal{B}_{\mu}e_a^{\nu}D_{\nu}\Psi +\frac{1}{2}\overline{\Psi}i\gamma^0\gamma^be_a^{\nu}(\nabla_{\mu}e_b^{\mu})D_{\nu}\Psi +\overline{\Psi}i\gamma^0\gamma^bD_{\nu}\Psi T_{ab}^{\nu}\\
\nonumber &-\overline{\Psi}i\gamma^0\gamma^be_a^{\nu}(\nabla_{\nu}e_b^{\mu})i\mathcal{B}_{\mu}\Psi -\overline{\Psi}i\gamma^0\gamma^be_a^{\nu}e_b^{\mu}i(\nabla_{\mu}\mathcal{B}_{\nu})\Psi +\frac{1}{2}\overline{\Psi}\gamma^0e_a^{\nu}\nabla_{\nu}i\gamma^be_b^{\mu}i\mathcal{B}_{\mu}\Psi \\
\nonumber &+\frac{1}{2}\overline{\Psi}\gamma^0e_a^{\nu}\nabla_{\nu}i\mathcal{B}_{\mu}i\gamma^be_b^{\mu}\Psi -\frac{1}{2}\overline{\Psi}\gamma^0e_a^{\nu}\nabla_{\nu}i\gamma^b(\nabla_{\mu}e_b^{\mu})\Psi -\overline{\Psi}\gamma^0\gamma^5e_b^{\mu}\gamma^b\Psi e_a^{\nu}\mathcal{F}_{\mu\nu}\\
\nonumber =&-\frac{1}{2}\overline{\Psi}\gamma^0i\mathcal{B}_{\mu}i\gamma^be_b^{\mu}e_a^{\nu}\nabla_{\nu}\Psi +\frac{1}{2}\overline{\Psi}\gamma^0i\mathcal{B}_{\mu}i\gamma^be_b^{\mu}e_a^{\nu}i\mathcal{B}_{\nu}\Psi -\frac{1}{2}\overline{\Psi}i\gamma^0\gamma^be_b^{\mu}i\mathcal{B}_{\mu}e_a^{\nu}\nabla_{\nu}\Psi +\frac{1}{2}\overline{\Psi}i\gamma^0\gamma^be_b^{\mu}i\mathcal{B}_{\mu}e_a^{\nu}i\mathcal{B}_{\nu}\Psi \\
\nonumber &+\frac{1}{2}\overline{\Psi}i\gamma^0\gamma^be_a^{\nu}(\nabla_{\mu}e_b^{\mu})\nabla_{\nu}\Psi -\frac{1}{2}\overline{\Psi}i\gamma^0\gamma^be_a^{\nu}(\nabla_{\mu}e_b^{\mu})i\mathcal{B}_{\nu}\Psi +\overline{\Psi}i\gamma^0\gamma^bD_{\nu}\Psi T_{ab}^{\nu} -\overline{\Psi}i\gamma^0\gamma^be_a^{\nu}(\nabla_{\nu}e_b^{\mu})i\mathcal{B}_{\mu}\Psi \\
\nonumber &-\overline{\Psi}i\gamma^0\gamma^be_a^{\nu}e_b^{\mu}i(\nabla_{\mu}\mathcal{B}_{\nu})\Psi +\frac{1}{2}\overline{\Psi}\gamma^0e_a^{\nu}i\gamma^b(\nabla_{\nu}e_b^{\mu})i\mathcal{B}_{\mu}\Psi +\frac{1}{2}\overline{\Psi}\gamma^0e_a^{\nu}i\gamma^be_b^{\mu}i(\nabla_{\nu}\mathcal{B}_{\mu})\Psi +\frac{1}{2}\overline{\Psi}\gamma^0e_a^{\nu}i\gamma^be_b^{\mu}i\mathcal{B}_{\mu}\nabla_{\nu}\Psi \\
\nonumber &+\frac{1}{2}\overline{\Psi}\gamma^0e_a^{\nu}i(\nabla_{\nu}\mathcal{B}_{\mu})i\gamma^be_b^{\mu}\Psi +\frac{1}{2}\overline{\Psi}\gamma^0e_a^{\nu}i\mathcal{B}_{\mu}i\gamma^b(\nabla_{\nu}e_b^{\mu})\Psi +\frac{1}{2}\overline{\Psi}\gamma^0e_a^{\nu}i\mathcal{B}_{\mu}i\gamma^be_b^{\mu}\nabla_{\nu}\Psi \\
\nonumber &-\frac{1}{2}\overline{\Psi}\gamma^0e_a^{\nu}i\gamma^b(\nabla_{\nu}\nabla_{\mu}e_b^{\mu})\Psi -\frac{1}{2}\overline{\Psi}i\gamma^0\gamma^be_a^{\nu}(\nabla_{\mu}e_b^{\mu})\nabla_{\nu}\Psi -\overline{\Psi}\gamma^0\gamma^5e_b^{\mu}\gamma^b\Psi e_a^{\nu}\mathcal{F}_{\mu\nu}
\end{align}
and
\begin{align}
\nonumber B_a=&-\overline{\Psi}\gamma^0\overset{\leftarrow}{\nabla}_{\nu}\overset{\leftarrow}{\nabla}_{\mu}i\gamma^be_a^{\nu}e_b^{\mu}\Psi -\overline{\Psi}\gamma^0i(\nabla_{\mu}\mathcal{B}_{\nu})i\gamma^be_a^{\nu}e_b^{\mu}\Psi -\overline{\Psi}\gamma^0\overset{\leftarrow}{D}_{\nu}i\gamma^b(\nabla_{\mu}e_a^{\nu})e_b^{\mu}\Psi \\
\nonumber &-\overline{\Psi}\gamma^0\overset{\leftarrow}{D}_{\nu}i\gamma^be_a^{\nu}(\nabla_{\mu}e_b^{\mu})\Psi -\overline{\Psi}\gamma^0\overset{\leftarrow}{D}_{\nu}i\gamma^be_a^{\nu}e_b^{\mu}\nabla_{\mu}\Psi\\
\nonumber =&-\overline{\Psi}\gamma^0\overset{\leftarrow}{\nabla}_{\mu}\overset{\leftarrow}{\nabla}_{\nu}i\gamma^be_a^{\nu}e_b^{\mu}\Psi -\overline{\Psi}\gamma^0\gamma^5e_b^{\mu}\gamma^b\Psi e_a^{\nu}\mathcal{F}_{\mu\nu}-\overline{\Psi}\gamma^0i(\nabla_{\mu}\mathcal{B}_{\nu})i\gamma^be_a^{\nu}e_b^{\mu}\Psi -\overline{\Psi}\gamma^0\overset{\leftarrow}{D}_{\nu}i\gamma^b(\nabla_{\mu}e_a^{\nu})e_b^{\mu}\Psi \\
\nonumber &-\overline{\Psi}\gamma^0\overset{\leftarrow}{D}_{\nu}i\gamma^be_a^{\nu}(\nabla_{\mu}e_b^{\mu})\Psi -\frac{1}{2}\overline{\Psi}\gamma^0\overset{\leftarrow}{D}_{\nu}i\gamma^be_a^{\nu}e_b^{\mu}i\mathcal{B}_{\mu}\Psi -\frac{1}{2}\overline{\Psi}\gamma^0\overset{\leftarrow}{D}_{\nu}i\mathcal{B}_{\mu}i\gamma^be_a^{\nu}e_b^{\mu}\Psi +\frac{1}{2}\overline{\Psi}\gamma^0\overset{\leftarrow}{D}_{\nu}i\gamma^be_a^{\nu}(\nabla_{\mu}e_b^{\mu})\Psi\\
\nonumber =&\overline{\Psi}\gamma^0\overset{\leftarrow}{\nabla}_{\mu}i\gamma^be_a^{\nu}(\nabla_{\nu}e_b^{\mu})\Psi -\overline{\Psi}\gamma^0\overset{\leftarrow}{\nabla}_{\mu}e_b^{\mu}\overset{\leftarrow}{\nabla}_{\nu}i\gamma^be_a^{\nu}\Psi -\overline{\Psi}\gamma^0i(\nabla_{\mu}\mathcal{B}_{\nu})i\gamma^be_a^{\nu}e_b^{\mu}\Psi -\overline{\Psi}\gamma^0\overset{\leftarrow}{D}_{\nu}i\gamma^b(\nabla_{\mu}e_a^{\nu})e_b^{\mu}\Psi \\
\nonumber &-\frac{1}{2}\overline{\Psi}\gamma^0\overset{\leftarrow}{D}_{\nu}i\gamma^be_a^{\nu}(\nabla_{\mu}e_b^{\mu})\Psi -\frac{1}{2}\overline{\Psi}\gamma^0\overset{\leftarrow}{D}_{\nu}e_a^{\nu}i\gamma^be_b^{\mu}i\mathcal{B}_{\mu}\Psi -\frac{1}{2}\overline{\Psi}\gamma^0\overset{\leftarrow}{D}_{\nu}i\mathcal{B}_{\mu}e_a^{\nu}i\gamma^be_b^{\mu}\Psi -\overline{\Psi}\gamma^0\gamma^5e_b^{\mu}\gamma^b\Psi e_a^{\nu}\mathcal{F}_{\mu\nu}\\
\nonumber =&-\overline{\Psi}\gamma^0\overset{\leftarrow}{D}_{\nu}i\gamma^b\Psi T_{ab}^{\nu}-\overline{\Psi}\gamma^0i\mathcal{B}_{\mu}i\gamma^be_a^{\nu}(\nabla_{\nu}e_b^{\mu})\Psi -\overline{\Psi}\gamma^0i(\nabla_{\mu}\mathcal{B}_{\nu})i\gamma^be_a^{\nu}e_b^{\mu}\Psi -\frac{1}{2}\overline{\Psi}\gamma^0\overset{\leftarrow}{D}_{\nu}i\gamma^be_a^{\nu}(\nabla_{\mu}e_b^{\mu})\Psi \\
\nonumber &-\frac{1}{2}\overline{\Psi}\gamma^0\overset{\leftarrow}{D}_{\nu}e_a^{\nu}i\gamma^be_b^{\mu}i\mathcal{B}_{\mu}\Psi -\frac{1}{2}\overline{\Psi}\gamma^0\overset{\leftarrow}{D}_{\nu}i\mathcal{B}_{\mu}i\gamma^be_a^{\nu}e_b^{\mu}\Psi +\frac{1}{2}\overline{\Psi}\gamma^0i\mathcal{B}_{\mu}i\gamma^be_b^{\mu}\overset{\leftarrow}{\nabla}_{\nu}e_a^{\nu}\Psi \\
\nonumber &+\frac{1}{2}\overline{\Psi}\gamma^0i\gamma^b(\nabla_{\mu}e_b^{\mu})\overset{\leftarrow}{\nabla}_{\nu}e_a^{\nu}\Psi +\frac{1}{2}\overline{\Psi}\gamma^0i\gamma^be_b^{\mu}i\mathcal{B}_{\mu}\overset{\leftarrow}{\nabla}_{\nu}e_a^{\nu}\Psi -\overline{\Psi}\gamma^0\gamma^5e_b^{\mu}\gamma^b\Psi e_a^{\nu}\mathcal{F}_{\mu\nu}\\
\nonumber =&-\overline{\Psi}\gamma^0\overset{\leftarrow}{D}_{\nu}i\gamma^b\Psi T_{ab}^{\nu}-\overline{\Psi}\gamma^0i\mathcal{B}_{\mu}i\gamma^be_a^{\nu}(\nabla_{\nu}e_b^{\mu})\Psi -\overline{\Psi}\gamma^0i(\nabla_{\mu}\mathcal{B}_{\nu})i\gamma^be_a^{\nu}e_b^{\mu}\Psi -\frac{1}{2}\overline{\Psi}\gamma^0\overset{\leftarrow}{\nabla}_{\nu}i\gamma^be_a^{\nu}(\nabla_{\mu}e_b^{\mu})\Psi \\
\nonumber &-\frac{1}{2}\overline{\Psi}\gamma^0i\mathcal{B}_{\nu}i\gamma^be_a^{\nu}(\nabla_{\mu}e_b^{\mu})\Psi -\frac{1}{2}\overline{\Psi}\gamma^0\overset{\leftarrow}{\nabla}_{\nu}e_a^{\nu}i\gamma^be_b^{\mu}i\mathcal{B}_{\mu}\Psi -\frac{1}{2}\overline{\Psi}\gamma^0i\mathcal{B}_{\nu}e_a^{\nu}i\gamma^be_b^{\mu}i\mathcal{B}_{\mu}\Psi -\frac{1}{2}\overline{\Psi}\gamma^0\overset{\leftarrow}{\nabla}_{\nu}i\mathcal{B}_{\mu}i\gamma^be_a^{\nu}e_b^{\mu}\Psi \\
\nonumber &-\frac{1}{2}\overline{\Psi}\gamma^0i\mathcal{B}_{\nu}i\mathcal{B}_{\mu}i\gamma^be_a^{\nu}e_b^{\mu}\Psi +\frac{1}{2}\overline{\Psi}\gamma^0\overset{\leftarrow}{\nabla}_{\nu}i\mathcal{B}_{\mu}i\gamma^be_b^{\mu}e_a^{\nu}\Psi +\frac{1}{2}\overline{\Psi}\gamma^0i(\nabla_{\nu}\mathcal{B}_{\mu})i\gamma^be_b^{\mu}e_a^{\nu}\Psi +\frac{1}{2}\overline{\Psi}\gamma^0i\mathcal{B}_{\mu}i\gamma^b(\nabla_{\nu}e_b^{\mu})e_a^{\nu}\Psi
\end{align}
\begin{align}
\nonumber &+\frac{1}{2}\overline{\Psi}\gamma^0\overset{\leftarrow}{\nabla}_{\nu}i\gamma^b(\nabla_{\mu}e_b^{\mu})e_a^{\nu}\Psi +\frac{1}{2}\overline{\Psi}\gamma^0i\gamma^b(\nabla_{\nu}\nabla_{\mu}e_b^{\mu})e_a^{\nu}\Psi +\frac{1}{2}\overline{\Psi}\gamma^0\overset{\leftarrow}{\nabla}_{\nu}i\gamma^be_b^{\mu}i\mathcal{B}_{\mu}e_a^{\nu}\Psi \\
\nonumber &+\frac{1}{2}\overline{\Psi}\gamma^0i\gamma^b(\nabla_{\nu}e_b^{\mu})i\mathcal{B}_{\mu}e_a^{\nu}\Psi +\frac{1}{2}\overline{\Psi}\gamma^0i\gamma^be_b^{\mu}i(\nabla_{\nu}\mathcal{B}^{\mu})e_a^{\nu}\Psi -\overline{\Psi}\gamma^0\gamma^5e_b^{\mu}\gamma^b\Psi e_a^{\nu}\mathcal{F}_{\mu\nu}.
\end{align}
Collection, cancellation and rearrangement of terms finally leads to Eq. (\ref{geononconservationofemt1}).\par
The divergence of the spin tensor may be evaluated using of the equations of motion and the anti-commutation relations of the $\gamma$-matrices.
Consequently $\nabla_{\mu}S^{\mu}_{ab}=A_{ab}+B_{ab}+C_{ab}$ with
\begin{align}
A_{ab}=&\frac{i}{16}[\overline{\Psi}\gamma^0\overset{\leftarrow}{\nabla}_{\mu}]e_c^{\mu}\{\gamma^c,[\gamma_a,\gamma_b]\}\Psi\nonumber\\
=&\frac{i}{16}[\overline{\Psi}\gamma^0\overset{\leftarrow}{\nabla}_{\mu}]e_c^{\mu}(\gamma^c[\gamma_a,\gamma_b]+[\gamma_a,\gamma_b]\gamma^c)\Psi\nonumber\\
=&\frac{i}{16}[\overline{\Psi}\gamma^0\overset{\leftarrow}{\nabla}_{\mu}]e_c^{\mu}(2\gamma^c,[\gamma_a,\gamma_b]-4\gamma_b\delta_a^c+4\gamma_a\delta_b^c)\Psi\nonumber\\
=&-\frac{i}{16}\overline{\Psi}\gamma^0(\gamma^ce_c^{\mu}i\mathcal{B}_{\mu}+i\mathcal{B}_{\mu}\gamma^ce_c^{\mu}+\gamma^c(\nabla_{\mu}e_c^{\mu}))[\gamma_a,\gamma_b]\Psi +\frac{i}{4}[\overline{\Psi}\gamma^0\overset{\leftarrow}{\nabla}_{\mu}](\gamma_ae_b^{\mu}-\gamma_be_a^{\mu})\Psi \nonumber ,
\end{align}
\begin{align}
B_{ab}=&\frac{i}{16}\overline{\Psi}\gamma^0(\nabla_{\mu}e_c^{\mu})\{\gamma^c,[\gamma_a,\gamma_b]\}\Psi \nonumber
\end{align}
and
\begin{align}
C_{ab}=&\frac{i}{16}\overline{\Psi}\gamma^0e_c^{\mu}\{\gamma^c,[\gamma_a,\gamma_b]\}\nabla_{\mu}\Psi\nonumber\\
=&\frac{i}{16}\overline{\Psi}\gamma^0e_c^{\mu}(\gamma^c[\gamma_a,\gamma_b]+[\gamma_a,\gamma_b]\gamma^c)\nabla_{\mu}\Psi\nonumber\\
=&\frac{i}{16}\overline{\Psi}\gamma^0e_c^{\mu}(2[\gamma_a,\gamma_b]\gamma^c+4\gamma_b\delta_a^c-4\gamma_a\delta_b^c)\nabla_{\mu}\Psi\nonumber\\
=&\frac{i}{16}\overline{\Psi}\gamma^0[\gamma_a,\gamma_b](\gamma^ce_c^{\mu}i\mathcal{B}_{\mu}+i\mathcal{B}_{\mu}\gamma^ce_c^{\mu}-\gamma^c(\nabla_{\mu}e_c^{\mu}))\Psi +\frac{i}{4}\overline{\Psi}\gamma^0(\gamma_be_a^{\mu}-\gamma_ae_b^{\mu})\nabla_{\mu}\Psi \nonumber .
\end{align}
By comparison with Eq. (\ref{emt}) we find (moving $[\gamma_a,\gamma_b]$ next to $\mathcal{B}_{\mu}$) the result claimed in Eqs. (\ref{sdivergence}) and (\ref{sdivergenceextra}), respectvely.

\section{System of units}\label{app3}

In this work we make use of units where $\hbar =k_B=1$ as usual. In addition we impose the convenient choice $e=(v_{\parallel}v_{\perp}^2)^{\frac{1}{3}}=1$ where $e$ is the vierbein determinant instead of $c=1$ with $c$ being the velocity of light in vacuum. This reduction leaves only the energy scale or distance scale undetermined. These are usually measured in units of $eV$ and $(eV)^{-1}$. We define the first unitless ratio $d=\frac{c}{(v_{\parallel}v_{\perp}^2)^{\frac{1}{3}}}$. Then physical quantities and their units are given by ($\hbar =k_B=1$)
\begin{center}
\begin{tabular}{lclclclcl}
energy  & $eV$ & \,\,\,\,\,\,\,\,\,\,\, temperature & $eV$ \\
momentum & $\frac{eV}{c}$ & \,\,\,\,\,\,\,\,\,\,\, pressure & $\frac{(eV)^4}{c^3}$ \\
mass & $\frac{eV}{c^2}$ & \,\,\,\,\,\,\,\,\,\,\, entropy density & $\frac{(eV)^3}{c^3}$ \\
time & $\frac{1}{eV}$ & \,\,\,\,\,\,\,\,\,\,\, particle number density & $\frac{(eV)^3}{c^3}$ \\
position & $\frac{c}{eV}$ & \,\,\,\,\,\,\,\,\,\,\, angular momentum density & $\frac{(eV)^3}{c^3}$
\end{tabular}
\end{center}
where we left units of velocity and energy explicit. Natural units set $c\to 1$, while our choice of units implies $c\to d$. Going from units with $c=1$ to our units requires multiplication with a power of $d$ coincident with that of $c$ indicated above, while back transformation requires multiplication with the inverse power of $d$. In addition we set $\frac{v_{\perp}}{R}=1$ implying dimensionless units where $R$ is the transverse radius of the cylinder containing the chiral particles and possibly a vortex in the center. We define the second unitless ratio $f(R)=\frac{(eV)\cdot R}{v_{\perp}}$. Our dimensionless units finally imply the replacement $eV\to f(R)$.\par
In order to provide examples we show how physical quantities in natural units arise from unity in our dimensionless units 
\begin{center}
\begin{tabular}{lclclclcl}
energy  & $1\cong \frac{1}{f(R)}eV$ & \,\,\,\,\,\,\,\,\,\,\, temperature & $1\cong \frac{1}{f(R)}eV$ \\
momentum & $1\cong\frac{d}{f(R)}eV$ & \,\,\,\,\,\,\,\,\,\,\, pressure & $1\cong\frac{d^3}{f(R)^4}(eV)^4$ \\
mass & $1\cong\frac{d^2}{f(R)}eV$ & \,\,\,\,\,\,\,\,\,\,\, entropy density & $1\cong\frac{d^3}{f(R)^3}(eV)^3$ \\
time & $1\cong f(R) (eV)^{-1}$ & \,\,\,\,\,\,\,\,\,\,\, particle number density & $1\cong\frac{d^3}{f(R)^3}(eV)^3$ \\
position & $1\cong\frac{f(R)}{d}(eV)^{-1}$ & \,\,\,\,\,\,\,\,\,\,\, angular momentum density & $1\cong\frac{d^3}{f(R)^3}(eV)^3$
\end{tabular}
\end{center}
The usual conventions to reach dimensionless units involve the choice $G=1$ where $G$ is Newton's gravitational constant. This implies the transition from natural units to Planck units. Newton's gravitational constant may be written in the form (with $\hbar =k_B=1$) $G=C^2\cdot\frac{c^5}{(eV)^2}$ with a numerical dimensionless coefficient $C$. In order to express physical quantities in Planck units one sets $c$ and then $G$ to unity which means $c\to 1$ and then $eV\to C$. For completeness we explain how our dimensionless units are related to Planck units.
In order to get to our convention from Planck units, we multiply by a power of $d$ and $\frac{f(R)}{C}$ coinciding with that of $c$ and $eV$ indicated above. The reverse transformation from our units to Planck units requires again the choice of inverse powers for multiplication.

\bibliography{ZubarevHe.bib,QFTMacroMotion.bib,NiehYan.bib,QHE.bib}

\begin{thebibliography}{50}%
\makeatletter
\providecommand \@ifxundefined [1]{%
 \@ifx{#1\undefined}
}%
\providecommand \@ifnum [1]{%
 \ifnum #1\expandafter \@firstoftwo
 \else \expandafter \@secondoftwo
 \fi
}%
\providecommand \@ifx [1]{%
 \ifx #1\expandafter \@firstoftwo
 \else \expandafter \@secondoftwo
 \fi
}%
\providecommand \natexlab [1]{#1}%
\providecommand \enquote  [1]{``#1''}%
\providecommand \bibnamefont  [1]{#1}%
\providecommand \bibfnamefont [1]{#1}%
\providecommand \citenamefont [1]{#1}%
\providecommand \href@noop [0]{\@secondoftwo}%
\providecommand \href [0]{\begingroup \@sanitize@url \@href}%
\providecommand \@href[1]{\@@startlink{#1}\@@href}%
\providecommand \@@href[1]{\endgroup#1\@@endlink}%
\providecommand \@sanitize@url [0]{\catcode `\\12\catcode `\$12\catcode
  `\&12\catcode `\#12\catcode `\^12\catcode `\_12\catcode `\%12\relax}%
\providecommand \@@startlink[1]{}%
\providecommand \@@endlink[0]{}%
\providecommand \url  [0]{\begingroup\@sanitize@url \@url }%
\providecommand \@url [1]{\endgroup\@href {#1}{\urlprefix }}%
\providecommand \urlprefix  [0]{URL }%
\providecommand \Eprint [0]{\href }%
\providecommand \doibase [0]{https://doi.org/}%
\providecommand \selectlanguage [0]{\@gobble}%
\providecommand \bibinfo  [0]{\@secondoftwo}%
\providecommand \bibfield  [0]{\@secondoftwo}%
\providecommand \translation [1]{[#1]}%
\providecommand \BibitemOpen [0]{}%
\providecommand \bibitemStop [0]{}%
\providecommand \bibitemNoStop [0]{.\EOS\space}%
\providecommand \EOS [0]{\spacefactor3000\relax}%
\providecommand \BibitemShut  [1]{\csname bibitem#1\endcsname}%
\let\auto@bib@innerbib\@empty
%</preamble>
\bibitem [{\citenamefont {Volovik}(2003)}]{Volovik2003}%
  \BibitemOpen
  \bibfield  {author} {\bibinfo {author} {\bibfnamefont {G.~E.}\ \bibnamefont
  {Volovik}},\ }\href@noop {} {\emph {\bibinfo {title} {``The Universe in a
  Helium Droplet.''}}}\ (\bibinfo  {publisher} {Clarendon Press},\ \bibinfo
  {address} {Oxford},\ \bibinfo {year} {2003})\BibitemShut {NoStop}%
\bibitem [{\citenamefont {Vollhardt}\ and\ \citenamefont
  {W\"olfle}(1990)}]{VollhardtWolfle1990}%
  \BibitemOpen
  \bibfield  {author} {\bibinfo {author} {\bibfnamefont {D.}~\bibnamefont
  {Vollhardt}}\ and\ \bibinfo {author} {\bibfnamefont {P.}~\bibnamefont
  {W\"olfle}},\ }\href@noop {} {\emph {\bibinfo {title} {The superfluid phases
  of helium 3}}}\ (\bibinfo  {publisher} {Taylor and Francis},\ \bibinfo
  {address} {London},\ \bibinfo {year} {1990})\BibitemShut {NoStop}%
\bibitem [{\citenamefont {Volovik}\ and\ \citenamefont
  {Zubkov}(2014{\natexlab{a}})}]{VZ2014}%
  \BibitemOpen
  \bibfield  {author} {\bibinfo {author} {\bibfnamefont {G.}~\bibnamefont
  {Volovik}}\ and\ \bibinfo {author} {\bibfnamefont {M.}~\bibnamefont
  {Zubkov}},\ }\bibfield  {title} {\bibinfo {title} {Emergent weyl spinors in
  multi-fermion systems},\ }\href
  {https://doi.org/https://doi.org/10.1016/j.nuclphysb.2014.02.018} {\bibfield
  {journal} {\bibinfo  {journal} {Nuclear Physics B}\ }\textbf {\bibinfo
  {volume} {881}},\ \bibinfo {pages} {514} (\bibinfo {year}
  {2014}{\natexlab{a}})}\BibitemShut {NoStop}%
\bibitem [{\citenamefont {Volovik}\ and\ \citenamefont
  {Vachaspati}(1996)}]{VolovikVachaspati1996}%
  \BibitemOpen
  \bibfield  {author} {\bibinfo {author} {\bibfnamefont {G.~E.}\ \bibnamefont
  {Volovik}}\ and\ \bibinfo {author} {\bibfnamefont {T.}~\bibnamefont
  {Vachaspati}},\ }\bibfield  {title} {\bibinfo {title} {Aspects of $^3$he and
  the standard electroweak model},\ }\href@noop {} {\bibfield  {journal}
  {\bibinfo  {journal} {Int. J. Mod. Phys.}\ }\textbf {\bibinfo {volume} {B
  10}},\ \bibinfo {pages} {471} (\bibinfo {year} {1996})}\BibitemShut {NoStop}%
\bibitem [{\citenamefont {Volovik}\ and\ \citenamefont
  {Khazan}(1983)}]{VolovikKhazan1982}%
  \BibitemOpen
  \bibfield  {author} {\bibinfo {author} {\bibfnamefont {G.~E.}\ \bibnamefont
  {Volovik}}\ and\ \bibinfo {author} {\bibfnamefont {M.~V.}\ \bibnamefont
  {Khazan}},\ }\bibfield  {title} {\bibinfo {title} {Dynamics of the a-phase of
  $^3$he at low pressure},\ }\href@noop {} {\bibfield  {journal} {\bibinfo
  {journal} {JETP}\ }\textbf {\bibinfo {volume} {55}},\ \bibinfo {pages} {551}
  (\bibinfo {year} {1983})}\BibitemShut {NoStop}%
\bibitem [{\citenamefont {Volovik}(1990)}]{Volovik1990}%
  \BibitemOpen
  \bibfield  {author} {\bibinfo {author} {\bibfnamefont {G.~E.}\ \bibnamefont
  {Volovik}},\ }\bibfield  {title} {\bibinfo {title} {Symmetry in superfluid
  $^3$he}\ }(\bibinfo {year} {1990})\ pp.\ \bibinfo {pages}
  {27--134}\BibitemShut {NoStop}%
\bibitem [{\citenamefont {Nambu}(2010)}]{Nambu1985}%
  \BibitemOpen
  \bibfield  {author} {\bibinfo {author} {\bibfnamefont {Y.}~\bibnamefont
  {Nambu}},\ }\bibfield  {title} {\bibinfo {title} {Fermion - boson relations
  in bcs type theories, physica d {\bf 15}, 147--151 (1985); energy gap, mass
  gap, and spontaneous symmetry breaking, in},\ }in\ \href@noop {} {\emph
  {\bibinfo {booktitle} {BCS: 50 Years}}},\ \bibinfo {editor} {edited by\
  \bibinfo {editor} {\bibfnamefont {L.~N.}\ \bibnamefont {Cooper}}\ and\
  \bibinfo {editor} {\bibfnamefont {D.}~\bibnamefont {Feldman}}}\ (\bibinfo
  {publisher} {World Scientific},\ \bibinfo {year} {2010})\BibitemShut
  {NoStop}%
\bibitem [{\citenamefont {Volovik}\ and\ \citenamefont
  {Zubkov}(2015)}]{VZ2015}%
  \BibitemOpen
  \bibfield  {author} {\bibinfo {author} {\bibfnamefont {G.~E.}\ \bibnamefont
  {Volovik}}\ and\ \bibinfo {author} {\bibfnamefont {M.~A.}\ \bibnamefont
  {Zubkov}},\ }\bibfield  {title} {\bibinfo {title} {Scalar excitation with
  leggett frequency in $^3$he-b and the 125 gev higgs particle in top quark
  condensation models as pseudo-goldstone bosons},\ }\href
  {https://doi.org/10.1103/PhysRevD.92.055004} {\bibfield  {journal} {\bibinfo
  {journal} {Phys. Rev. D}\ }\textbf {\bibinfo {volume} {92}},\ \bibinfo
  {pages} {055004} (\bibinfo {year} {2015})}\BibitemShut {NoStop}%
\bibitem [{\citenamefont {Volovik}\ and\ \citenamefont
  {Zubkov}(2014{\natexlab{b}})}]{VolovikZubkov2014}%
  \BibitemOpen
  \bibfield  {author} {\bibinfo {author} {\bibfnamefont {G.~E.}\ \bibnamefont
  {Volovik}}\ and\ \bibinfo {author} {\bibfnamefont {M.~A.}\ \bibnamefont
  {Zubkov}},\ }\bibfield  {title} {\bibinfo {title} {Higgs bosons in particle
  physics and in condensed matter},\ }\href@noop {} {\bibfield  {journal}
  {\bibinfo  {journal} {J. Low Temp. Phys.}\ }\textbf {\bibinfo {volume}
  {175}},\ \bibinfo {pages} {486} (\bibinfo {year}
  {2014}{\natexlab{b}})}\BibitemShut {NoStop}%
\bibitem [{\citenamefont {Volovik}\ and\ \citenamefont
  {Zubkov}(2013)}]{VolovikZubkovHiggs}%
  \BibitemOpen
  \bibfield  {author} {\bibinfo {author} {\bibfnamefont {G.~E.}\ \bibnamefont
  {Volovik}}\ and\ \bibinfo {author} {\bibfnamefont {M.~A.}\ \bibnamefont
  {Zubkov}},\ }\bibfield  {title} {\bibinfo {title} {The nambu sum rule and the
  relation between the masses of composite higgs bosons},\ }\href@noop {}
  {\bibfield  {journal} {\bibinfo  {journal} {Phys. Rev. D}\ }\textbf {\bibinfo
  {volume} {87}},\ \bibinfo {pages} {301} (\bibinfo {year} {2013})}\BibitemShut
  {NoStop}%
\bibitem [{\citenamefont {Zubkov}(2016)}]{Z2016_1}%
  \BibitemOpen
  \bibfield  {author} {\bibinfo {author} {\bibfnamefont {M.~A.}\ \bibnamefont
  {Zubkov}},\ }\bibfield  {title} {\bibinfo {title} {Wigner transformation,
  momentum space topology, and anomalous transport},\ }\href@noop {} {\bibfield
   {journal} {\bibinfo  {journal} {Annals Phys.}\ }\textbf {\bibinfo {volume}
  {373}},\ \bibinfo {pages} {298} (\bibinfo {year} {2016})},\ \Eprint
  {https://arxiv.org/abs/1603.03665} {arXiv:1603.03665} \BibitemShut {NoStop}%
\bibitem [{\citenamefont {Zhang}\ and\ \citenamefont
  {Zubkov}(2021)}]{ZZ2019_0}%
  \BibitemOpen
  \bibfield  {author} {\bibinfo {author} {\bibfnamefont {C.~X.}\ \bibnamefont
  {Zhang}}\ and\ \bibinfo {author} {\bibfnamefont {M.~A.}\ \bibnamefont
  {Zubkov}},\ }\bibfield  {title} {\bibinfo {title} {Influence of interactions
  on the anomalous quantum hall effect},\ }\href@noop {} {\bibfield  {journal}
  {\bibinfo  {journal} {J. Phys. A: Math. Theor.}\ }\textbf {\bibinfo {volume}
  {53}},\ \bibinfo {pages} {195002} (\bibinfo {year} {2021})}\BibitemShut
  {NoStop}%
\bibitem [{\citenamefont {Zhang}\ and\ \citenamefont {Zubkov}(2022)}]{ZZ2021}%
  \BibitemOpen
  \bibfield  {author} {\bibinfo {author} {\bibfnamefont {C.~X.}\ \bibnamefont
  {Zhang}}\ and\ \bibinfo {author} {\bibfnamefont {M.~A.}\ \bibnamefont
  {Zubkov}},\ }\bibfield  {title} {\bibinfo {title} {Influence of interactions
  on integer quantum hall effect},\ }\href@noop {} {\bibfield  {journal}
  {\bibinfo  {journal} {Annals of Physics}\ }\textbf {\bibinfo {volume}
  {444}},\ \bibinfo {pages} {169016} (\bibinfo {year} {2022})},\ \Eprint
  {https://arxiv.org/abs/2011.04030} {arXiv:2011.04030} \BibitemShut {NoStop}%
\bibitem [{\citenamefont {Zubkov}\ and\ \citenamefont {Wu}(2020)}]{ZW2019}%
  \BibitemOpen
  \bibfield  {author} {\bibinfo {author} {\bibfnamefont {M.~A.}\ \bibnamefont
  {Zubkov}}\ and\ \bibinfo {author} {\bibfnamefont {X.}~\bibnamefont {Wu}},\
  }\bibfield  {title} {\bibinfo {title} {Topological invariant in terms of the
  green functions for the quantum hall effect in the presence of varying
  magnetic field},\ }\href@noop {} {\bibfield  {journal} {\bibinfo  {journal}
  {Annals of Physics}\ }\textbf {\bibinfo {volume} {418}},\ \bibinfo {pages}
  {168179} (\bibinfo {year} {2020})}\BibitemShut {NoStop}%
\bibitem [{\citenamefont {M.Selch}\ \emph {et~al.}(2021)\citenamefont
  {M.Selch}, \citenamefont {Zubkov},\ and\ \citenamefont {Zhang}}]{SZZ2020}%
  \BibitemOpen
  \bibfield  {author} {\bibinfo {author} {\bibfnamefont {M.~S.}\ \bibnamefont
  {M.Selch}}, \bibinfo {author} {\bibfnamefont {M.~A.}\ \bibnamefont
  {Zubkov}},\ and\ \bibinfo {author} {\bibfnamefont {C.~X.}\ \bibnamefont
  {Zhang}},\ }\href@noop {} {\emph {\bibinfo {title} {Hall conductivity as the
  topological invariant in magnetic Brillouin zone}}},\ \bibinfo {type}
  {preprint}\ (\bibinfo {year} {2021})\ \Eprint
  {https://arxiv.org/abs/2112.03974} {arXiv:2112.03974} \BibitemShut {NoStop}%
\bibitem [{\citenamefont {Nieh}\ and\ \citenamefont {Yan}(1982)}]{NiehYan1982}%
  \BibitemOpen
  \bibfield  {author} {\bibinfo {author} {\bibfnamefont {H.}~\bibnamefont
  {Nieh}}\ and\ \bibinfo {author} {\bibfnamefont {M.}~\bibnamefont {Yan}},\
  }\bibfield  {title} {\bibinfo {title} {Quantized dirac field in curved
  riemann-cartan background. i. symmetry properties, green's function},\
  }\href@noop {} {\bibfield  {journal} {\bibinfo  {journal} {Annals of
  Physics}\ }\textbf {\bibinfo {volume} {138}},\ \bibinfo {pages} {237}
  (\bibinfo {year} {1982})}\BibitemShut {NoStop}%
\bibitem [{\citenamefont {Huang}\ \emph {et~al.}(2020)\citenamefont {Huang},
  \citenamefont {Han},\ and\ \citenamefont {Stone}}]{HuangHanStone2020}%
  \BibitemOpen
  \bibfield  {author} {\bibinfo {author} {\bibfnamefont {Z.-M.}\ \bibnamefont
  {Huang}}, \bibinfo {author} {\bibfnamefont {B.}~\bibnamefont {Han}},\ and\
  \bibinfo {author} {\bibfnamefont {M.}~\bibnamefont {Stone}},\ }\bibfield
  {title} {\bibinfo {title} {Nieh-yan anomaly: Torsional landau levels, central
  charge, and anomalous thermal hall effect},\ }\href@noop {} {\bibfield
  {journal} {\bibinfo  {journal} {Physical Review B}\ }\textbf {\bibinfo
  {volume} {101}},\ \bibinfo {pages} {125201} (\bibinfo {year}
  {2020})}\BibitemShut {NoStop}%
\bibitem [{\citenamefont {Khaidukov}\ and\ \citenamefont
  {Zubkov}(2018)}]{KhaidukovZubkov2018}%
  \BibitemOpen
  \bibfield  {author} {\bibinfo {author} {\bibfnamefont {Z.~V.}\ \bibnamefont
  {Khaidukov}}\ and\ \bibinfo {author} {\bibfnamefont {M.}~\bibnamefont
  {Zubkov}},\ }\bibfield  {title} {\bibinfo {title} {Chiral torsional effect},\
  }\href@noop {} {\bibfield  {journal} {\bibinfo  {journal} {JETP letters}\
  }\textbf {\bibinfo {volume} {108}},\ \bibinfo {pages} {670} (\bibinfo {year}
  {2018})}\BibitemShut {NoStop}%
\bibitem [{\citenamefont {Nissinen}\ and\ \citenamefont
  {Volovik}(2019)}]{NissinenVolovik2019}%
  \BibitemOpen
  \bibfield  {author} {\bibinfo {author} {\bibfnamefont {J.}~\bibnamefont
  {Nissinen}}\ and\ \bibinfo {author} {\bibfnamefont {G.~E.}\ \bibnamefont
  {Volovik}},\ }\bibfield  {title} {\bibinfo {title} {On thermal nieh-yan
  anomaly in topological weyl materials},\ }\href@noop {} {\bibfield  {journal}
  {\bibinfo  {journal} {JETP Letters}\ }\textbf {\bibinfo {volume} {110}},\
  \bibinfo {pages} {789} (\bibinfo {year} {2019})}\BibitemShut {NoStop}%
\bibitem [{\citenamefont {Nissinen}\ and\ \citenamefont
  {Volovik}(2020)}]{NissinenVolovik2020}%
  \BibitemOpen
  \bibfield  {author} {\bibinfo {author} {\bibfnamefont {J.}~\bibnamefont
  {Nissinen}}\ and\ \bibinfo {author} {\bibfnamefont {G.}~\bibnamefont
  {Volovik}},\ }\bibfield  {title} {\bibinfo {title} {Thermal nieh-yan anomaly
  in weyl superfluids},\ }\href@noop {} {\bibfield  {journal} {\bibinfo
  {journal} {Physical Review Research}\ }\textbf {\bibinfo {volume} {2}},\
  \bibinfo {pages} {033269} (\bibinfo {year} {2020})}\BibitemShut {NoStop}%
\bibitem [{\citenamefont {Nissinen}(2020)}]{Nissinen2020}%
  \BibitemOpen
  \bibfield  {author} {\bibinfo {author} {\bibfnamefont {J.}~\bibnamefont
  {Nissinen}},\ }\bibfield  {title} {\bibinfo {title} {Emergent spacetime and
  gravitational nieh-yan anomaly in chiral p+ ip weyl superfluids and
  superconductors},\ }\href@noop {} {\bibfield  {journal} {\bibinfo  {journal}
  {Physical review letters}\ }\textbf {\bibinfo {volume} {124}},\ \bibinfo
  {pages} {117002} (\bibinfo {year} {2020})}\BibitemShut {NoStop}%
\bibitem [{\citenamefont {Laurila}\ and\ \citenamefont
  {Nissinen}(2020)}]{LaurilaNissinen2020}%
  \BibitemOpen
  \bibfield  {author} {\bibinfo {author} {\bibfnamefont {S.}~\bibnamefont
  {Laurila}}\ and\ \bibinfo {author} {\bibfnamefont {J.}~\bibnamefont
  {Nissinen}},\ }\bibfield  {title} {\bibinfo {title} {Torsional landau levels
  and geometric anomalies in condensed matter weyl systems},\ }\href@noop {}
  {\bibfield  {journal} {\bibinfo  {journal} {Physical Review B}\ }\textbf
  {\bibinfo {volume} {102}},\ \bibinfo {pages} {235163} (\bibinfo {year}
  {2020})}\BibitemShut {NoStop}%
\bibitem [{\citenamefont {Zubarev}\ \emph {et~al.}(1979)\citenamefont
  {Zubarev}, \citenamefont {Prozorkevich},\ and\ \citenamefont
  {Smolyanskii}}]{Zubarev1979}%
  \BibitemOpen
  \bibfield  {author} {\bibinfo {author} {\bibfnamefont {D.~N.}\ \bibnamefont
  {Zubarev}}, \bibinfo {author} {\bibfnamefont {A.~V.}\ \bibnamefont
  {Prozorkevich}},\ and\ \bibinfo {author} {\bibfnamefont {S.~A.}\ \bibnamefont
  {Smolyanskii}},\ }\bibfield  {title} {\bibinfo {title} {``derivation of
  nonlinear generalized equations of quantum relativistic hydrodynamics.''},\
  }\href@noop {} {\bibfield  {journal} {\bibinfo  {journal} {Theoretical and
  Mathematical Physics}\ }\textbf {\bibinfo {volume} {40}},\ \bibinfo {pages}
  {821} (\bibinfo {year} {1979})}\BibitemShut {NoStop}%
\bibitem [{\citenamefont {Selch}\ \emph {et~al.}(2024)\citenamefont {Selch},
  \citenamefont {Abramchuk},\ and\ \citenamefont {Zubkov}}]{AZZ2023}%
  \BibitemOpen
  \bibfield  {author} {\bibinfo {author} {\bibfnamefont {M.}~\bibnamefont
  {Selch}}, \bibinfo {author} {\bibfnamefont {R.~A.}\ \bibnamefont
  {Abramchuk}},\ and\ \bibinfo {author} {\bibfnamefont {M.}~\bibnamefont
  {Zubkov}},\ }\bibfield  {title} {\bibinfo {title} {Effective lagrangian for
  the macroscopic motion of fermionic matter},\ }\href@noop {} {\bibfield
  {journal} {\bibinfo  {journal} {Physical Review D}\ }\textbf {\bibinfo
  {volume} {109}},\ \bibinfo {pages} {016003} (\bibinfo {year}
  {2024})}\BibitemShut {NoStop}%
\bibitem [{\citenamefont {Alonso}\ and\ \citenamefont
  {Popov}(1977{\natexlab{a}})}]{He3}%
  \BibitemOpen
  \bibfield  {author} {\bibinfo {author} {\bibfnamefont {V.}~\bibnamefont
  {Alonso}}\ and\ \bibinfo {author} {\bibfnamefont {V.~N.}\ \bibnamefont
  {Popov}},\ }\bibfield  {title} {\bibinfo {title} {Functional for the
  hydrodynamic action and the bose spectrum of superfluid fermi systems of the
  he3 type},\ }\href@noop {} {\bibfield  {journal} {\bibinfo  {journal} {Zh.
  Eksp. Teor. Fiz.}\ }\textbf {\bibinfo {volume} {73}},\ \bibinfo {pages}
  {1445} (\bibinfo {year} {1977}{\natexlab{a}})}\BibitemShut {NoStop}%
\bibitem [{\citenamefont {Brusov}\ and\ \citenamefont
  {Popov}(1980{\natexlab{a}})}]{He3gauss}%
  \BibitemOpen
  \bibfield  {author} {\bibinfo {author} {\bibfnamefont {P.~N.}\ \bibnamefont
  {Brusov}}\ and\ \bibinfo {author} {\bibfnamefont {V.~N.}\ \bibnamefont
  {Popov}},\ }\bibfield  {title} {\bibinfo {title} {Stability of the bose
  spectrum of superfluid systems of the he3 type},\ }\href@noop {} {\bibfield
  {journal} {\bibinfo  {journal} {Zh. Eksp. Teor. Fiz.}\ }\textbf {\bibinfo
  {volume} {78}},\ \bibinfo {pages} {234} (\bibinfo {year}
  {1980}{\natexlab{a}})}\BibitemShut {NoStop}%
\bibitem [{\citenamefont {Brusov}\ and\ \citenamefont
  {Popov}(1980{\natexlab{b}})}]{He3B}%
  \BibitemOpen
  \bibfield  {author} {\bibinfo {author} {\bibfnamefont {P.~N.}\ \bibnamefont
  {Brusov}}\ and\ \bibinfo {author} {\bibfnamefont {V.~N.}\ \bibnamefont
  {Popov}},\ }\bibfield  {title} {\bibinfo {title} {Nonphonon branches of the
  bose spectrum in the b phase of systems of the he3 type},\ }\href@noop {}
  {\bibfield  {journal} {\bibinfo  {journal} {JETP}\ }\textbf {\bibinfo
  {volume} {51}},\ \bibinfo {pages} {1217} (\bibinfo {year}
  {1980}{\natexlab{b}})}\BibitemShut {NoStop}%
\bibitem [{\citenamefont {Brusov}\ and\ \citenamefont
  {Popov}(1980{\natexlab{c}})}]{BrusovPopov1980}%
  \BibitemOpen
  \bibfield  {author} {\bibinfo {author} {\bibfnamefont {P.~N.}\ \bibnamefont
  {Brusov}}\ and\ \bibinfo {author} {\bibfnamefont {V.~N.}\ \bibnamefont
  {Popov}},\ }\bibfield  {title} {\bibinfo {title} {Zero-phonon branches of the
  bose spectrum in the a phase of a system of the he3 type},\ }\href@noop {}
  {\bibfield  {journal} {\bibinfo  {journal} {JETP}\ }\textbf {\bibinfo
  {volume} {52}},\ \bibinfo {pages} {945} (\bibinfo {year}
  {1980}{\natexlab{c}})}\BibitemShut {NoStop}%
\bibitem [{\citenamefont {Brusov}\ and\ \citenamefont
  {Brusov}(2010)}]{Brusovs}%
  \BibitemOpen
  \bibfield  {author} {\bibinfo {author} {\bibfnamefont {P.}~\bibnamefont
  {Brusov}}\ and\ \bibinfo {author} {\bibfnamefont {P.}~\bibnamefont
  {Brusov}},\ }\bibfield  {title} {\bibinfo {title} {Collective excitations in
  unconventional superconductors and superfluids},\ }\href@noop {} {\bibfield
  {journal} {\bibinfo  {journal} {World Scientific Publishing Co. Pte. Ltd}\ }
  (\bibinfo {year} {2010})}\BibitemShut {NoStop}%
\bibitem [{\citenamefont {Buzzegoli}\ \emph {et~al.}(2017)\citenamefont
  {Buzzegoli}, \citenamefont {Grossi},\ and\ \citenamefont
  {Becattini}}]{Buzzegoli2017cqy}%
  \BibitemOpen
  \bibfield  {author} {\bibinfo {author} {\bibfnamefont {M.}~\bibnamefont
  {Buzzegoli}}, \bibinfo {author} {\bibfnamefont {E.}~\bibnamefont {Grossi}},\
  and\ \bibinfo {author} {\bibfnamefont {F.}~\bibnamefont {Becattini}},\
  }\bibfield  {title} {\bibinfo {title} {{General equilibrium second-order
  hydrodynamic coefficients for free quantum fields}},\ }\href
  {https://doi.org/10.1007/JHEP10(2017)091} {\bibfield  {journal} {\bibinfo
  {journal} {JHEP}\ }\textbf {\bibinfo {volume} {10}},\ \bibinfo {pages}
  {091}},\ \bibinfo {note} {[Erratum: JHEP 07, 119 (2018)]},\ \Eprint
  {https://arxiv.org/abs/1704.02808} {arXiv:1704.02808 [hep-th]} \BibitemShut
  {NoStop}%
\bibitem [{\citenamefont {Becattini}\ \emph {et~al.}(2021)\citenamefont
  {Becattini}, \citenamefont {Buzzegoli},\ and\ \citenamefont
  {Palermo}}]{Becattini2020qol}%
  \BibitemOpen
  \bibfield  {author} {\bibinfo {author} {\bibfnamefont {F.}~\bibnamefont
  {Becattini}}, \bibinfo {author} {\bibfnamefont {M.}~\bibnamefont
  {Buzzegoli}},\ and\ \bibinfo {author} {\bibfnamefont {A.}~\bibnamefont
  {Palermo}},\ }\bibfield  {title} {\bibinfo {title} {{Exact equilibrium
  distributions in statistical quantum field theory with rotation and
  acceleration: scalar field}},\ }\href
  {https://doi.org/10.1007/JHEP02(2021)101} {\bibfield  {journal} {\bibinfo
  {journal} {JHEP}\ }\textbf {\bibinfo {volume} {02}},\ \bibinfo {pages}
  {101}},\ \Eprint {https://arxiv.org/abs/2007.08249} {arXiv:2007.08249
  [hep-th]} \BibitemShut {NoStop}%
\bibitem [{\citenamefont {Buzzegoli}(2021)}]{Buzzegoli2020ycf}%
  \BibitemOpen
  \bibfield  {author} {\bibinfo {author} {\bibfnamefont {M.}~\bibnamefont
  {Buzzegoli}},\ }\bibfield  {title} {\bibinfo {title} {{Thermodynamic
  equilibrium of massless fermions with vorticity, chirality and
  electromagnetic field}},\ }\href
  {https://doi.org/10.1007/978-3-030-71427-7_3} {\bibfield  {journal} {\bibinfo
   {journal} {Lect. Notes Phys.}\ }\textbf {\bibinfo {volume} {987}},\ \bibinfo
  {pages} {59} (\bibinfo {year} {2021})},\ \Eprint
  {https://arxiv.org/abs/2011.09974} {arXiv:2011.09974 [hep-th]} \BibitemShut
  {NoStop}%
\bibitem [{\citenamefont {Bravina}\ \emph {et~al.}(2021)\citenamefont
  {Bravina}, \citenamefont {Bugaev}, \citenamefont {Vitiuk},\ and\
  \citenamefont {Zabrodin}}]{Bravina2021arj}%
  \BibitemOpen
  \bibfield  {author} {\bibinfo {author} {\bibfnamefont {L.~V.}\ \bibnamefont
  {Bravina}}, \bibinfo {author} {\bibfnamefont {K.~A.}\ \bibnamefont {Bugaev}},
  \bibinfo {author} {\bibfnamefont {O.}~\bibnamefont {Vitiuk}},\ and\ \bibinfo
  {author} {\bibfnamefont {E.~E.}\ \bibnamefont {Zabrodin}},\ }\bibfield
  {title} {\bibinfo {title} {{Transport Model Approach to $\Lambda$ and $\bar
  \Lambda$ Polarization in Heavy-Ion Collisions}},\ }\href
  {https://doi.org/10.3390/sym13101852} {\bibfield  {journal} {\bibinfo
  {journal} {Symmetry}\ }\textbf {\bibinfo {volume} {13}},\ \bibinfo {pages}
  {1852} (\bibinfo {year} {2021})}\BibitemShut {NoStop}%
\bibitem [{\citenamefont {Selch}\ \emph {et~al.}(2025)\citenamefont {Selch},
  \citenamefont {Zubkov}, \citenamefont {Pramanik},\ and\ \citenamefont
  {Lewkowicz}}]{selch2025nonrenormalizationfractionalquantumhall}%
  \BibitemOpen
  \bibfield  {author} {\bibinfo {author} {\bibfnamefont {M.}~\bibnamefont
  {Selch}}, \bibinfo {author} {\bibfnamefont {M.~A.}\ \bibnamefont {Zubkov}},
  \bibinfo {author} {\bibfnamefont {S.}~\bibnamefont {Pramanik}},\ and\
  \bibinfo {author} {\bibfnamefont {M.}~\bibnamefont {Lewkowicz}},\ }\href
  {https://arxiv.org/abs/2502.04047} {\bibinfo {title} {Non-renormalization of
  the fractional quantum hall conductivity by interactions}} (\bibinfo {year}
  {2025}),\ \Eprint {https://arxiv.org/abs/2502.04047} {arXiv:2502.04047
  [cond-mat.mes-hall]} \BibitemShut {NoStop}%
\bibitem [{\citenamefont {Yamamoto}\ and\ \citenamefont
  {Hirono}(2013)}]{Yamamoto2013zwa}%
  \BibitemOpen
  \bibfield  {author} {\bibinfo {author} {\bibfnamefont {A.}~\bibnamefont
  {Yamamoto}}\ and\ \bibinfo {author} {\bibfnamefont {Y.}~\bibnamefont
  {Hirono}},\ }\bibfield  {title} {\bibinfo {title} {{Lattice QCD in rotating
  frames}},\ }\href {https://doi.org/10.1103/PhysRevLett.111.081601} {\bibfield
   {journal} {\bibinfo  {journal} {Phys. Rev. Lett.}\ }\textbf {\bibinfo
  {volume} {111}},\ \bibinfo {pages} {081601} (\bibinfo {year} {2013})},\
  \Eprint {https://arxiv.org/abs/1303.6292} {arXiv:1303.6292 [hep-lat]}
  \BibitemShut {NoStop}%
\bibitem [{\citenamefont {Braguta}\ \emph {et~al.}(2020)\citenamefont
  {Braguta}, \citenamefont {Kotov}, \citenamefont {Kuznedelev},\ and\
  \citenamefont {Roenko}}]{Braguta2020biu}%
  \BibitemOpen
  \bibfield  {author} {\bibinfo {author} {\bibfnamefont {V.~V.}\ \bibnamefont
  {Braguta}}, \bibinfo {author} {\bibfnamefont {A.~Y.}\ \bibnamefont {Kotov}},
  \bibinfo {author} {\bibfnamefont {D.~D.}\ \bibnamefont {Kuznedelev}},\ and\
  \bibinfo {author} {\bibfnamefont {A.~A.}\ \bibnamefont {Roenko}},\ }\bibfield
   {title} {\bibinfo {title} {{Study of the Confinement/Deconfinement Phase
  Transition in Rotating Lattice SU(3) Gluodynamics}},\ }\href
  {https://doi.org/10.31857/S1234567820130029} {\bibfield  {journal} {\bibinfo
  {journal} {Pisma Zh. Eksp. Teor. Fiz.}\ }\textbf {\bibinfo {volume} {112}},\
  \bibinfo {pages} {9} (\bibinfo {year} {2020})}\BibitemShut {NoStop}%
\bibitem [{\citenamefont {Braguta}\ \emph {et~al.}(2021)\citenamefont
  {Braguta}, \citenamefont {Kotov}, \citenamefont {Kuznedelev},\ and\
  \citenamefont {Roenko}}]{Braguta2021jgn}%
  \BibitemOpen
  \bibfield  {author} {\bibinfo {author} {\bibfnamefont {V.~V.}\ \bibnamefont
  {Braguta}}, \bibinfo {author} {\bibfnamefont {A.~Y.}\ \bibnamefont {Kotov}},
  \bibinfo {author} {\bibfnamefont {D.~D.}\ \bibnamefont {Kuznedelev}},\ and\
  \bibinfo {author} {\bibfnamefont {A.~A.}\ \bibnamefont {Roenko}},\ }\bibfield
   {title} {\bibinfo {title} {{Influence of relativistic rotation on the
  confinement-deconfinement transition in gluodynamics}},\ }\href
  {https://doi.org/10.1103/PhysRevD.103.094515} {\bibfield  {journal} {\bibinfo
   {journal} {Phys. Rev. D}\ }\textbf {\bibinfo {volume} {103}},\ \bibinfo
  {pages} {094515} (\bibinfo {year} {2021})},\ \Eprint
  {https://arxiv.org/abs/2102.05084} {arXiv:2102.05084 [hep-lat]} \BibitemShut
  {NoStop}%
\bibitem [{\citenamefont {Chernodub}\ \emph {et~al.}(2023)\citenamefont
  {Chernodub}, \citenamefont {Goy},\ and\ \citenamefont
  {Molochkov}}]{Chernodub2022veq}%
  \BibitemOpen
  \bibfield  {author} {\bibinfo {author} {\bibfnamefont {M.~N.}\ \bibnamefont
  {Chernodub}}, \bibinfo {author} {\bibfnamefont {V.~A.}\ \bibnamefont {Goy}},\
  and\ \bibinfo {author} {\bibfnamefont {A.~V.}\ \bibnamefont {Molochkov}},\
  }\bibfield  {title} {\bibinfo {title} {{Inhomogeneity of a rotating gluon
  plasma and the Tolman-Ehrenfest law in imaginary time: Lattice results for
  fast imaginary rotation}},\ }\href
  {https://doi.org/10.1103/PhysRevD.107.114502} {\bibfield  {journal} {\bibinfo
   {journal} {Phys. Rev. D}\ }\textbf {\bibinfo {volume} {107}},\ \bibinfo
  {pages} {114502} (\bibinfo {year} {2023})},\ \Eprint
  {https://arxiv.org/abs/2209.15534} {arXiv:2209.15534 [hep-lat]} \BibitemShut
  {NoStop}%
\bibitem [{\citenamefont {Braguta}\ \emph
  {et~al.}(2023{\natexlab{a}})\citenamefont {Braguta}, \citenamefont {Kudrov},
  \citenamefont {Roenko}, \citenamefont {Sychev},\ and\ \citenamefont
  {Chernodub}}]{Braguta2023kwl}%
  \BibitemOpen
  \bibfield  {author} {\bibinfo {author} {\bibfnamefont {V.~V.}\ \bibnamefont
  {Braguta}}, \bibinfo {author} {\bibfnamefont {I.~E.}\ \bibnamefont {Kudrov}},
  \bibinfo {author} {\bibfnamefont {A.~A.}\ \bibnamefont {Roenko}}, \bibinfo
  {author} {\bibfnamefont {D.~A.}\ \bibnamefont {Sychev}},\ and\ \bibinfo
  {author} {\bibfnamefont {M.~N.}\ \bibnamefont {Chernodub}},\ }\bibfield
  {title} {\bibinfo {title} {{Lattice Study of the Equation of State of a
  Rotating Gluon Plasma}},\ }\href {https://doi.org/10.1134/S0021364023600830}
  {\bibfield  {journal} {\bibinfo  {journal} {JETP Lett.}\ }\textbf {\bibinfo
  {volume} {117}},\ \bibinfo {pages} {639} (\bibinfo {year}
  {2023}{\natexlab{a}})}\BibitemShut {NoStop}%
\bibitem [{\citenamefont {Braguta}\ \emph
  {et~al.}(2023{\natexlab{b}})\citenamefont {Braguta}, \citenamefont
  {Chernodub}, \citenamefont {Roenko},\ and\ \citenamefont
  {Sychev}}]{Braguta2023yjn}%
  \BibitemOpen
  \bibfield  {author} {\bibinfo {author} {\bibfnamefont {V.~V.}\ \bibnamefont
  {Braguta}}, \bibinfo {author} {\bibfnamefont {M.~N.}\ \bibnamefont
  {Chernodub}}, \bibinfo {author} {\bibfnamefont {A.~A.}\ \bibnamefont
  {Roenko}},\ and\ \bibinfo {author} {\bibfnamefont {D.~A.}\ \bibnamefont
  {Sychev}},\ }\bibfield  {title} {\bibinfo {title} {{Negative moment of
  inertia and rotational instability of gluon plasma}},\ }\href@noop {} {\
  (\bibinfo {year} {2023}{\natexlab{b}})},\ \Eprint
  {https://arxiv.org/abs/2303.03147} {arXiv:2303.03147 [hep-lat]} \BibitemShut
  {NoStop}%
\bibitem [{\citenamefont {Morales-Tejera}\ and\ \citenamefont
  {Landsteiner}(2020)}]{Landsteiner2020xuv}%
  \BibitemOpen
  \bibfield  {author} {\bibinfo {author} {\bibfnamefont {S.}~\bibnamefont
  {Morales-Tejera}}\ and\ \bibinfo {author} {\bibfnamefont {K.}~\bibnamefont
  {Landsteiner}},\ }\bibfield  {title} {\bibinfo {title} {{Out of equilibrium
  chiral vortical effect in holography}},\ }\href
  {https://doi.org/10.1103/PhysRevD.102.106020} {\bibfield  {journal} {\bibinfo
   {journal} {Phys. Rev. D}\ }\textbf {\bibinfo {volume} {102}},\ \bibinfo
  {pages} {106020} (\bibinfo {year} {2020})},\ \Eprint
  {https://arxiv.org/abs/2006.16031} {arXiv:2006.16031 [hep-th]} \BibitemShut
  {NoStop}%
\bibitem [{\citenamefont {Prokhorov}\ \emph {et~al.}(2023)\citenamefont
  {Prokhorov}, \citenamefont {Teryaev},\ and\ \citenamefont
  {Zakharov}}]{Prokhorov2023dfg}%
  \BibitemOpen
  \bibfield  {author} {\bibinfo {author} {\bibfnamefont {G.~Y.}\ \bibnamefont
  {Prokhorov}}, \bibinfo {author} {\bibfnamefont {O.~V.}\ \bibnamefont
  {Teryaev}},\ and\ \bibinfo {author} {\bibfnamefont {V.~I.}\ \bibnamefont
  {Zakharov}},\ }\bibfield  {title} {\bibinfo {title} {{Novel phase transition
  at the Unruh temperature}},\ }\href@noop {} {\  (\bibinfo {year} {2023})},\
  \Eprint {https://arxiv.org/abs/2304.13151} {arXiv:2304.13151 [hep-th]}
  \BibitemShut {NoStop}%
\bibitem [{\citenamefont {Khakimov}\ \emph {et~al.}(2023)\citenamefont
  {Khakimov}, \citenamefont {Prokhorov}, \citenamefont {Teryaev},\ and\
  \citenamefont {Zakharov}}]{Khakimov2023emy}%
  \BibitemOpen
  \bibfield  {author} {\bibinfo {author} {\bibfnamefont {R.~V.}\ \bibnamefont
  {Khakimov}}, \bibinfo {author} {\bibfnamefont {G.~Y.}\ \bibnamefont
  {Prokhorov}}, \bibinfo {author} {\bibfnamefont {O.~V.}\ \bibnamefont
  {Teryaev}},\ and\ \bibinfo {author} {\bibfnamefont {V.~I.}\ \bibnamefont
  {Zakharov}},\ }\bibfield  {title} {\bibinfo {title} {{Unruh effect in curved
  space-time and hydrodynamics}},\ }\href@noop {} {\  (\bibinfo {year}
  {2023})},\ \Eprint {https://arxiv.org/abs/2308.08647} {arXiv:2308.08647
  [hep-th]} \BibitemShut {NoStop}%
\bibitem [{\citenamefont {Chernodub}\ \emph {et~al.}(2024)\citenamefont
  {Chernodub}, \citenamefont {Pochinok}, \citenamefont {Molochkov},
  \citenamefont {Stepanov},\ and\ \citenamefont {Goy}}]{Chernodubconf}%
  \BibitemOpen
  \bibfield  {author} {\bibinfo {author} {\bibfnamefont {M.}~\bibnamefont
  {Chernodub}}, \bibinfo {author} {\bibfnamefont {A.~S.}\ \bibnamefont
  {Pochinok}}, \bibinfo {author} {\bibfnamefont {A.}~\bibnamefont {Molochkov}},
  \bibinfo {author} {\bibfnamefont {D.~V.}\ \bibnamefont {Stepanov}},\ and\
  \bibinfo {author} {\bibfnamefont {V.}~\bibnamefont {Goy}},\ }\bibfield
  {title} {\bibinfo {title} {``gluon matter under weak acceleration: lattice
  results .'' poster at the 8th international conference on chirality,
  vorticity, and magnetic field in quantum matter, july 22 - 26, 2024, at the
  west university of timișoara, romania.},\ }\href@noop {} {\  (\bibinfo
  {year} {2024})}\BibitemShut {NoStop}%
\bibitem [{\citenamefont {Hayata}\ \emph {et~al.}(2015)\citenamefont {Hayata},
  \citenamefont {Hidaka}, \citenamefont {Noumi},\ and\ \citenamefont
  {Hongo}}]{H1}%
  \BibitemOpen
  \bibfield  {author} {\bibinfo {author} {\bibfnamefont {T.}~\bibnamefont
  {Hayata}}, \bibinfo {author} {\bibfnamefont {Y.}~\bibnamefont {Hidaka}},
  \bibinfo {author} {\bibfnamefont {T.}~\bibnamefont {Noumi}},\ and\ \bibinfo
  {author} {\bibfnamefont {M.}~\bibnamefont {Hongo}},\ }\bibfield  {title}
  {\bibinfo {title} {Relativistic hydrodynamics from quantum field theory on
  the basis of the generalized gibbs ensemble method},\ }\href
  {https://doi.org/10.1103/PhysRevD.92.065008} {\bibfield  {journal} {\bibinfo
  {journal} {Phys. Rev. D}\ }\textbf {\bibinfo {volume} {92}},\ \bibinfo
  {pages} {065008} (\bibinfo {year} {2015})}\BibitemShut {NoStop}%
\bibitem [{\citenamefont {Hongo}(2017)}]{H2}%
  \BibitemOpen
  \bibfield  {author} {\bibinfo {author} {\bibfnamefont {M.}~\bibnamefont
  {Hongo}},\ }\bibfield  {title} {\bibinfo {title} {Path-integral formula for
  local thermal equilibrium},\ }\bibfield  {journal} {\bibinfo  {journal}
  {Annals Phys.}\ }\textbf {\bibinfo {volume} {383}},\ \href
  {https://doi.org/10.1016/j.aop.2017.04.004} {10.1016/j.aop.2017.04.004}
  (\bibinfo {year} {2017})\BibitemShut {NoStop}%
\bibitem [{\citenamefont {Alonso}\ and\ \citenamefont
  {Popov}(1977{\natexlab{b}})}]{alonsopopov}%
  \BibitemOpen
  \bibfield  {author} {\bibinfo {author} {\bibfnamefont {V.}~\bibnamefont
  {Alonso}}\ and\ \bibinfo {author} {\bibfnamefont {V.~N.}\ \bibnamefont
  {Popov}},\ }\bibfield  {title} {\bibinfo {title} {``functional for the
  hydrodynamic action and the bose spectrum of superfluid fermi systemsof the
  he3 type.''},\ }\href@noop {} {\bibfield  {journal} {\bibinfo  {journal} {Zh.
  Eksp. Teor. Fiz.}\ }\textbf {\bibinfo {volume} {73}},\ \bibinfo {pages}
  {1445} (\bibinfo {year} {1977}{\natexlab{b}})}\BibitemShut {NoStop}%
\bibitem [{\citenamefont {Volovik}(1992)}]{volovik1993exotic}%
  \BibitemOpen
  \bibfield  {author} {\bibinfo {author} {\bibnamefont {Volovik}},\ }\href@noop
  {} {\bibinfo {title} {Exotic properties of superfluid he 3}} (\bibinfo {year}
  {1992})\BibitemShut {NoStop}%
\bibitem [{\citenamefont {Fujii}\ and\ \citenamefont
  {Nishida}(2018)}]{fujii2018low}%
  \BibitemOpen
  \bibfield  {author} {\bibinfo {author} {\bibfnamefont {K.}~\bibnamefont
  {Fujii}}\ and\ \bibinfo {author} {\bibfnamefont {Y.}~\bibnamefont
  {Nishida}},\ }\bibfield  {title} {\bibinfo {title} {Low-energy effective
  field theory of superfluid 3he-b and its gyromagnetic and hall responses},\
  }\href@noop {} {\bibfield  {journal} {\bibinfo  {journal} {Annals of
  Physics}\ }\textbf {\bibinfo {volume} {395}},\ \bibinfo {pages} {170}
  (\bibinfo {year} {2018})}\BibitemShut {NoStop}%
\bibitem [{\citenamefont {Furusawa}\ \emph {et~al.}(2021)\citenamefont
  {Furusawa}, \citenamefont {Fujii},\ and\ \citenamefont
  {Nishida}}]{furusawa2021hall}%
  \BibitemOpen
  \bibfield  {author} {\bibinfo {author} {\bibfnamefont {T.}~\bibnamefont
  {Furusawa}}, \bibinfo {author} {\bibfnamefont {K.}~\bibnamefont {Fujii}},\
  and\ \bibinfo {author} {\bibfnamefont {Y.}~\bibnamefont {Nishida}},\
  }\bibfield  {title} {\bibinfo {title} {Hall viscosity in the a phase of
  superfluid he 3},\ }\href@noop {} {\bibfield  {journal} {\bibinfo  {journal}
  {Physical Review B}\ }\textbf {\bibinfo {volume} {103}},\ \bibinfo {pages}
  {064506} (\bibinfo {year} {2021})}\BibitemShut {NoStop}%
\end{thebibliography}%

\end{document}